# On the Construction of Low-Energy Cislunar and Trans-lunar Transfers Based on the Libration Points

Ming Xu Yan Wei Shijie Xu

**Abstract**: There exist cislunar and trans-lunar libration points near the Moon, which are referred as the $LL_1$ and $LL_2$ points respectively and can generate the different types of low-energy trajectories transferring from Earth to Moon. The time-dependent analytic model including the gravitational forces from the Sun, Earth and Moon is employed to investigate the energy-minimal and practical transfer trajectories. However, different from the circular restricted three-body problem, the equivalent gravitational equilibria are defined according to the geometry of instantaneous Hill's boundary due to the gravitational perturbation from the Sun. The relationship between the altitudes of periapsis and eccentricities is achieved from the Poincaré mapping for all the lunar captured trajectories, which presents the statistical feature of the fuel cost and captured orbital elements rather than generating a specified Moon-captured segment. The minimum energy required by the captured trajectory on a lunar circular orbit is deduced in the spatial bi-circular model. It is presented that the asymptotical behaviors of invariant manifolds approaching to/from the libration points or halo orbits are destroyed by the solar perturbation. In fact, the energy-minimal cislunar transfer trajectory is acquired by transiting $LL_1$ point, while the energy-minimal trans-lunar transfer trajectory is obtained by transiting $LL_2$ point. Finally, the transfer opportunities for the practical trajectories escaped from the Earth and captured by the Moon are yielded by transiting halo orbits near $LL_1$ and $LL_2$ points, which can be used to generate the whole trajectories.

M. Xu (✉) Y. Wei S. Xu
Department of Aerospace Engineering
School of Astronautics
Beihang University, Beijing 100191, China
e-mail: xuming@buaa.edu.cn

## 1. Introduction

Previous researches on cislunar transfer trajectories from the Earth to Moon in the context of two-body dynamics reached the conclusion that the spacecraft has to be accelerated up to the hyperbolic velocity so as to escape the Earth's gravitational force; while some recent researches from the viewpoint of the restricted circular three-body problem (abbr. CR3BP) showed that the hyperbolic velocity is not the necessary condition for the cislunar transfer (**Koon et al. 2007**). Compared with the Hohmann transfer, the ballistic trajectory captured trajectory known as one type of low-energy transfer trajectories (**Xu and Xu 2009**), which is obtained within the context of CR3BP, has lower fuel consumption but longer transfer duration.

Conley studied the local dynamical behavior of planar CR3BP near the collinear libration point and classified all the trajectories into four different types as: periodical orbit (named as Lyapunov orbit), stable/unstable manifolds of periodic orbit, transiting and non-transiting trajectories (**Conley 1968**). It is concluded from Conley's work that the invariant manifolds of periodic orbits will separate transiting and non-transiting trajectories, and only the transiting ones can be employed to generate the low-energy cislunar transfer trajectories.

McGehee investigated the global dynamical behavior of CR3BP and achieved the similar results, i.e., the stable and unstable manifolds of Lyapunov orbit form a 2-dimensional hyper-surface in the 3-dimensional Euclidean space which may play a significant role in understanding the transiting trajectories (**McGehee 1969**). Based on the preliminary work of Conley and McGehee, Marsden and Ross extended and thoroughly investigated the dynamical structure near the libration point, and denoted the invariant manifolds as *Conley–McGehee tubes* (abbr. C-M tube) in order to memorize their contributions (**Marsden and Ross 2006**). Yamato demonstrated that most of the tubes are distorted but few of them are preserved by small perturbations from the perturbed gravitation of the third celestial body (**Yamato 2003**).

Several scholars were devoting to the topic on some transiting trajectories near $LL_1$ point, since Conley had achieved the low-energy cislunar trajectories from the viewpoint of $LL_1$ point (**Conley 1969**). Bolt and Meiss obtained a low-energy cislunar transfer trajectory by the shooting method developed in chaotic dynamics with the total fuel consumption of $\Delta V$ =750m/s and the flight duration of $\Delta t$ =748 days (**Bolt and Meiss 1995**). Schroer and Ott improved the shooting method to achieve the transfer trajectory with similar fuel consumption but cutting off half of the transfer time ($\Delta t$ =377.5 days) (**Schroer and Ott 1997**). Macau gained a transfer trajectory with a little more fuel consumption but much less transfer time than Schroer and Ott, i.e., $\Delta V$ =767m/s and $\Delta t$ =284 days (**Macau 1998**). Ross and Koon optimized the transfer time and fuel consumption to yield the better result, i.e., $\Delta V$ =860m/s and $\Delta t$ =65 days (**Ross and Koon 2003**). Topputo and Vasile employed the Lambert equation in CR3BP to solve the two-point boundary problems and obtained the similar result with Ross and Koon (**Topputo and Vasile 2005**). Xu et al. investigated the occurrence condition for low-energy transfer and discovered that the transiting trajectories near $LL_1$ point are preferred to generate the low-thrust cislunar trajectory (**Xu et al. 2012**).

On the other hands, Belbruno et al. raised a new type of trans-lunar trajectories by the numerical method, which has a great application in rescuing Japanese lunar spacecraft "Hiten" in 1991 and then is referred as the weak stability boundary (abbr. WSB) trajectory (**Belbruno and Miller 1993**; **Belbruno 2004**). The WSB trajectory is considered as a significant contribution to celestial mechanics, and more analytic or semi-analytic investigations were implemented on this theory by **Circi and Teofilatto** (**2001**), **Yagasaki** (**2004**), **Parker and Lo** (**2005**), and **García and Gómez** (**2007**).

Koon et al. investigated the long-term evolutions of C-M tubes under the gravitational perturbation from the Sun, and divided the restricted four body problem into two different CR3BPs, i.e., the Sun-Earth/Moon system and the Earth-Moon system (**Koon et al. 2001**). A magic result was achieved that a Belbruno's WSB trajectory can be generated from the stable manifolds near $EL_1$ (or $EL_2$) point and the unstable manifolds near the trans-lunar $LL_2$ point, with the assist of the numerical tool of Poincaré mapping.

Different from the above researches focusing on only one specified Earth-to-Moon transfer trajectories, a systematic discussion on both cislunar and trans-lunar trajectories are implemented in the context of restricted four-body dynamics in this paper. The statistical features of the fuel cost and captured orbital elements, like altitude of periapsis and eccentricity, are investigated by the tool of Poincaré mapping rather than a specified Moon-captured segment. Compared to CR3BP and Hill's model, both the cislunar and trans-lunar trajectories with the minimum energy are deduced in a spatial analytical four-body model including the gravitational forces from the Sun, Earth and Moon. It is presented that the asymptotical behaviors of invariant manifolds approaching to/from libration points or halo orbits are destroyed in the time-independent model. The energy-minimal and practical cislunar transfer trajectories are acquired by transiting $LL_1$ point and halo orbits near the point respectively; however, the energy-minimal and practical trans-lunar transfers are obtained by transiting $LL_2$ point and halo orbits near the point. Furthermore, the transfer opportunities for the practical trajectories escaped from the Earth and captured by the Moon are yielded by transiting halo orbits near $LL_1$ and $LL_2$ points, which can be used to generate the whole trajectories.

## 2. Lunar Capturing Trajectories in Spatial Bi-Circular Model

Compared with the Hohmann direct transfer employed by Apollo (NASA) and Chang'E (China) missions, the low-energy WSB transfer requires more fuels in the accelerating maneuvers, and then much less fuel cost in the decelerating maneuvers, which will make the WSB type of lunar transfer trajectories more economic than the Hohmann type. Therefore, **Belbruno and Miller** (**1993**), and **García and Gómez** (**2007**) proposed the concept of lunar temporary capturing trajectories to measure the opportunity of a spacecraft to transferring from the Earth to Moon, which owns the somewhat higher energy than the libration point $LL_1$ or $LL_2$. When the spacecraft on the Hohmann trajectory arrives at the Moon, its flight velocity is hyperbolical and its osculating eccentricity is greater than 1, hence the spacecraft owns much higher energy than $LL_2$ point. However, the low-energy trajectories are elliptical since their osculating eccentricities are less than 1 during the flight, so that the spacecraft will keep orbiting the Earth with several loops before transiting the libation point, and also keep orbiting the Moon after transiting the point. Thus, the fuel cost of the lunar temporary capture turning into temporary capture is smaller than the Hohmann transfer.

An analytic spatial bi-circular model (abbr. SBCM) including the gravitational forces from the Sun, Earth and Moon is developed in this section, and then a systematical discussion on Moon-captured energy in this model is implemented by the tool of numerical Poincare mapping; however, no specific trajectory is referred in this section.

### 2.1 The definition of SBCM

The SBCM originates from the planar bi-circular model developed by **Koon et al.** (**2001**) and the quasi bi-circular model by **Andreu** (**1999**); specially, the SBCM shows significant improvements in the inclination between the ecliptic and lunar planes. Compared with the three models referred above, the SBCM has the following assumptions: *i*) The Earth and Moon act as different simple gravitational points, and move around their barycenter in Kepler circular motions with their eccentricities ignored; *ii*) The barycenter of the Earth-Moon system stays circumsolar in the ecliptic plane with its eccentricity ignored; *iii*) The inclination of the lunar plane relative to the ecliptic plane is considered with an average angle of 5°9'.

In order to reduce the computational work in Kepler circular motions under the Sun-Earth/Moon and Earth-Moon systems, three different coordinates are introduced in this paper, as shown in Fig.1.

The inertial $\mathbf{I}_{\text{S-E/M}}$ frame with its components ($X$, $Y$, $Z$) is defined as following: the origin is fixed at the barycenter of the Sun-Earth/Moon system, and the axis $X$ is along the intersection of the ecliptic and lunar planes, which follows an inertial direction in the system, and the axis $Z$ is perpendicular to the lunar plane and along the revolution axis of the Earth-Moon system, and the axis $Y$ is determined by the right-hand-side rule. Inheriting from $\mathbf{I}_{\text{S-E/M}}$, a new inertial frame $\mathbf{I}_{\text{E-M}}$ has the same definition of the three axes, but fixes its origin at the barycenter of the Earth-Moon system. The syzygy $\mathbf{S}_{\text{S-E/M}}$ frame with its components ($\xi$, $\eta$, $\zeta$) is defined as following: the origin is fixed at the barycenter of the Sun-Earth/Moon system, and the axis $\xi$ points from the Sun to the barycenter of the Earth and Moon, the axis $\zeta$ is perpendicular to the ecliptic orbital plane, and the axis $\eta$ is determined by the right-hand-side rule. The syzygy $\mathbf{S}_{\text{E-M}}$ frame with its components ($x$, $y$, $z$) is defined as following: the origin is fixed at the barycenter of the Earth-Moon system, and the axis $x$ points from the Earth to the Moon, and the axis $z$ is perpendicular to the lunar plane and along the revolution axis of the Earth-Moon system, and the axis $y$ is determined by the right-hand-side rule.

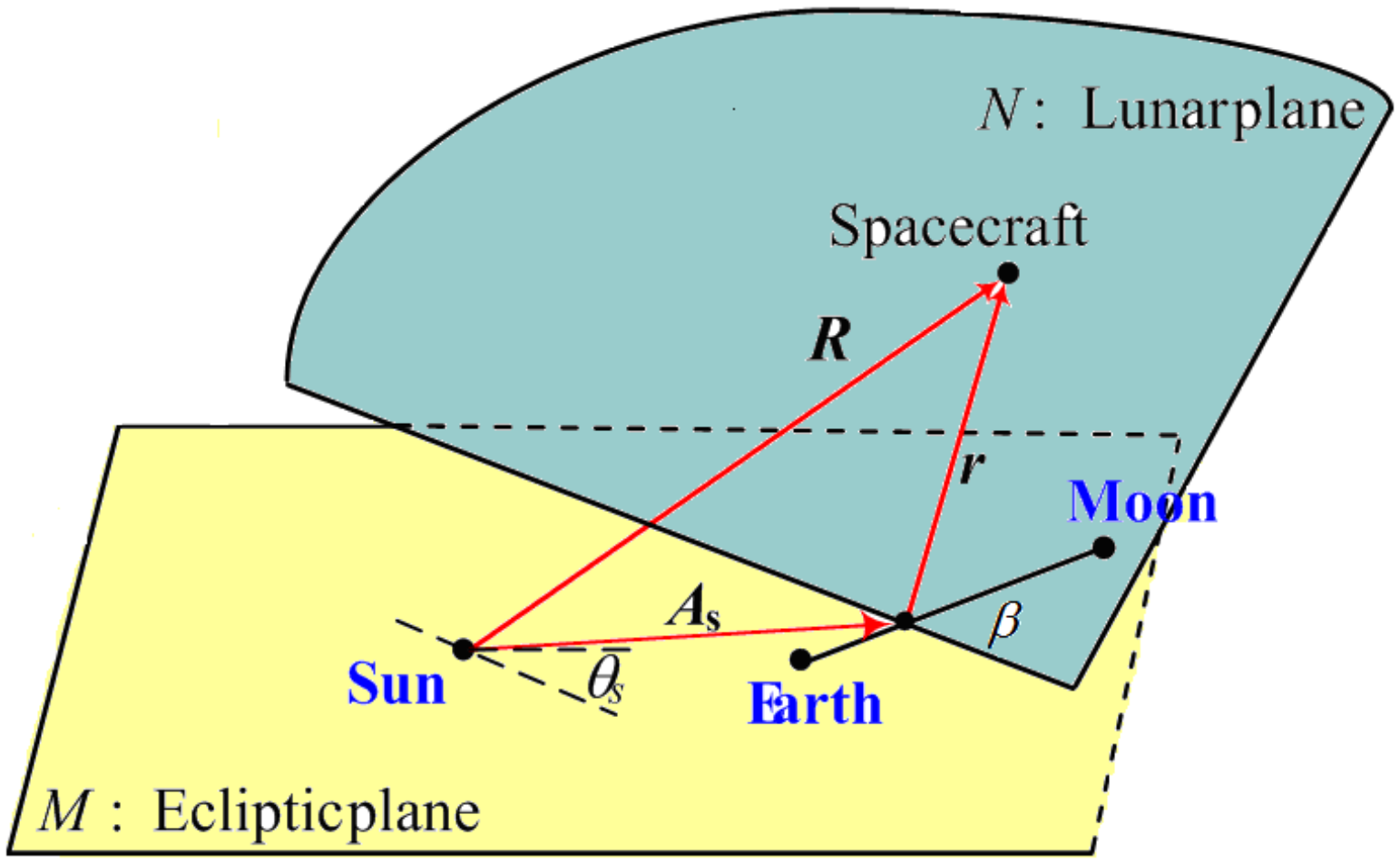

**Fig.1 The geometrical view of the SBCM model**: the inclination of the lunar plane relative to the ecliptic plane is considered with an average angle of 5°9'; the solar phasic angle $\beta$ measures the included angle between the line from the Earth to Moon and the intersecting line of the ecliptic and lunar planes; the lunar phasic angle $\theta_s$ measures the included angle between the line from the Sun to the barycenter of the Earth-Moon system and the intersecting line of the ecliptic and lunar planes; the ecliptic plane is painted in yellow color, while the lunar plane is painted in green color.

The equations derived in this paper can be normalized by means of the characteristic length, time and mass, as following:

$$\begin{cases} [L] = L_{E-M}\text{, average distance between the Earth and Moon} \\ [M] = m_E + m_M\text{, total mass in the Earth - Moon system} \\ [T] = \left[R^3_{E-M} / G(m_E + m_M)\right]^{1/2} \end{cases}$$

where $m_E$ and $m_M$ are the mass of the Earth and Moon, respectively; $L_{E\text{-}M}$ is the average distance between the Earth and Moon; $G$ is the universal gravitation constant.

$\boldsymbol{R}_I = [X \quad Y \quad Z]^T$, $\Re = [\xi \quad \eta \quad \zeta]^T$ and $\boldsymbol{r} = [x \quad y \quad z]^T$ are defined as the position vector of the spacecraft in the rotating frames $\mathbf{I}_{\mathbf{S\text{-}E/M}}$, $\mathbf{S}_{\mathbf{S\text{-}E/M}}$ and $\mathbf{S}_{\mathbf{E\text{-}M}}$, respectively. Therefore, the position vector from the Sun to the origin of the barycenter in the frame $\mathbf{I}_{\mathbf{S\text{-}E/M}}$ can be expressed as $\boldsymbol{A_S} = (1-\mu_S)a_s[\cos\theta_S \quad \sin\theta_S \quad 0]^T$, where $a_s$ (=388.81114 in the length unit normalization mentioned above) is the average distance between the heliocenter and the barycenter of the Earth-Moon system, and $\mu_s$ (=3.040357143×$10^{-6}$) is the mass ratio of the Earth-Moon system with respect to the full Sun-Earth-Moon system. $i$ (=5°9') is the inclination between the ecliptic and lunar planes. $\theta_s$ is defined as the lunar phasic angle measured between the line from the Sun to the barycenter of the Earth-Moon system and the intersecting line of the ecliptic and lunar planes, and $\beta$ is defined as the solar phasic angle measured between the line from the Earth to Moon and the intersecting line. In this paper, the initial lunar angle $\theta_{s0}$ is set as $0^0$ at the moment $t_0$=0, while the initial value of the solar angle $\beta_0$ is selected as the time variable to investigate Earth-to-Moon transfers in the time-dependent SBCM model in the following sections.

According to the defined coordinate systems and SBCM assumptions, the required relationship between the spacecraft's position vector $\boldsymbol{r}$, $\Re$ and $\boldsymbol{R}_I$ are listed as:

$$\Re = \boldsymbol{R}_z(\theta_S)\boldsymbol{R}_I \tag{1}$$

$$\boldsymbol{R_I} = \boldsymbol{R}_x(-i)\boldsymbol{R}_z(-\beta)\boldsymbol{r} + \boldsymbol{A_S} \tag{2}$$

where $\boldsymbol{R}_z(\theta)$ and $\boldsymbol{R}_x(\theta)$ are the elementary transformation matrixes around the $Z$ (or $z$) and $X$ (or $x$) axes respectively. Thus, the Newtonian dynamics in the unit normalization is formulized as following:

$$\ddot{\boldsymbol{R}}_I = -(1-\mu)\cdot\frac{\boldsymbol{r}-\boldsymbol{r}_E}{\|\boldsymbol{r}-\boldsymbol{r}_E\|^3} - \mu\cdot\frac{\boldsymbol{r}-\boldsymbol{r}_M}{\|\boldsymbol{r}-\boldsymbol{r}_M\|^3} - m_S\frac{\Re-\Re_S}{\|\Re-\Re_S\|^3} \tag{3}$$

where $\mu$ (=0.0121516) is the mass ratio of the Moon with respective to the Earth-Moon system, and $\Re_S = a_S\cdot[-\mu_S \quad 0 \quad 0]^T$ is the position vector of the Sun in the frame $\mathbf{S}_{\mathbf{S\text{-}E/M}}$. The kinematics formulized by Eqs. (1) and (2) can be used to deduce the dynamical equation, as

$$\begin{bmatrix}\ddot{x}\\ \ddot{y}\\ \ddot{z}\end{bmatrix}=-2\begin{bmatrix}-\dot{y}\\ \dot{x}\\ 0\end{bmatrix}+\begin{bmatrix}x\\ y\\ 0\end{bmatrix}-(1-\mu)\cdot\frac{\boldsymbol{r}-\boldsymbol{r}_E}{\|\boldsymbol{r}-\boldsymbol{r}_E\|^3}-\mu\cdot\frac{\boldsymbol{r}-\boldsymbol{r}_M}{\|\boldsymbol{r}-\boldsymbol{r}_M\|^3}+\omega_S^2\boldsymbol{R}_z(\beta)\boldsymbol{R}_x(i)\boldsymbol{A}_S-\boldsymbol{R}_z(\beta)\boldsymbol{R}_x(i)\boldsymbol{R}_z(-\theta_S)m_S\frac{\Re-\Re_S}{\|\Re-\Re_S\|^3} \tag{4}$$

where $\omega_s$ (=0.0748 in the time unit normalization) is the angular velocity of the Earth-Moon system with respect to the inertial reference system, and $m_s$ (=328900.54) is the Sun's mass in the mass unit normalization, and the position vectors of the Earth and Moon in $\mathbf{S_{E/M}}$ can be expressed as $\boldsymbol{r}_E=[-\mu \quad 0 \quad 0]^T$ and $\boldsymbol{r}_M=[1-\mu \quad 0 \quad 0]^T$.

Both of the CR3BP and SBCM models are classified as the conservative Hamiltonian systems without any external forces. Thus, the Newtonian dynamics can be deduced from the Hamiltonian function $H_1$ equivalently as

$$\begin{cases}\dot{\boldsymbol{r}}=\dfrac{\partial H_1}{\partial \boldsymbol{p}}\\ \dot{\boldsymbol{p}}=-\dfrac{\partial H_1}{\partial \boldsymbol{r}}\end{cases} \tag{5}$$

where $\boldsymbol{p}=[p_x \quad p_y \quad p_z]^T$ is the generalized momentum, defined by the position and velocity vectors $\boldsymbol{r}$ and $\dot{\boldsymbol{r}}$ as

$$\begin{cases}p_x=\dot{x}-y\\ p_y=\dot{y}+x\\ p_z=\dot{z}\end{cases}. \tag{6}$$

Thus the Hamiltonian function $H_1$ can be resolved from Eqs. (4) and (5), as

$$H_1=H_0-\frac{m_S}{\|\Re-\Re_S\|}-\omega_S^2\cdot\boldsymbol{A}_S^T\cdot\boldsymbol{R}_x(-i)\cdot\boldsymbol{R}_z(-\beta)\cdot\boldsymbol{r}+\frac{m_S}{a_S} \tag{7}$$

where $H_0$ is the Hamiltonian function modeling the dynamics of CR3BP, and the other terms are considered as the perturbation from the solar gravity, and the term $\|\Re-\Re_S\|$ can be reproduced as:

$$\|\Re-\Re_S\|=\left\|a_S[1 \quad 0 \quad 0]^T+\boldsymbol{R}_z(\theta_s)\boldsymbol{R}_x(-i)\boldsymbol{R}_z(-\beta)\boldsymbol{r}\right\|=\sqrt{a_S^2+\boldsymbol{r}^T\boldsymbol{r}+2a_S\cdot r_C} \tag{8}$$

where $r_c$ is defined as $r_C=[\cos\theta_S \quad \sin\theta_S \quad 0]\cdot\boldsymbol{R}_x(-i)\boldsymbol{R}_z(-\beta)\boldsymbol{r}$, which has the same order of magnitude as $\|\boldsymbol{r}\|$, but is smaller than $\|\boldsymbol{r}\|$ (i.e., $r_C\le\|\boldsymbol{r}\|$).

Moreover, $H_0$, which can be found in textbooks, has the general form:

$$H_0 = \frac{1}{2}\left(p_x^2 + p_y^2 + p_z^2\right) - xp_y + yp_x - \frac{1-\mu}{\left\|\boldsymbol{r} - \boldsymbol{r}_E\right\|} - \frac{\mu}{\left\|\boldsymbol{r} - \boldsymbol{r}_M\right\|}. \tag{9}$$

The solar gravity brings periodic perturbation into the dynamics of CR3BP, which can be characterized by the difference between the two Hamiltonian functions. For a spacecraft flying inside the Earth-Moon system with its distance $\|\boldsymbol{r}\|$ from the system barycenter much shorter than $a_s$, i.e., $\|\boldsymbol{r}\| << a_s$, the difference $\Delta H$ has the following expression as

$$\Delta H = m_S\left[\frac{1}{a_S} - \frac{1}{\left\|\Re - \Re_S\right\|}\right] - \omega_S^2 \boldsymbol{A}_S^T \boldsymbol{R}_x(-i)\boldsymbol{R}_z(-\beta)\boldsymbol{r}. \tag{10}$$

where the second term can be simplified as

$$\frac{1}{\left\|\Re - \Re_S\right\|} = \frac{1}{a_S}\left(1 + 2\frac{r_C}{a_S} + \frac{\|\boldsymbol{r}\|^2}{a_S^2}\right)^{-\frac{1}{2}} = \frac{1}{a_S}\left[1 - \frac{\|\boldsymbol{r}\|^2}{2a_S^2} - \frac{r_C}{a_S} + \frac{3}{2}\cdot\frac{r_C^2}{a_S^2} + O\left(\frac{\|\boldsymbol{r}\|^3}{a_S^3}\right)\right]. \tag{11}$$

and the third term can be simplified by the Kepler's third law, as

$$\omega_S^2 \cdot a_S = \frac{m_S}{a_S^2}. \tag{12}$$

Therefore, the difference $\Delta H$ can be refined as

$$\Delta H = \frac{m_S}{a_S^2}\left[\frac{\|\boldsymbol{r}\|^2}{2a_S} - \frac{3}{2}\cdot\frac{r_C^2}{a_S^3} + O\left(\frac{\|\boldsymbol{r}\|^3}{a_S^4}\right)\right] = \frac{m_S}{a_S^2} O\left(\frac{\|\boldsymbol{r}\|^2}{a_S}\right). \tag{13}$$

For the trajectories inside the Earth-Moon system discussed in this paper, the magnitude of $\boldsymbol{r}$ is close to 1 according to the length unit normalization, i.e., $\|\boldsymbol{r}\| \approx 1$. Hence, for the halo orbits employed in this paper, the following fact can be obtained from their Hamiltonian values of $H_0$ ($\approx$-1.6 presented in Section 2.2) that $\Delta H/H_0$ is of $10^{-3}$ order of magnitude, which can be considered as a small perturbation onto the Hamiltonian system $H_0$.

## 2.2 The definition of equivalent libration points in SBCM

The SBCM dynamics is time-dependent due to the periodic perturbation from the solar gravitation, compared to the time-independent CR3BP dynamics (**Koon et al. 2007**; **Belbruno 2004**). Consequently, there are no equilibrium point existing in this gravitational fields. Nevertheless, the gravitational equivalent equilibria will be defined in this section according to the geometry of instantaneous Hill's boundary.

For the trajectories flying inside the Earth-Moon system, the instantaneous Hill's region defined by the Hamiltonian function $H_1$ has the similar geometry with the constant Hamiltonian function $H_0$. From the geometrical point of view, $LL_1$ point is essentially the critical point connecting the two gravitational fields around the Earth and Moon, while $LL_2$ point is the critical point connecting the interior and the forbidden regions. Therefore, the equivalent cislunar $LL_1$ point and trans-lunar $LL_2$ point are defined respectively as the geometrical critical points of the instantaneous Hill's boundary for a specified solar phasic angle $\beta$, marked as $\begin{bmatrix} x_{LL_1} & 0 & 0 \end{bmatrix}^T$ and $\begin{bmatrix} x_{LL_2} & 0 & 0 \end{bmatrix}^T$ respectively.

Mathematically, the procedure to compute $x_{LL_1}$ and $x_{LL_2}$ is demonstrated as follows. The Hamiltonian function is an integral of motion written in position and velocity form formulized by Eq. (7), and its potential function with only the position term is formulized as:

$$U_1 = H_1\big|_{\dot{\boldsymbol{r}}=\boldsymbol{0}} = -\frac{1}{2}\left(x^2 + y^2\right) - \frac{1-\mu}{\left\|\boldsymbol{r} - \boldsymbol{r}_E\right\|} - \frac{\mu}{\left\|\boldsymbol{r} - \boldsymbol{r}_M\right\|} - \frac{m_S}{\left\|\Re - \Re_S\right\|} - \omega_S^2 \cdot \boldsymbol{A}_S^T \cdot \boldsymbol{R}_x\left(-i\right) \cdot \boldsymbol{R}_z\left(-\beta\right) \cdot \boldsymbol{r} + \frac{m_S}{a_S} \quad (14)$$

Moreover, the Hill's boundary is dominated by the potential function with its velocity $\dot{\boldsymbol{r}} = \boldsymbol{0}$, known also as the zero velocity surface. According to the definition of equivalent equilibrium mentioned above, the locations of $LL_1$ and $LL_2$ points can be solved from the partial derivative of $U_1$ with respect to the $x$ component, i.e., $\frac{\partial U_1}{\partial x}$.

The geometry of time-dependent Hill's boundaries, the locations of equivalent equilibria and their Hamiltonian values are respectively shown in Fig.2, 3 and 4, where the equivalent cislunar $LL_1$ or $LL_2$ point is denoted as SBCM-$LL_1$ or SBCM-$LL_2$, compared with CR3BP-$LL_1$ or CR3BP-$LL_2$ in this unperturbed model. Due to the solar perturbation, the locations and the Hamiltonian values of equivalent equilibria are depending on $\beta$, i.e., $x_{LL_i} = x_{LL_i}(\beta), i = 1,2$ and $H_1^{LL_i} = H_1^{LL_i}(\beta), i = 1,2$ .Thus, the initial lunar phasic angle at the epoch time ($t$=0) is set as $\theta_{s0}=0^0$ to produce these figures, and the solar phasic angle $\beta$ ranges from $0^0$ to $360^0$.

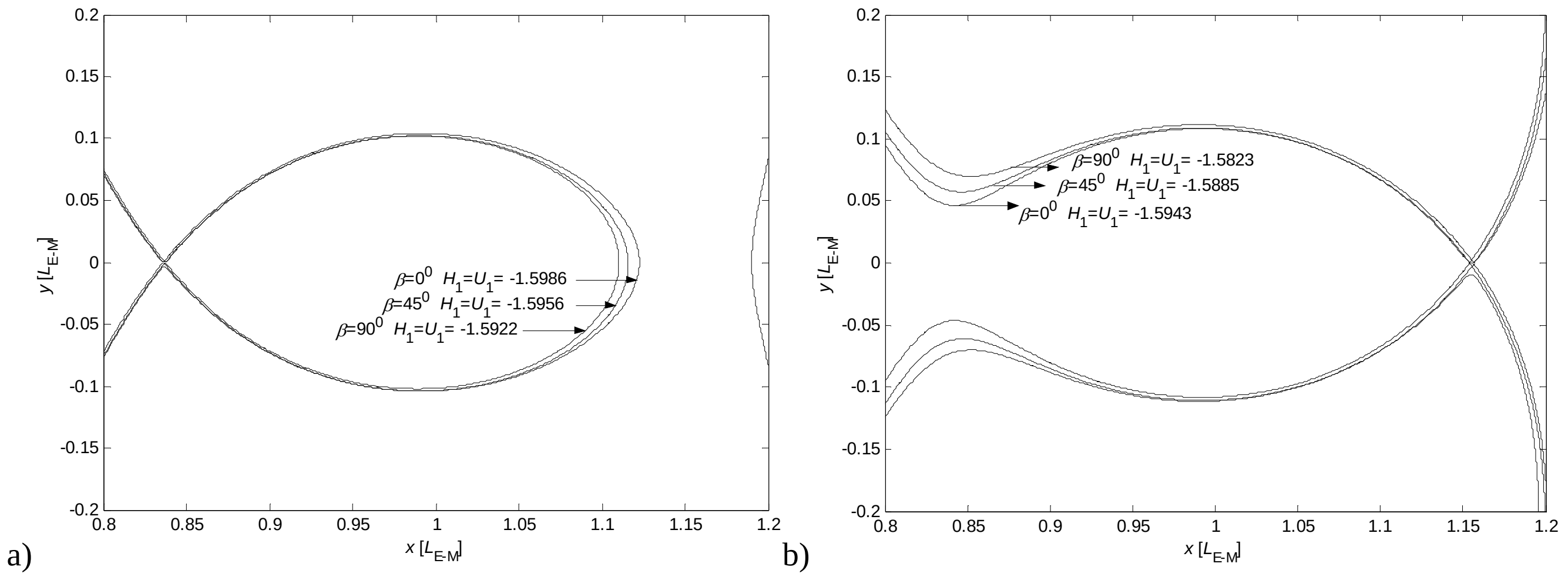

**Fig.2 Time-dependent Hill's boundaries and equivalent equilibria**: a) the equivalent $LL_1$ point and its Hill's boundary; b) the equivalent $LL_2$ point and its Hill's boundary; the initial lunar phasic angle at the epoch time ($t$=0) is $\theta_{s0}=0^0$.

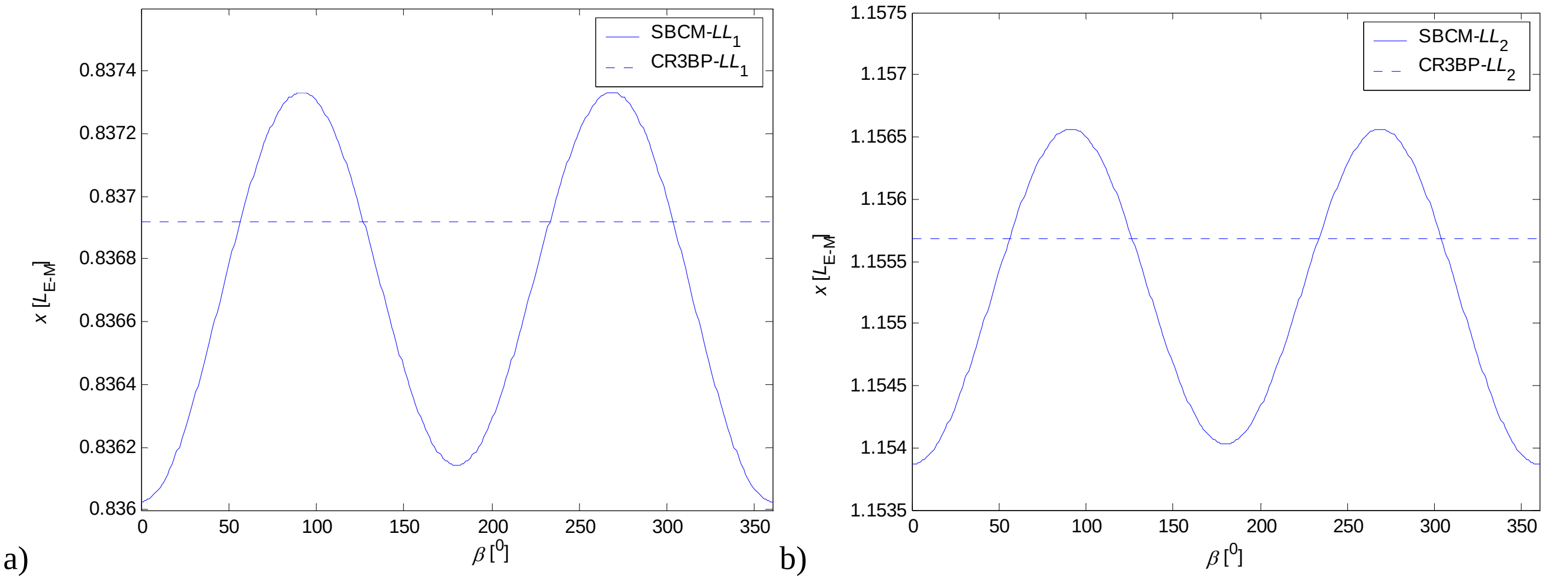

**Fig.3 The relationship between the location of equivalent equilibrium and $\beta$**: a) for the equivalent $LL_1$ point case, the location varies from 0.836 to 0.8374 $L_{\text{E-M}}$; b) for the equivalent $LL_2$ point case, the location varies from 1.1535 to 1.1565 $L_{\text{E-M}}$; the equivalent cislunar $LL_1$ or $LL_2$ point is denoted as SBCM-$LL_1$ or SBCM-$LL_2$, compared with CR3BP-$LL_1$ or CR3BP-$LL_2$ in this unperturbed model; the initial lunar phasic angle at the epoch time ($t$=0) is $\theta_{s0}=0^0$.

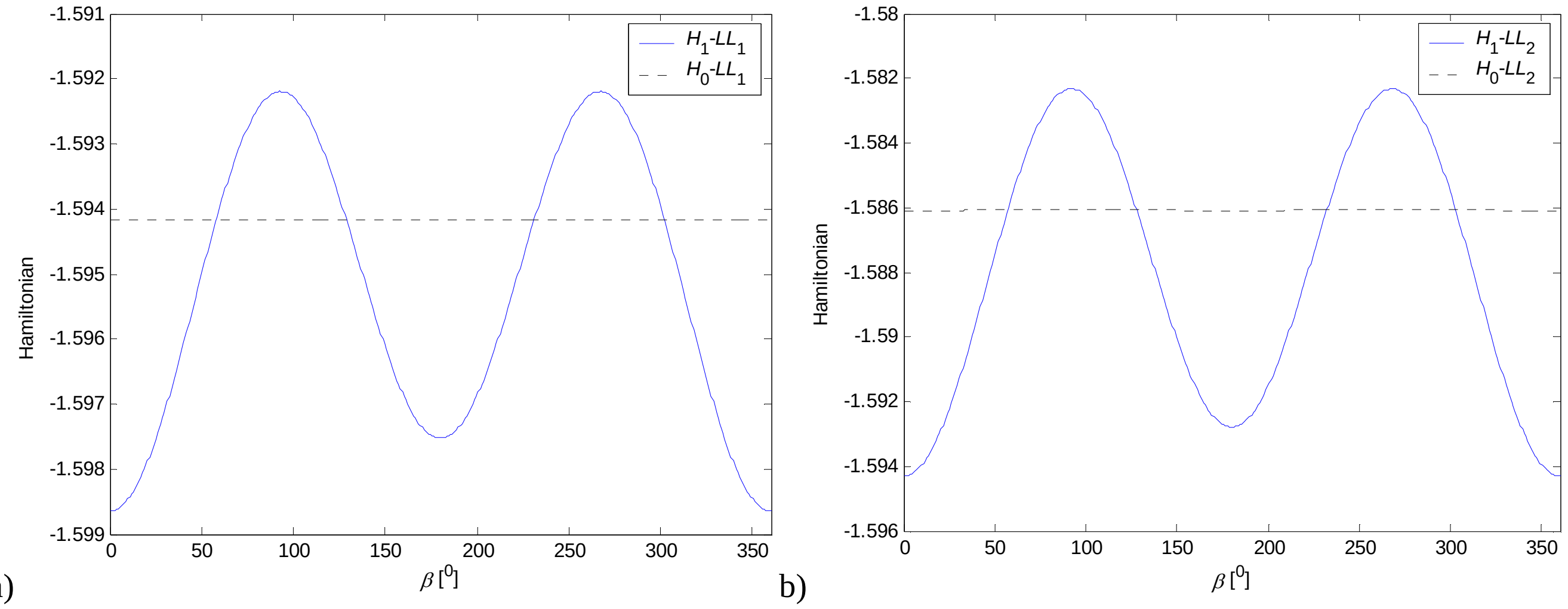

**Fig.4 The relationship between the Hamiltonian value of equivalent equilibrium and *β***: a) for the equivalent $LL_1$ point case, the Hamiltonian value varies from -1.599 to -1.592 compared with the constant value of -1.594 in CR3BP model; b) for the equivalent $LL_2$ point case, the Hamiltonian value varies from -1.594 to -1.582 compared with the constant value of -1.586 in CR3BP model; the initial lunar phasic angle at the epoch time ($t$=0) is $\theta_{s0}=0^0$.

From these figures above, the location of $LL_1$ point varies from 0.836 to 0.8374 $L_{E\text{-}M}$, and that of $LL_2$ point varies from 1.1535 to 1.1565 $L_{E\text{-}M}$. For the equivalent $LL_1$ point case, the Hamiltonian value varies from -1.599 to -1.592 compared with the constant value of -1.594 in CR3BP model; while, for the equivalent $LL_2$ point case, the Hamiltonian value varies from -1.594 to -1.582 compared with the constant value of -1.586 in CR3BP model.

## 2.3 Poincaré map for lunar captured trajectories

Villac and Scheeres investigated escaping trajectories in the Hill's three-body problem and then concluded that the deceleration at the periapsis can reach the minimum energy for the transiting trajectories from $LL_1$ or $LL_2$ point to the Moon (**Villac and Scheeres 2002**). Moreover, the impulse maneuver $\Delta V$ to decelerate the spacecraft on a lunar circular orbit can be estimated by the Hamiltonian value $H_1$ and the radius of periapsis $r_p$ of the targeting orbit (being equal to the sum of the radius of lunar surface and the altitude of periapsis), as following (**Mengali and Quarta 2005**):

$$\Delta V\left(r_p, H_1\right) = -\sqrt{\frac{\mu}{r_p}} + \sqrt{\left(1-\mu\right)^2 + r_p^2 + \frac{2\mu}{r_p} + \frac{2\left(1-\mu\right)}{1-r_p} - 2\left(1-\mu\right)r_p + H_1} \,. \qquad (15)$$

However, for a specified transiting trajectory, $H_1$ and $r_p$ are dependent on each other, and their relationship will be investigated by Poincaré map in the following section.

The technique of Poincaré map is employed to investigate preliminarily the statistical features of the fuel cost and captured elements rather than a specified Moon-captured segment transiting $LL_1$ point, and the similar case can be implemented for $LL_2$ point.

Mathematically, the procedure to compute the Poincaré map is presented as follows. For any initial value $\beta_0$ at the epoch time, two sections $\Sigma_1$ and $\Sigma_2$ are used to define the following Poincaré mapping, where the oriented section $\Sigma_1$ is located on the hyper-surface $x = x_{LL_1}$ with the Hamiltonian flow defined by Eq. (4) from left to right, and is formulized as

$$\Sigma_1^{LL_1} : x = x_{LL_1}(\beta), \dot{x} > 0 \ . \tag{16}$$

Thus, all the transiting trajectories crossing this section and dominated by the identical Hamiltonian value $H_1$ and Eq. (7), are parameterized by the remaining four dimensional coordinates. In this paper, this parameterization is implemented by $y_0$ and $z_0$, and two direction angles ($\delta$, $\phi$) ranging within the interval $[-\pi/2, \pi/2]$ of the velocity vector, whose magnitude $v_0$ is determined by the Hamiltonian value $H_1$. Therefore, the initial conditions on $\Sigma_1$ can be written as

$$\begin{cases} x|_{t=0} = x_{LL_1} & \dot{x}|_{t=0} = v_0 \cos\theta \cos\delta \\ y|_{t=0} = y_0 & \dot{y}|_{t=0} = v_0 \cos\theta \sin\delta \\ z|_{t=0} = z_0 & \dot{z}|_{t=0} = v_0 \sin\theta \end{cases} . \tag{17}$$

The procedure to produce the initial conditions is: (*i*) based on the restriction by Eq. (7), refine the interval $[y_{\min}, y_{\max}]$ for the variable $y_0$ from the case that $z$ and $v$ are set to be zeros temporarily, and then select randomly a value of $y_0$ from its interval; (*ii*) refine the interval $[z_{\min}, z_{\max}]$ for the variable $z_0$ from the case that $v$ is set to be zeros temporarily, and then select randomly a value of $z_0$ from its interval; (*iii*) calculate the rest variable $v_0$ from Eq. (7) once $y_0$ and $z_0$ are chosen in the steps mentioned above, i.e.,

$$v = \sqrt{\left(x_{LL_1}^2 + y_0^2\right) + 2\left(H_1 + \frac{1-\mu}{\|\boldsymbol{r} - \boldsymbol{r}_E\|} + \frac{\mu}{\|\boldsymbol{r} - \boldsymbol{r}_M\|} + \frac{m_S}{\|\Re - \Re_S\|} + \omega_S^2 \cdot \boldsymbol{A}_S^T \cdot \boldsymbol{R}_x(-i) \cdot \boldsymbol{R}_z(-\beta) \cdot \boldsymbol{r} - \frac{m_S}{a_S}\right)} \ ; \tag{18}$$

(*iv*) select randomly the direction angles $\delta$ and $\phi$ from their interval $[-\pi/2, \pi/2]$.

In consequence, the initial conditions given by the section $\Sigma_1$ are integrated forwards until the second section $\Sigma_2$ defined as

$$\Sigma_2 : \dot{r}_M = 0, \ddot{r}_M > 0 \tag{19}$$

where $r_M$ is the distance between the Hamiltonian flow and the Moon, which illustrates that the section $\Sigma_2$ terminates the integration routine at the periapsis of the integrated trajectory. Moreover, define a new position vector from the flow to the Moon, as

$$\tilde{\boldsymbol{r}} = [x + \mu - 1 \quad y \quad z]^T \tag{20}$$

and then rewrite Eq. (17) equivalently, as

$$\Sigma_2 : \tilde{\boldsymbol{r}}^T \dot{\boldsymbol{r}} = 0, \dot{\boldsymbol{r}}^T \dot{\boldsymbol{r}} + \tilde{\boldsymbol{r}}^T \boldsymbol{a} - \frac{(\tilde{\boldsymbol{r}}^T \dot{\boldsymbol{r}})(\tilde{\boldsymbol{r}}^T \dot{\boldsymbol{r}})}{\tilde{\boldsymbol{r}}^T \tilde{\boldsymbol{r}}} > 0 . \tag{21}$$

where $\boldsymbol{a}$ is the acceleration in the rotating frame. In the numerical computations performed, all the four dimensional coordinates $(y_0, z_0, \theta, \delta)$ are selected independently in the feasible areas, and each of the coordinates involves 300 random points.

The Poincaré map defined by the flow between the two sections, i.e., $\Sigma_1 \rightarrow \Sigma_2$, gives a mapping relating all the transiting trajectories from the region near $LL_1$ point defined by the section $\Sigma_1^{LL_1}$ forward to their first periapsis defined by the section $\Sigma_2$. Due to the dimensional reduction by Poincaré mapping, all the captured trajectories are refined as the sections of $\Sigma_a$ and $\Sigma_b$, shown as in Fig.5 for $\Delta H_1 = 3.5 \times 10^{-3}$ and $\beta = 0^0$. Essentially, $\Sigma_b$ is confined close to $LL_1$ point and corresponds to the center manifolds of halo or lissajous orbits near $LL_1$ point, and $\Sigma_a$ is confined close to the Moon and corresponds to the unstable manifolds of halo or lissajous orbits. Only the transiting trajectories corresponded by $\Sigma_a$ are discussed in this paper as the main topic on Earth-to-Moon transfers.

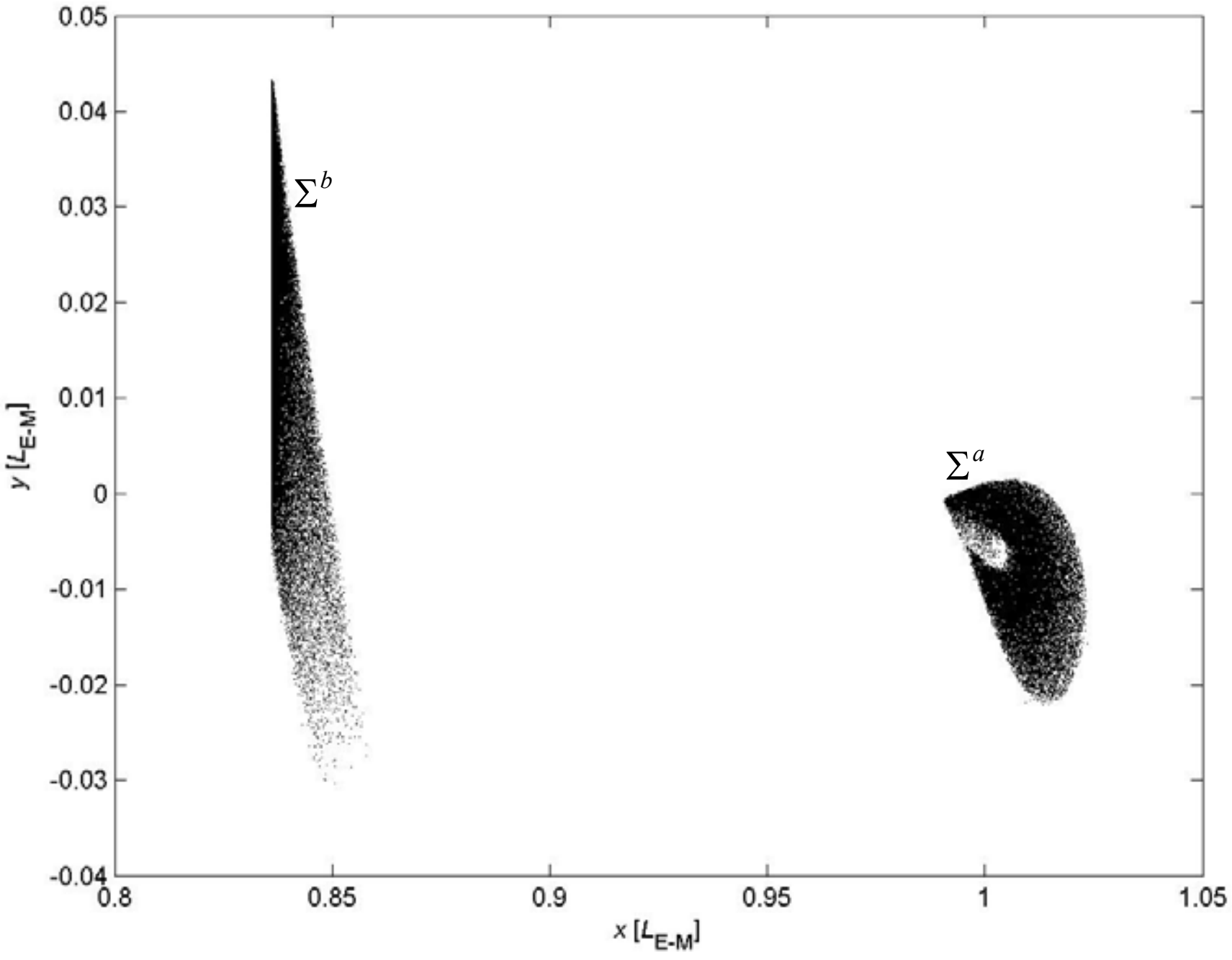

**Fig.5 Example of a Poincaré map parameterized by the Cartesian coordinates for $\Delta H_1$=3.5×10$^{-3}$ and $\beta$=0$^0$ for $LL_1$ point case**: all the captured trajectories are refined as the sections of $\Sigma_a$ and $\Sigma_b$. $\Sigma_b$ is confined close to $LL_1$ point and corresponds to the center manifolds of halo or lissajous orbits near $LL_1$ point; $\Sigma_a$ is confined close to the Moon and corresponds to the unstable manifolds of halo or lissajous orbits; only the transiting trajectories corresponded by $\Sigma_a$ are discussed in this paper as the main topic on Earth-to-Moon transfers; the initial lunar and solar phasic angles at the epoch time ($t$=0) are $\theta_{s0}$=0$^0$ and $\beta$=0$^0$, and the Hamiltonian value of 300 sampling points is set as $\Delta H_1$=3.5×10$^{-3}$.

The consequence of all the trajectories corresponded by $\Sigma_a$ could be deduced by their parameterization on the Cartesian coordinates or classical orbital elements, which will be presented in the following figures. The series of Poincaré map $\Sigma_a$ are illustrated in the following sixteen subgraphes of Fig.6 and 7 as a function of the Hamiltonian value $\Delta H_1$ and the solar phasic angle $\beta$, where the circles illustrate the lunar surface, and the chaotic points illustrate all the capturing trajectories mapped numerically from 300×300×300×300 random points selected on the section $\Sigma_1^{LL_1}$ or $\Sigma_1^{LL_2}$.

The initial conditions are listed as following: for Fig.5, the initial lunar and solar phasic angles at the epoch time ($t$=0) are $\theta_{s0}$=0$^0$ and $\beta$=0$^0$ respectively, and the Hamiltonian value of 300 sampling points is set as $\Delta H_1$=3.5×10$^{-3}$. For Figs.6 and 7, the initial lunar and solar phasic angles at the epoch time ($t_0$=0) are $\theta_{s0}$=0$^0$ and $\beta$=0$^0$, and the subgraphs in a row have the same solar phasic angle ranked in 0$^0$, 90$^0$, 180$^0$ and 270$^0$, and the subgraphs in a column have the same Hamiltonian value $\Delta H_1$ ranked in 5×10$^{-6}$, 1×10$^{-4}$, 1×10$^{-3}$ and 5×10$^{-3}$.

Compared with the chaotic points located on the right hand of the Moon in rotating $\mathbf{S_{E\text{-}M}}$ frame for $LL_1$ point case, the chaotic locate on the left hand for $LL_2$ point case, because all the captured trajectories reach their first periapsis on the opposite hand of the initial leaving section $\Sigma_1^{LL_1}$ or $\Sigma_1^{LL_2}$. The fact above is in accordance with the theory of Keplerian hyperbolic or parabolic orbit that the periapsis of the transfer trajectory locates at the opposite hand of the initial leaving velocity at infinity $V_\infty$ relative to the targeting planet.

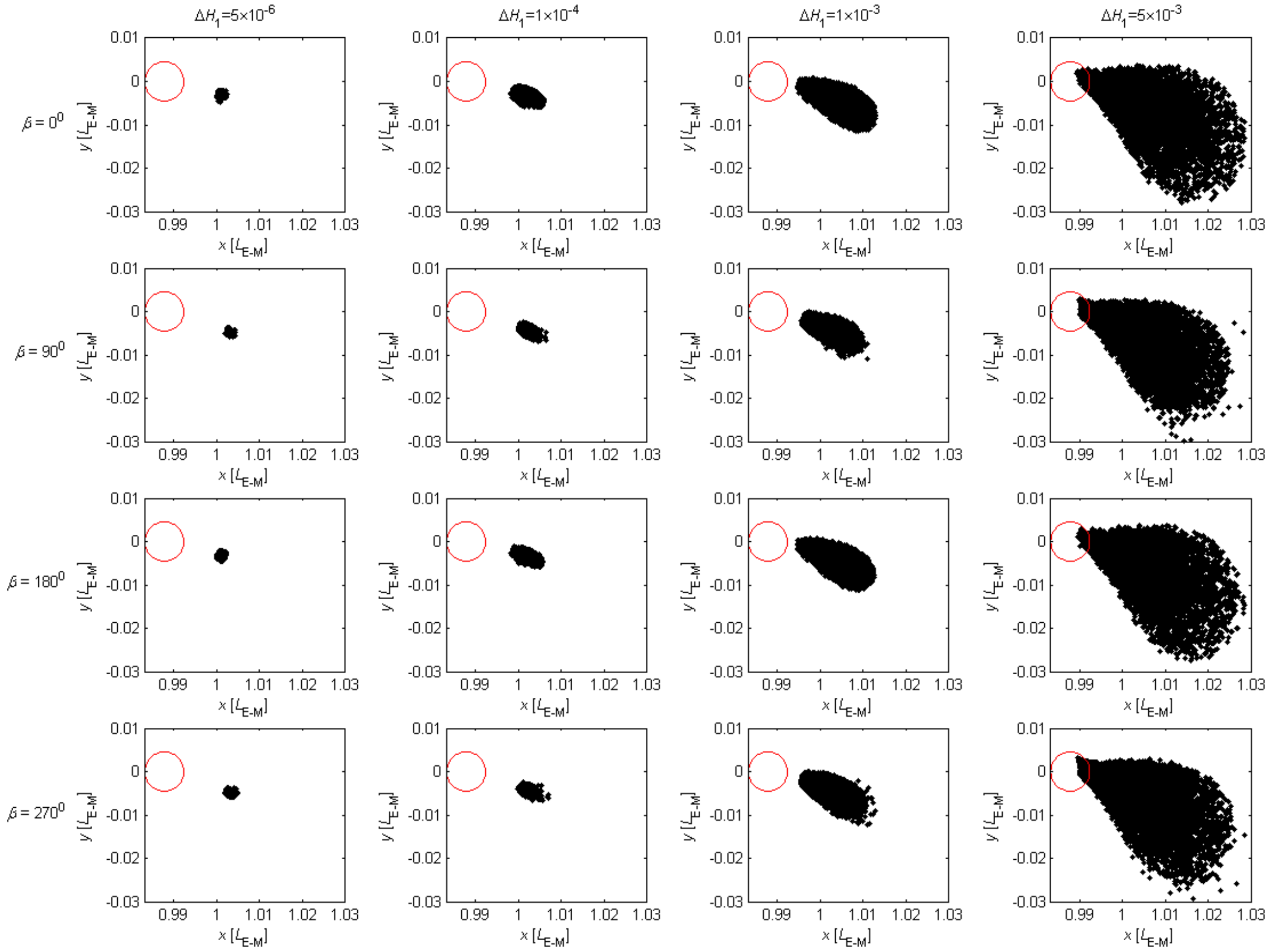

**Fig.6 Series of Poincaré map $\Sigma_a$ as a function of $\Delta H_1$ and $\beta$ for $LL_1$ point case**: the circles illustrate the lunar surface, and the chaotic points illustrate all the capturing trajectories mapped numerically from random points selected on the section $\Sigma_1^{LL_1}$; the Hamiltonian value $\Delta H_1$ has much more effects on the extrema than the solar phasic angle $\beta$, verified by the Poincaré map in Fig.8; the initial lunar and solar phasic angles at the epoch time ($t_0$=0) are $\theta_{s0}=0^0$ and $\beta=0^0$, and the subgraphs in a row have the same solar phasic angle ranked in $0^0$, $90^0$, $180^0$ and $270^0$, and the subgraphs in a column have the same Hamiltonian value $\Delta H_1$ ranked in $5\times10^{-6}$, $1\times10^{-4}$, $1\times10^{-3}$ and $5\times10^{-3}$.

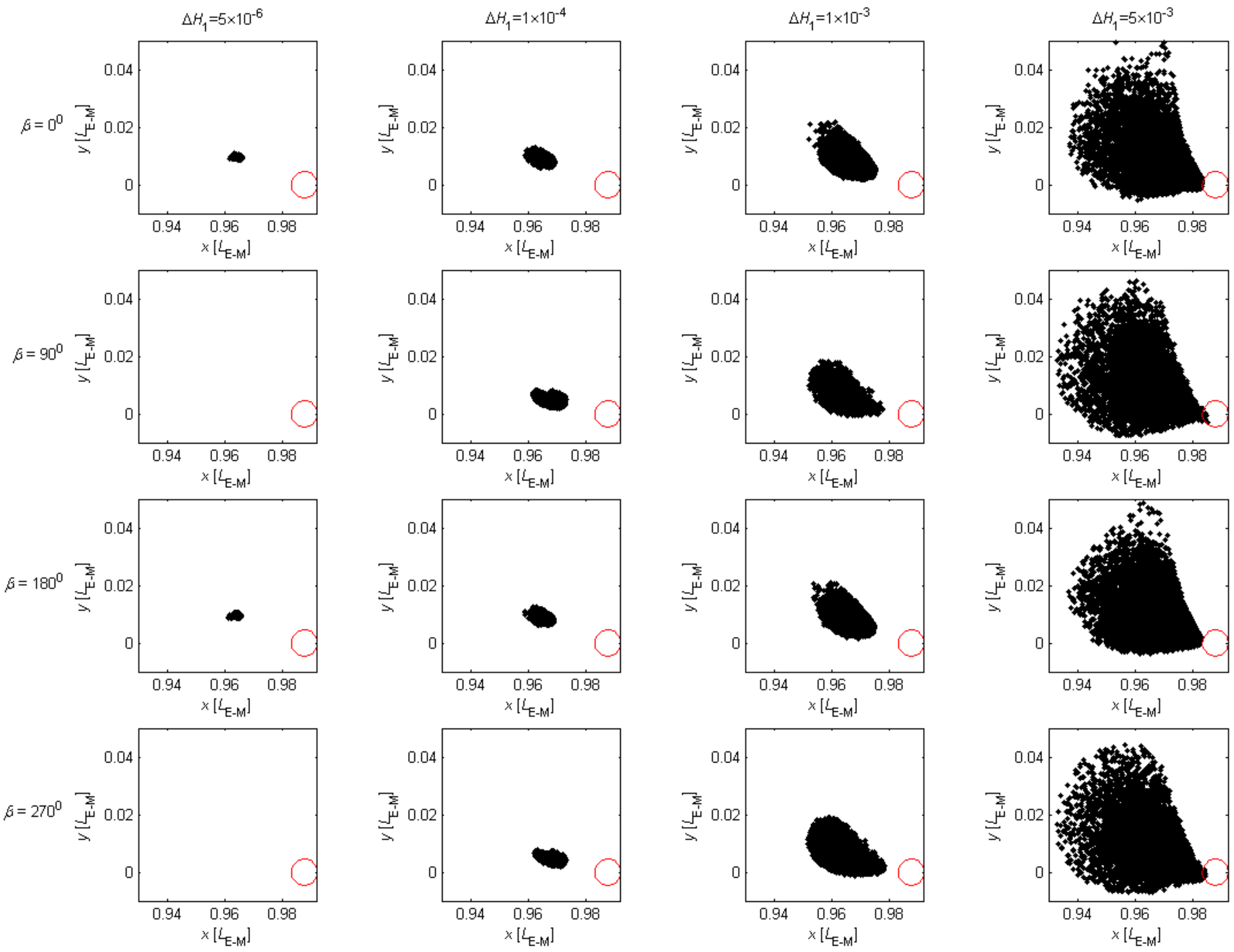

**Fig.7 Series of Poincaré map $\Sigma_a$ as a function of $\Delta H_1$ and $\beta$ for $LL_2$ point case**: the circles illustrate the lunar surface, and the chaotic points illustrate all the capturing trajectories mapped numerically from random points selected on the section $\Sigma_1^{LL_2}$; the Hamiltonian value $\Delta H_1$ has much more effects on the extrema than the solar phasic angle $\beta$ as well as $LL_1$ point case; the initial lunar and solar phasic angles at the epoch time ($t_0$=0) are $\theta_{s0}=0^0$ and $\beta=0^0$, and the subgraphs in a row have the same solar phasic angle ranked in $0^0$, $90^0$, $180^0$ and $270^0$, and the subgraphs in a column have the same Hamiltonian value $\Delta H_1$ ranked in $5\times10^{-6}$, $1\times10^{-4}$, $1\times10^{-3}$ and $5\times10^{-3}$.

The characteristics of the Poincaré maps vary as $\Delta H_1$ and $\beta$ vary, which are captured by the extremum surfaces of the altitude of periapsis and eccentricity for all the transiting trajectories in Figs.8 and 9. Inherited from the Poincaré map in Figs.6 and 7, the Hamiltonian value $\Delta H_1$ has much more effects on the extremum than the solar phasic angle $\beta$. Moreover, the maximum and minimum are illustrated respectively by the top and bottom branches of the extremum surface, and any altitude of periapsis or eccentricity inside the two branches is available for a specified captured trajectory, demonstrated in shallow-painted areas in the projection subgraphes ***b*** and ***d*** of Figs.8 and 9. In particular, the maximum and minimum will be equal to each other at some specified values of $\Delta H_1$ and $\beta$ when the top and bottom branches encounter with each other smoothly at the left edge of the extremum surface. By the subgraphes ***b*** and ***d***, it is verified that the Hamiltonian value $\Delta H_1$ has much more effects on the extremum than the solar phasic angle $\beta$ inherited from the Poincaré map in Figs.6 and 7.

The procedure to produce the characteristics of the altitude of periapsis and eccentricity of captured trajectories is presented as following: (*i*) collect the position and velocity ($\boldsymbol{r}$, $\dot{\boldsymbol{r}}$) of all the captured trajectories at their first periapsis based on the procedure to produce Figs.5, 6 and 7; (*ii*) transform the state ($\boldsymbol{r}$, $\dot{\boldsymbol{r}}$) from the syzygy $\mathbf{S}_{\text{E-M}}$ frame to the Moon-center inertial frame, which has the same coordinate axis definition as that of the inertial frame $\mathbf{I}_{\text{S-E/M}}$ or $\mathbf{I}_{\text{E-M}}$, but fixes its origin at the barycenter of the Earth-Moon system; (*iii*) convert the updated inertial states into the classical orbital elements, including $r_p$ and $e$, based on the lunar gravitational coefficients and Keplerian two body theory (the conversion between classical orbital elements and Cartesian coordinates is common and can be found in textbooks); (*iv*) plot the extremum surfaces in Figs.8 and 9 for $LL_1$ and $LL_2$ points respectively.

The initial conditions are listed as following: the initial lunar phasic angle is $\theta_{s0}=0^0$, and the solar phasic angle $\beta$ ranges from $0^0$ to $360^0$, and the Hamiltonian value $\Delta H_1$ ranges from 0 to $5\times10^{-3}$ (for $LL_1$ point) or from 0 to $1.1\times10^{-3}$ (for $LL_2$ point).

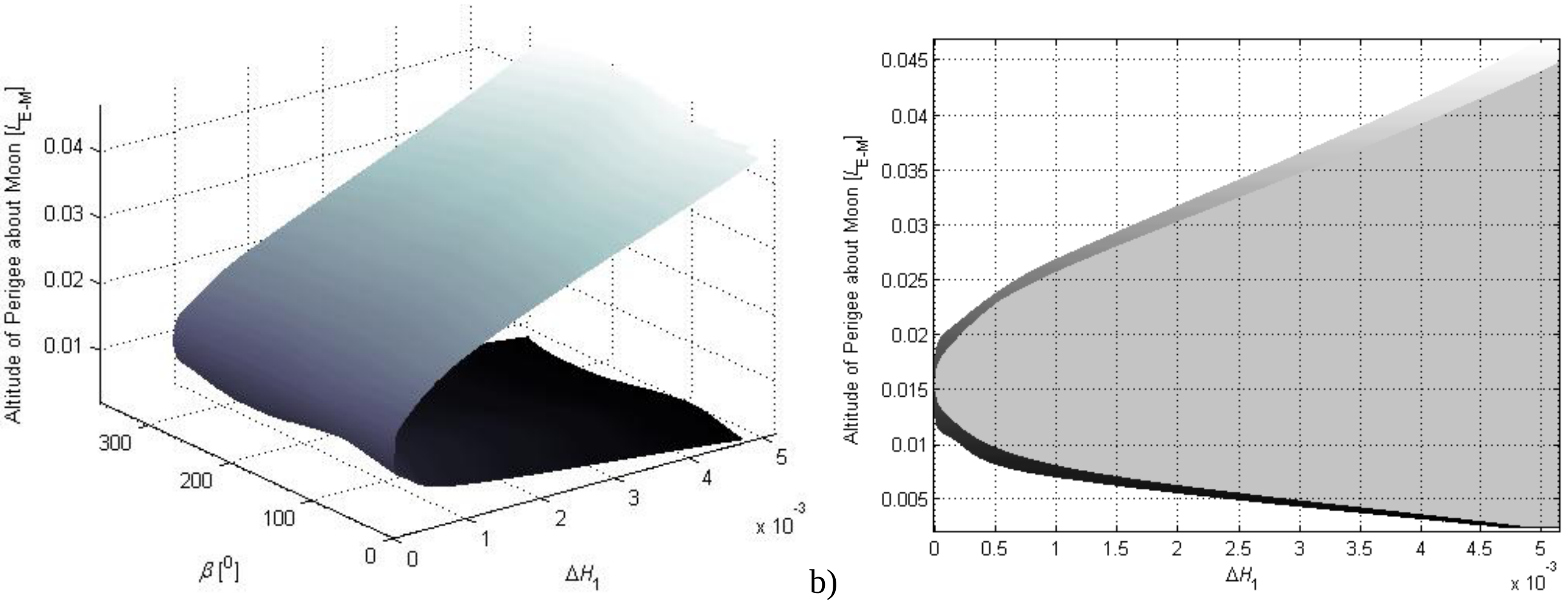

a) b)

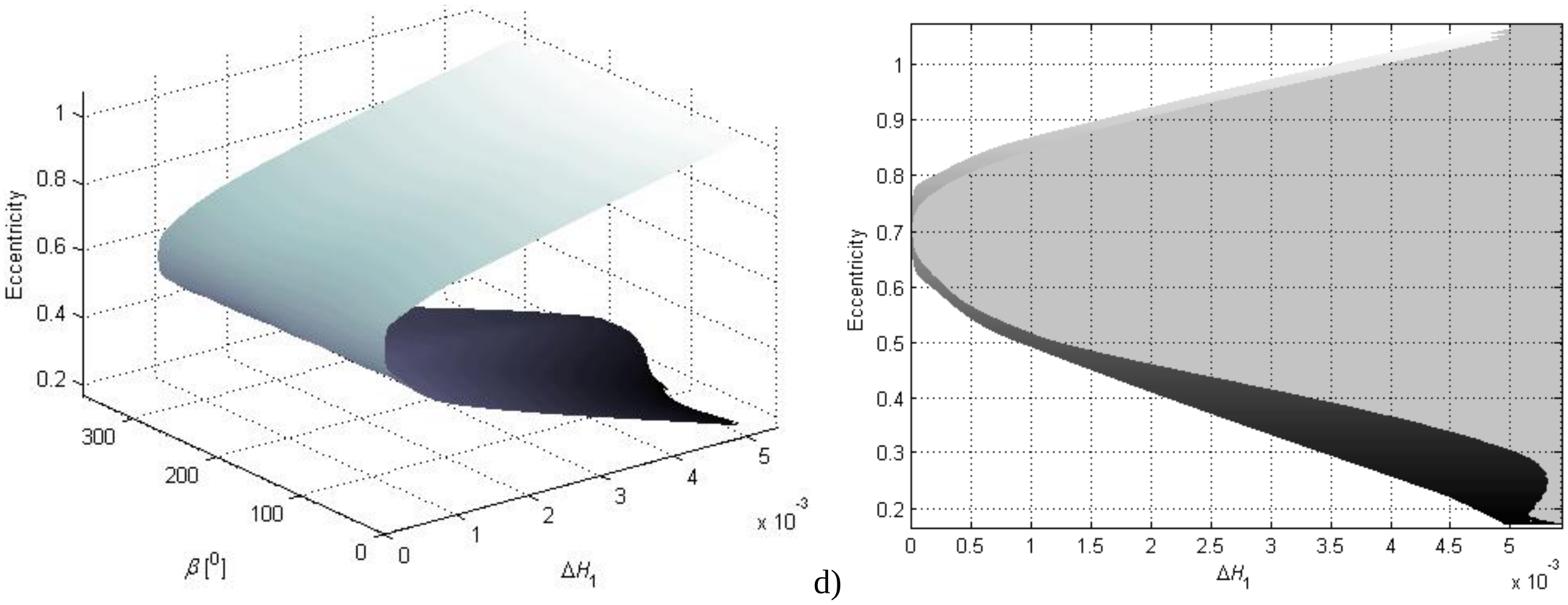

c) d)

**Fig.8 Extremum surfaces of altitude of periapsis and eccentricity for all the captured trajectories for $LL_1$ point case**: a) 3-D illustration of extremum surface of altitude of periapsis $r_p$ expressed as a function of $\Delta H_1$ and $\beta$; b) 2-D projection onto the ($\Delta H_1$, $r_p$) space of the extremum; c) 3-D illustration of extremum surface of eccentricity $e$ expressed as a function of $\Delta H_1$ and $\beta$; d) 2-D projection onto the ($\Delta H_1$, $e$) space of the extremum; the Hamiltonian value $\Delta H_1$ has much more effects on the extrema than the solar phasic angle $\beta$, inherited from the Poincaré map in Fig.6; the extremum includes both the maximum and minimum, which are illustrated respectively by the top and bottom branches of the extremum surface; any altitude of periapsis or eccentricity inside the two branches is available for a specified transiting trajectory, demonstrated in the shallow-painted areas in the 2-D illustration ***b*** and ***d***; the maximum and minimum are equal to each other at some specified values of $\Delta H_1$ and $\beta$ when the top and bottom branches encounter with each other smoothly at the left edge of the extremum surface; the initial lunar phasic angle is $\theta_{s0}=0^0$, the solar phasic angle $\beta$ ranges from $0^0$ to $360^0$, and the Hamiltonian value $\Delta H_1$ ranges from 0 to $5\times10^{-3}$.

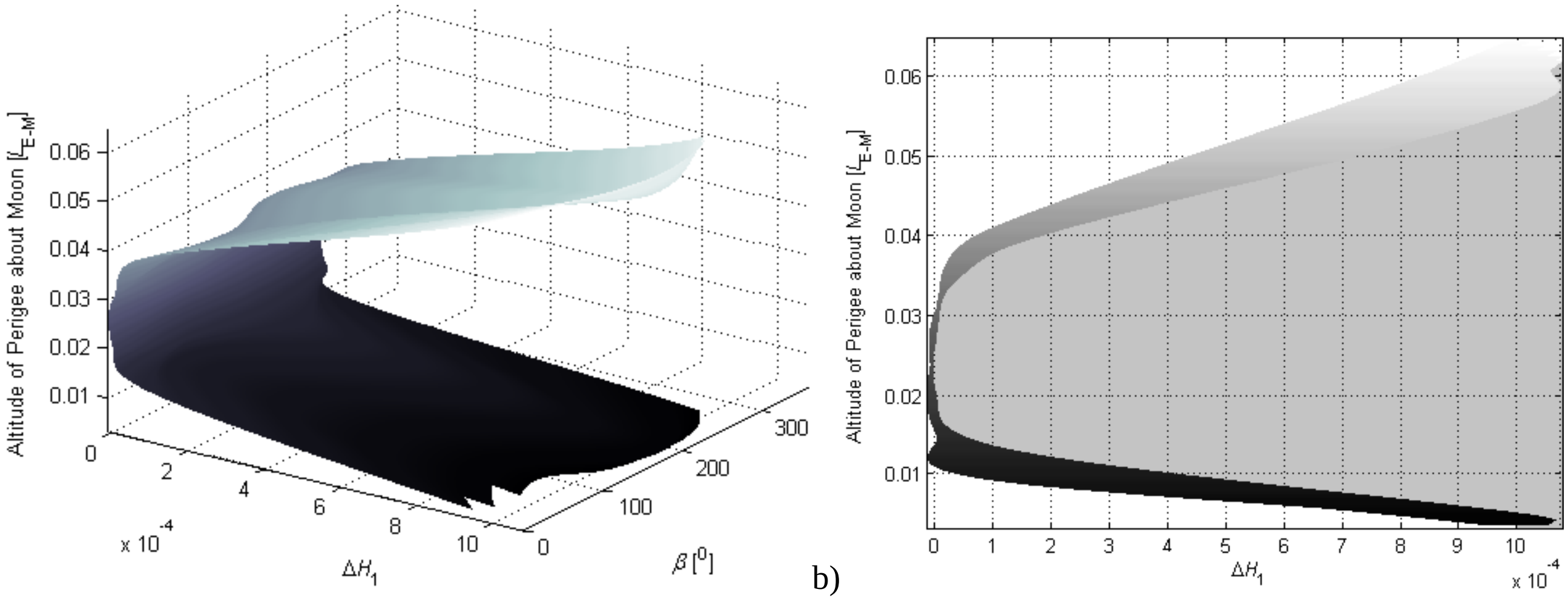

a) b)

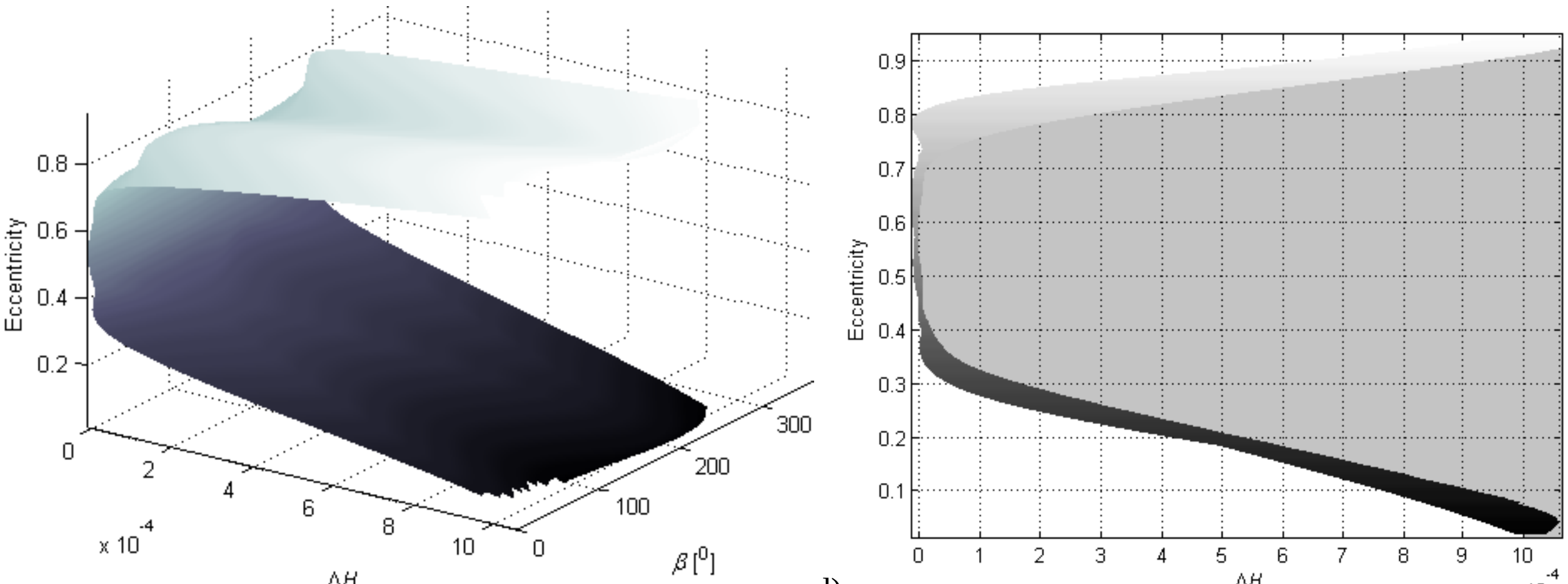

c) d)

**Fig.9 Extremum surfaces of altitude of periapsis and eccentricity for all the captured trajectories for $LL_2$ point case**: a) 3-D illustration of extremum surface of altitude of periapsis $r_p$ expressed as a function of $\Delta H_1$ and $\beta$; b) 2-D projection onto the ($\Delta H_1$, $r_p$) space of the extremum; c) 3-D presentation of extremum surface of eccentricity $e$ expressed as a function of $\Delta H_1$ and $\beta$; d) 2-D projection onto the ($\Delta H_1$, $e$) space of the extremum; the Hamiltonian value $\Delta H_1$ has much more effects on the extrema than the solar phasic angle $\beta$, inherited from the Poincaré map in Fig.7; the extremum includes both the maximum and minimum, which are illustrated respectively by the top and bottom branches of the extremum surface; any altitude of periapsis or eccentricity inside the two branches is available for a specified transiting trajectory, demonstrated in the shallow-painted areas in the 2-D illustration ***b*** and ***d***; the maximum and minimum are equal to each other at some specified values of $\Delta H_1$ and $\beta$ when the top and bottom branches encounter with each other smoothly at the left edge of the extremum surface; the initial lunar phasic angle is $\theta_{s0}=0^0$, the solar phasic angle $\beta$ ranges from $0^0$ to $360^0$, and the Hamiltonian value $\Delta H_1$ ranges from 0 to $1.1\times10^{-3}$.

A tangential burn $\Delta V$ is required to capture a circular orbit about the Moon, which is also regarded as the criterion to measure some candidate trajectories from the viewpoint of energy. Considering a fixed radius of periapsis captured by the Moon, e.g., $r_\mathrm{p}$=1838km (i.e., the altitude of periapsis is equal to 100km), only a specified Hamiltonian value $\Delta H_1$ is refined from the shallow-painted areas shown in Figs. 10a and 10b for an arbitrary $\beta\in[0, 2\pi]$, and then the minimum $\Delta V_\mathrm{min}$ can be obtained from Eq. (14) by the refined minimum value of $\Delta H_1$. Thus, the improved Poincaré mapping with a fixed $r_\mathrm{p}$ establishes the relationship between $\Delta V_\mathrm{min}$ and $\beta$, as illustrated in Fig.10. Compared with the captured $\Delta V_\mathrm{min}$ of 695.7m/s yielded by Keplerian two-body model, 656.8m/s by the Hill's model, and 649.2m/s ($LL_1$ point) and 652.9m/s ($LL_2$ point) in the CR3BP model (**Villac and Scheeres 2002**; **He and Xu 2007**), the SBCM model can reach the minimum value of 642.9m/s ($LL_1$ point) and 646.7m/s ($LL_2$ point).

Mathematically, the procedures to produce Figs.8 and 9 establish a mapping (or function) from $H_1$ and $\beta$ to $r_p$, which is formulized as $r_p = \Gamma(H_1, \beta)$; however, for a fixed radius of periapsis $r_p^*$=1838km, $H_1$ is parameterized by the only variable $\beta$, i.e., $H_1 = \Gamma_{r_p^*}^{-1}(\beta)$, which can be solved from numerical procedures of Fig.8 and 9. Subsequently, the minimum $\Delta V_{\min}$ can be obtained from Eq. (14) by the refined minimum value of $H_1$. The initial conditions to produce Fig.10 are listed as following: the initial lunar phasic angle is $\theta_{s0}$=$0^0$, and the solar phasic angle $\beta$ ranges from $0^0$ to $360^0$.

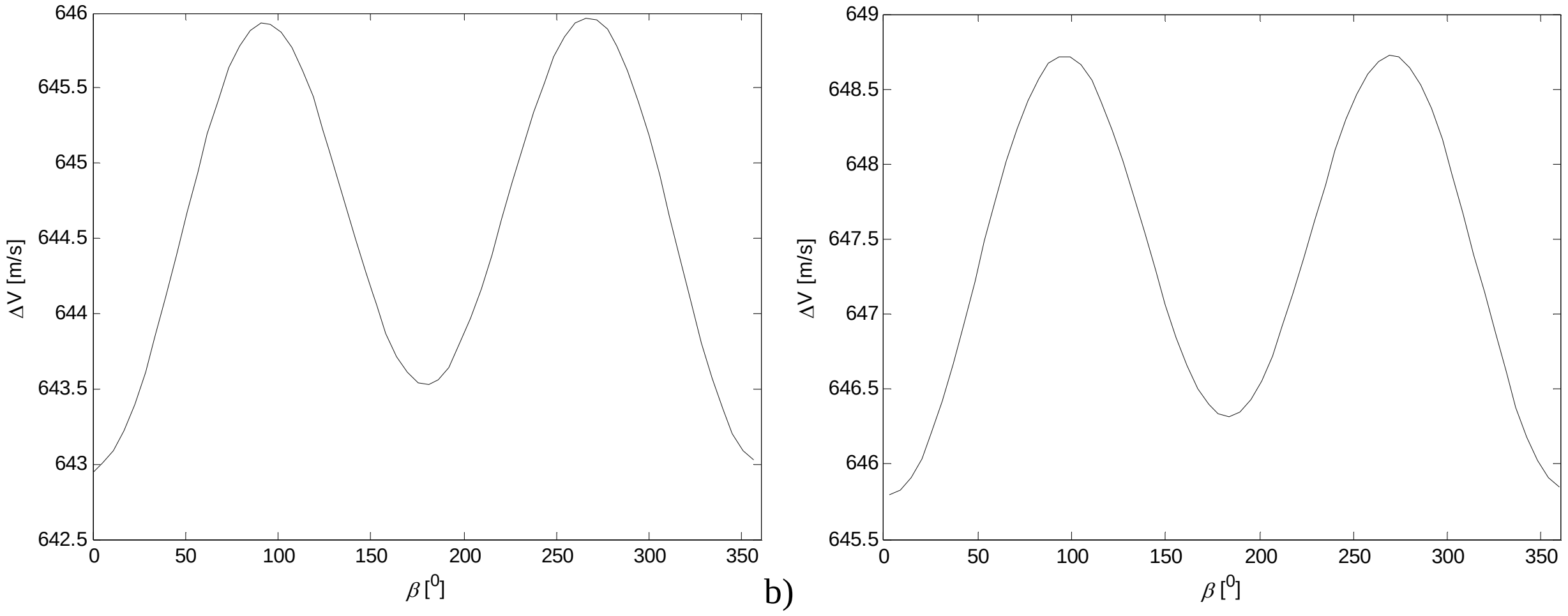

**Fig.10 The relationship between the minimal captured $\Delta V_{\min}$ and $\beta$ produced by the improved Poincaré mapping with a fixed altitude of periapsis $r_p$=1838km**: a) the relationship for $LL_1$ point case; b) the relationship for $LL_2$ point case; compared with the captured $\Delta V_{\min}$ of 695.7m/s yielded by Keplerian two-body model, 656.8m/s by the Hill's model, and 649.2m/s ($LL_1$ point) and 652.9m/s ($LL_2$ point) in the CR3BP model (**Villac and Scheeres 2002**; **He and Xu 2007**), the SBCM model can reach the minimal value of 642.9m/s ($LL_1$ point) and 646.7m/s ($LL_2$ point).

## 3. Low-energy Transfers by Transiting Equivalent Libration Points

Compared with the statistical features of captured orbital elements discussed in the section above, the minimum-energy cislunar and trans-lunar trajectories are yielded by transiting $LL_1$ and $LL_2$ points in this section. It is presented that the asymptotical behaviors of invariant manifolds approaching to or from the libration points or halo orbits are destroyed by the solar perturbation. Moreover, the transfer opportunities measured by the solar phasic angle $\beta$ are achieved for the Earth-escaping and Moon-captured segments, respectively.

### 3.1 Low-energy transfers by transiting $LL_1$ point

For the CR3BP model, the minimum energy trajectory transiting $LL_1$ point is essentially two branches of the invariant manifolds originating from this equilibrium point. Thus, the spacecraft may follow the stable manifold from the interior region dominated by the Earth's gravity to $LL_1$ point, and then leave along the unstable manifold for the exterior region dominated by the lunar gravity. However, this transfer trajectory is not practical because its duration is infinite, which is inheriting from the fact that the invariant manifolds approach or leave $LL_1$ point asymptotically in an infinite duration.

The perturbation of the solar gravity employed by the SBCM model will change topologically the invariant manifolds to fail in transiting $LL_1$ point for some phasic angles $\beta\in[0, 2\pi]$; however, the transiting manifolds are preserved for the other values of $\beta$, inheriting from the time-invariant CR3BP model. For the interval of $\beta$ transiting $LL_1$ point, the asymptotical infinite durations are cut down to finite ones by the perturbation, which is quite beneficial to Earth-to-Moon transfers. For the interval of $\beta$ not transiting $LL_1$ point, the perturbed manifolds will lose the phase of orbiting the Earth or Moon, i.e., there is no periapsis about the Earth or Moon in this case. Therefore, the gaps between the altitudes of periapsis about the Earth and Moon are presented by the phasic angle $\beta$ in Fig.11. Only the intersection between $\beta$s intervals transiting from the Earth to $LL_1$ point and another intervals transiting from $LL_1$ point to the Moon, i.e., $[77^0, 109^0]U[285^0, 342^0]$, can drive the trajectories to orbit successively the Earth and Moon, and can be also considered as the cislunar transfer opportunities which is bounded by the vertical dashed lines in Fig.11.

The procedure to produce the cislunar transfer opportunities measured by the solar phasic angle $\beta$ is presented as follows. Vary $\beta$ in the interval of $[0^0, 360^0]$ to integrate backwards the SBCM dynamics formulized by the differential Eq. (4) backwards to yield the transfer opportunities for Earth-escaping segment, and integrate forwards to yield the transfer opportunities for Moon-captured segment. The two integrations (forwards and backwards) have the same initial condition of $[x_{LL_1}, 0, 0, 0, 0, 0]^T$. Only several subintervals of $\beta$ can make the integrated trajectories closer to the Earth or Moon, which are considered as cislunar or trans-lunar transfer opportunities. The initial conditions are: the initial lunar phasic angle is $\theta_{s0}=0^0$, and the solar phasic angle $\beta$ ranges from $0^0$ to $360^0$, and the initial values to integrate forwards and backwards Eq.(4) are equally $[x_{LL_1}, 0, 0, 0, 0, 0]^T$.

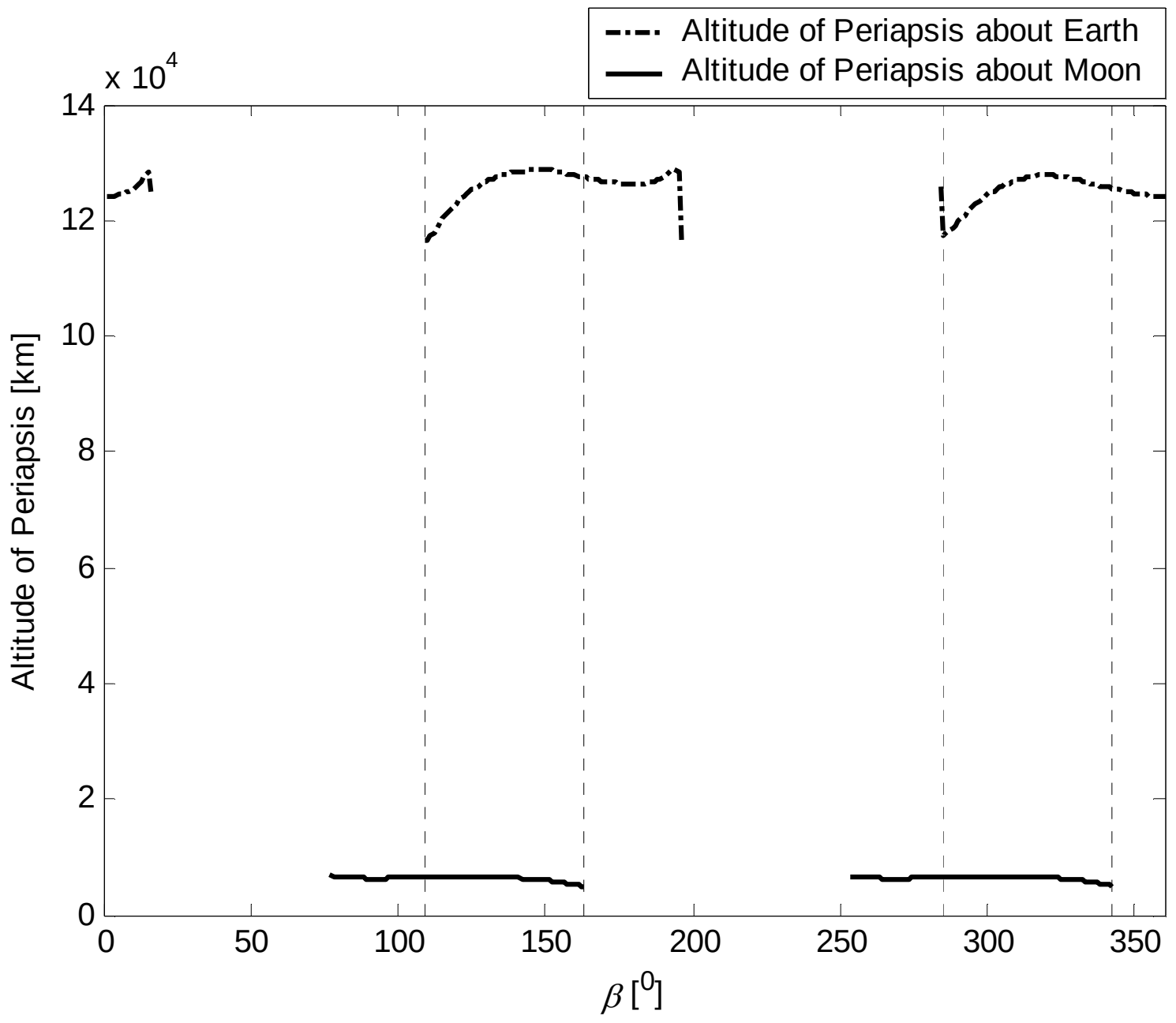

**Fig.11 Cislunar transfer opportunities measured by the solar phasic angle *β***: the solid lines illustrate the available altitudes of periapsis about the Earth, and the dash-dotted lines illustrate the available altitudes of periapsis about the Moon; the intersection of the intervals transiting from the Earth to $LL_1$ point and from $LL_1$ point to the Moon is bounded by the vertical dashed lines, i.e., $[77^0, 109^0] \cup [285^0, 342^0]$, which drives the trajectories to orbit successively the Earth and Moon; the gaps between the altitudes of periapsis are caused by some $\beta$s intervals failing the invariant manifolds in transiting $LL_1$ point.

The three-dimensional cislunar trajectory is presented in $\mathbf{S}_{\mathbf{E\text{-}M}}$ (Fig. 12) and $\mathbf{I}_{\mathbf{E\text{-}M}}$ frames (Fig. 13). From the two figures, it is deduced that the $z$ component ranges between $-2\times10^{-3}$~$+2\times10^{-3}$ ($L_{\text{E-M}}$), while the $x$ and $y$ components range respectively between -0.9~+1.2 and -0.8~+0.8 ($L_{\text{E-M}}$). This conclusion can also be summarized from the trajectories transiting $LL_2$ point or halos orbits near the two libration points. Thus, for all the cislunar and trans-lunar trajectories discussed in this paper, the $z$ component is much smaller than the other components (the $z$ component is only about one thousandth of $x$ or $y$ component), which indicates the spatial perturbation has few effects on the low-energy transfer.

The procedure to produce typical cislunar transfer trajectory transiting $LL_1$ point in the rotating $\mathbf{S}_{\mathbf{E\text{-}M}}$ frame is: for some specified value of $\beta$, integrate Eq.(4) backwards from the equivalent equilibrium to obtain the Earth-escaping segment and forwards to achieve the Moon-captured segment in the rotating $\mathbf{S}_{\mathbf{E\text{-}M}}$ frame. The transfer trajectories $\boldsymbol{R}_I$ in the inertial $\mathbf{I}_{\text{E-M}}$ frame in Fig.13 are converted from the integrated trajectories $\boldsymbol{r}$ in the rotating $\mathbf{S}_{\text{E-M}}$ frame in Fig.12, based on the transition matrix of $\boldsymbol{R}_I = \boldsymbol{R}_x(-i)\boldsymbol{R}_z(-\beta)\boldsymbol{r}$ . The initial conditions are: the initial lunar phasic angle at the epoch time ($t$=0) is $\theta_{s0}=0^0$ for the four figures, and the initial solar phasic angle $\beta$ at the epoch time ($t$=0) is $286^0$ for Figs.12 and 13, and the integral initial values to produce Figs.12 and 13 are $[x_{LL_1}, 0, 0, 0, 0, 0]^T$.

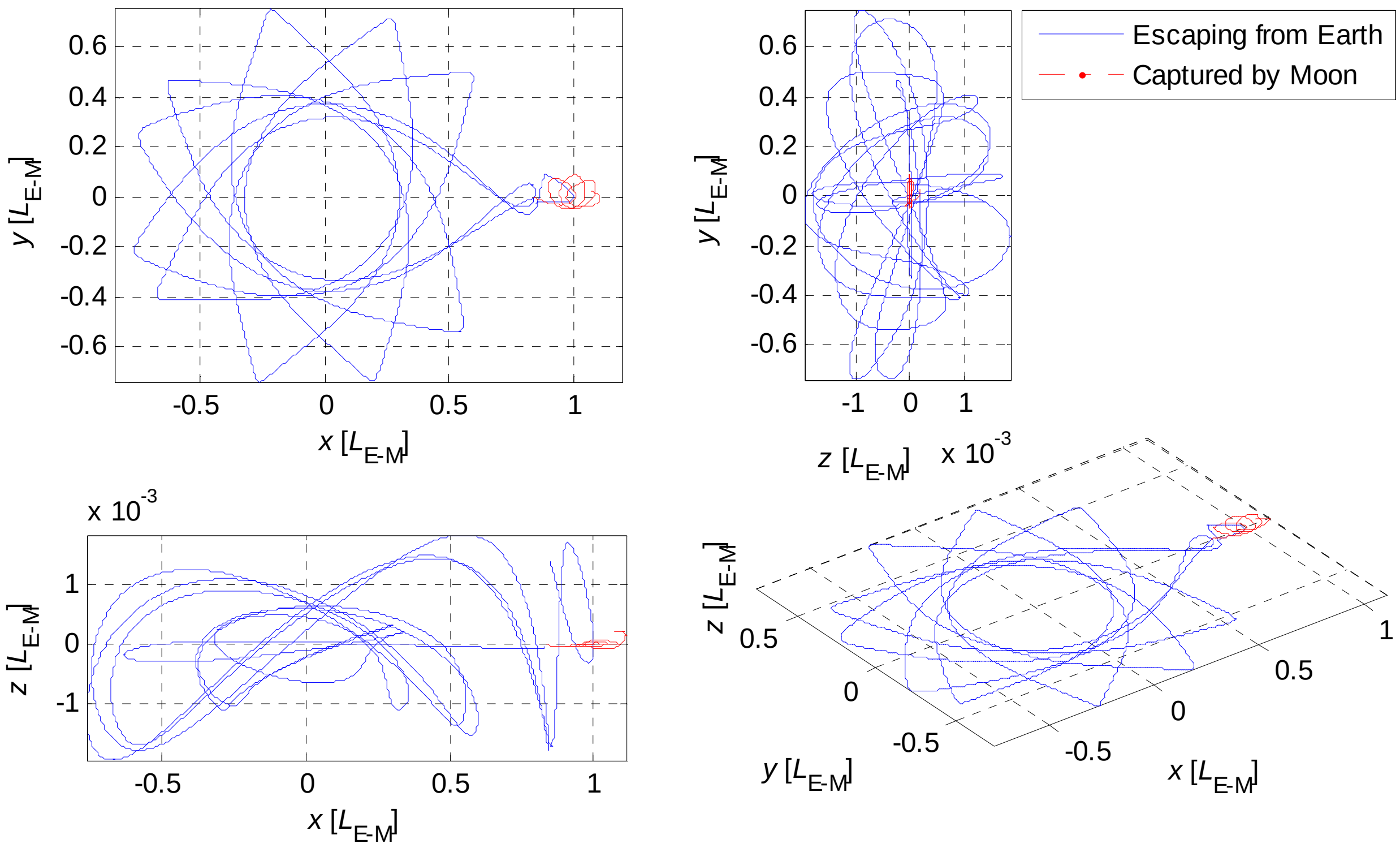

**Fig.12 Typical cislunar transfer trajectory transiting $LL_1$ point in the rotating $S_{E\text{-}M}$ frame for $\beta$=$286^0$**: the blue solid lines illustrate the segment escaping from the Earth, and the red dash-dotted lines illustrate the segment captured by the Moon; the $z$ component ranges between $-2\times10^{-3}$~$+2\times10^{-3}$ ($L_{E\text{-}M}$), while the $x$ and $y$ components range respectively between -0.9~+1.2 and -0.8~+0.8 ($L_{E\text{-}M}$), which indicates that the spatial perturbation has few effects on the low-energy transfer.

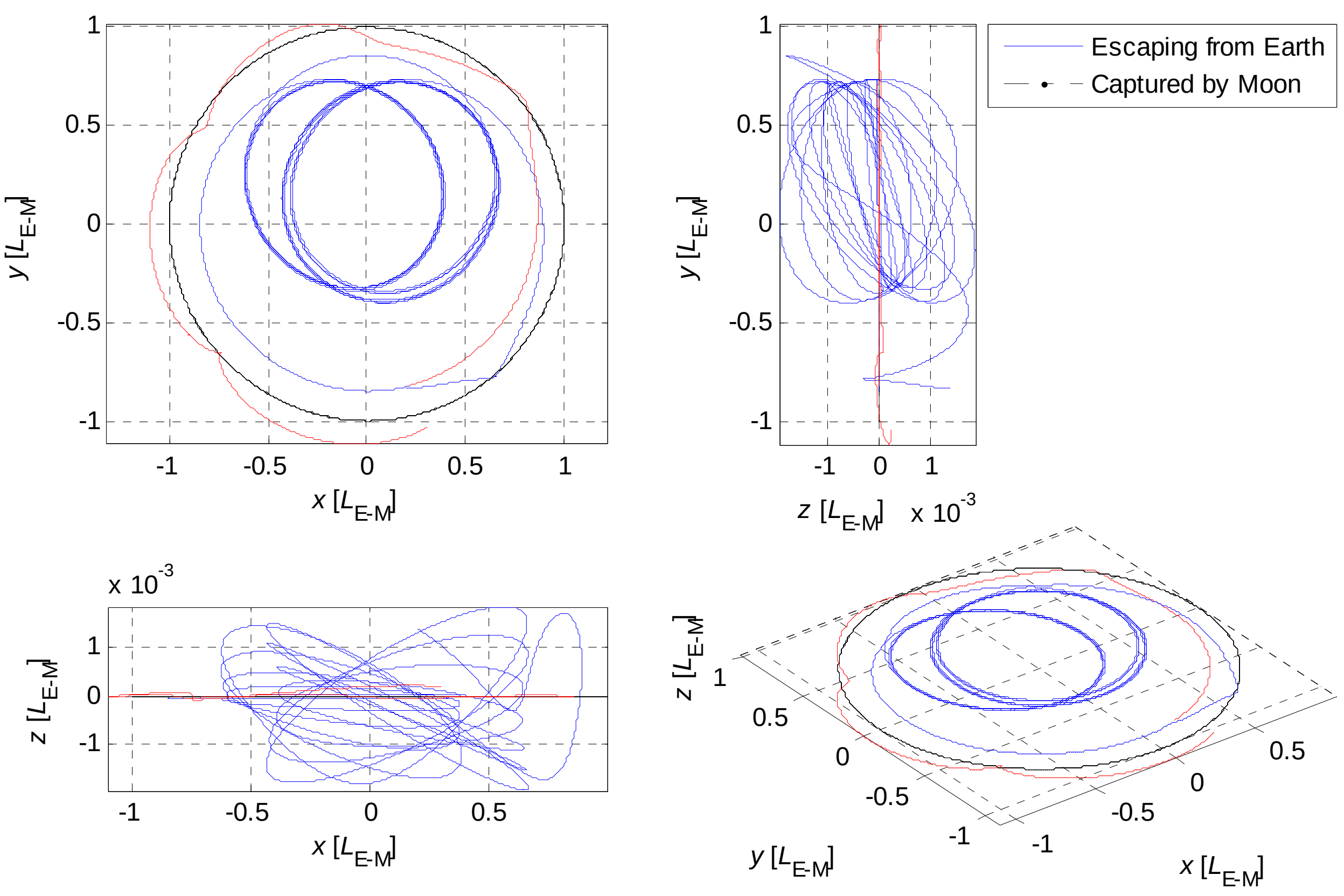

**Fig.13 Typical cislunar transfer trajectory transiting $LL_1$ point in the inertial $I_{E\text{-}M}$ frame for $\beta$=286$^0$**: the blue solid lines illustrate the segment escaping from the Earth, and the red dash-dotted lines illustrate the segment captured by the Moon; the $z$ component ranges between $-2\times10^{-3}$~$+2\times10^{-3}$ ($L_{E\text{-}M}$), while the $x$ and $y$ components range respectively between -0.9~+1.2 and -0.8~+0.8 ($L_{E\text{-}M}$), which indicates that the spatial perturbation has few effects on the low-energy transfer.

The semi-major axis and eccentricity of the typical cislunar low-energy trajectory for $\beta$=286$^0$ are illustrated in Fig.14. Due to the solar gravitational perturbation, the duration of the transiting manifold is finite. For a cislunar trajectory, the time epoch ($t$=0) is set as the moment of passing through $LL_1$ point, and its stable manifold will orbit the Earth before this epoch (i.e., $t$<0), while its unstable manifold will orbit the Moon after this epoch (i.e., $t$>0). Thus, the osculating semi-major axis and eccentricity before the epoch ($t$<0) should be conversed from the position and velocity in the Earth-center inertial frame based on the Keplerian restricted two body theory, while the osculating semi-major axis and eccentricity after the epoch ($t$>0) should be conversed in the Moon-center inertial frame. In this case, the jumps at the epoch ($t$=0) are caused by the fact that the orbital elements before and after this epoch are conversed in two different inertial frames, i.e., the former is in the Earth-center frame but the latter is in the Moon-center one.

Furthermore, the orbital elements after the epoch have more considerable variation in amplitude than before the epoch, especially for the eccentricity. It is because the osculating orbital elements are achieved only based on the Keplerian restricted two body theory; however, the perturbation from the other celestial body's gravity affects the elements greatly, which is referred as the second body in the CR3BP or SBCM model. Compared with the lunar perturbation before the epoch, the Earth has more perturbation on the osculating orbital elements conversed in the Moon-center inertial frame after the epoch, which accounts for more jumps on eccentricity (in the right subgraph of Fig.14) after the epoch than before the epoch. This cislunar transiting trajectory is classified as low-energy transfer because both the eccentricities before and after the epoch are less than 1, compared to the hyperbolical velocity captured by the Moon in classical Hohmann transfer (like Apollo (NASA) and Chang'E (China) missions).

The procedure to produce Fig.14 is: (*i*) integrate Eq.(4) backwards to obtain the Earth-escaping segment and forwards to achieve the Moon-captured segment in the rotating $\mathbf{S}_{\text{E-M}}$ frame both from the same initial condition of $[x_{LL_1}, 0, 0, 0, 0, 0]^T$; (*ii*) transform the state ($\boldsymbol{r}$, $\dot{\boldsymbol{r}}$) from the syzygy $\mathbf{S}_{\text{E-M}}$ frame to the Earth-center inertial frame for the Earth-escaping segment, and transform the state ($\boldsymbol{r}$, $\dot{\boldsymbol{r}}$) from the syzygy $\mathbf{S}_{\text{E-M}}$ frame to the Moon-center inertial frame for the Moon-captured segment; (*iii*) convert the semi-major axis $a$ and eccentricity $e$ before the epoch ($t$<0) from the Earth-escaping segment based on the Earth's gravitational coefficients, and convert $a$ and $e$ after the epoch ($t$>0) from the Moon-captured segment based on the lunar gravitational coefficients. The initial conditions are: the initial lunar phasic angle at the epoch time ($t$=0) is $\theta_{s0}$=$0^0$ for the four figures, and the initial solar phasic angle $\beta$ at the epoch time ($t$=0) is $286^0$, and the integral initial values to produce Fig.14 is $[x_{LL_1}, 0, 0, 0, 0, 0]^T$.

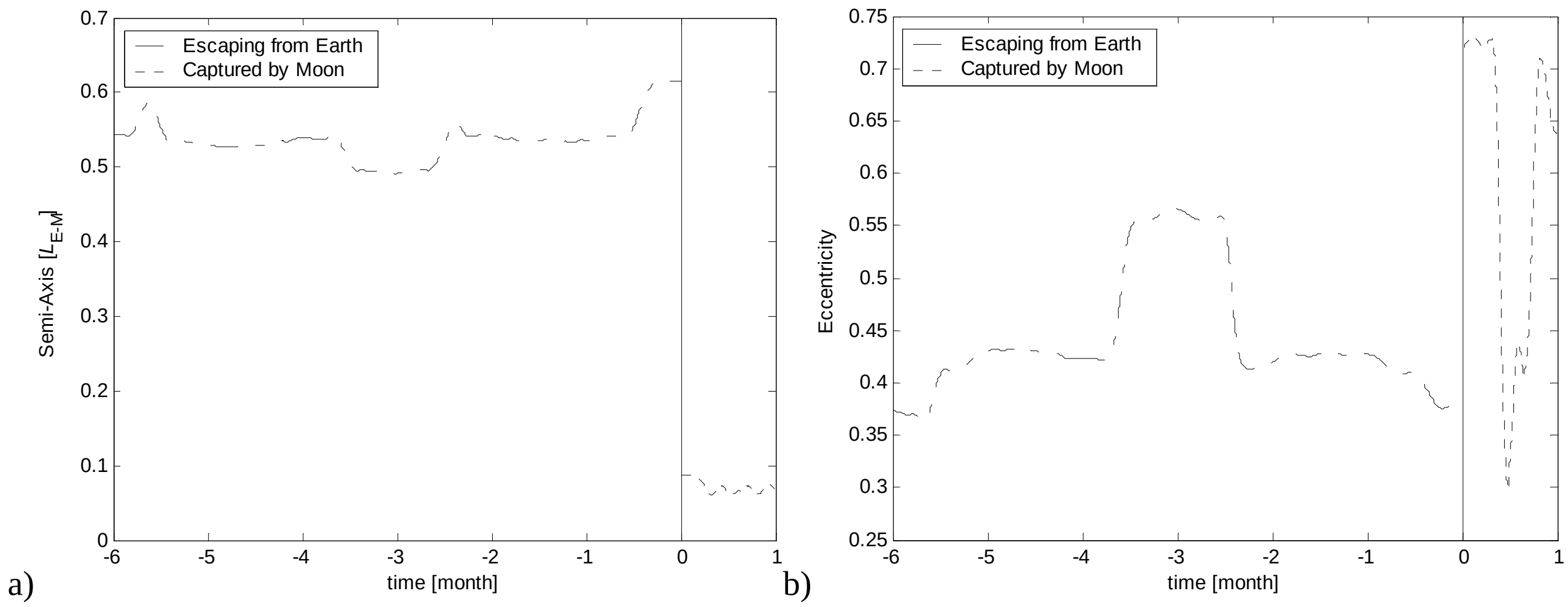

**Fig.14 Osculating semi-major axis and eccentricity of a cislunar transfer trajectory for $\beta$=286$^0$: *a*)** the history of the osculating semi-major axis; ***b***) the history of the osculating eccentricity. The dash-dotted lines illustrate the elements during escaping from the Earth, and the dashed lines illustrate the elements during being captured by the Moon, and the vertical solid lines indicate the time epoch ($t$=0); the osculating semi-major axis and eccentricity before the epoch ($t$<0) are conversed from the position and velocity in the Earth-center inertial frame, and the osculating elements after the epoch ($t$<0) are conversed in the Moon-center inertial frame; compared with the lunar perturbation before the epoch, the Earth has more perturbation on the osculating orbital elements conversed in the Moon-center inertial frame; this cislunar transiting trajectory is classified as low-energy transfer because the eccentricities before and after the epoch are less than 1, compared to the hyperbolical velocity captured by the Moon in classical Hohmann transfer (like Apollo (NASA) and Chang'E (China) missions).

The low-energy cislunar transfers transiting $LL_1$ point have the minimum energy, because the $LL_1$ point has the minimum energy in itself compared to $LL_2$ point and periodic orbits near the two equivalent equilibria.

### 3.2 Low-energy transfers by transiting $LL_2$ point

Similar to transiting $LL_1$ point, the solar perturbation will change topologically the invariant manifolds to fail in transiting $LL_2$ point for some phasic angles $\beta\in[0, 2\pi]$; however, the transiting manifolds are preserved for the other values of $\beta$. For the available interval of $\beta$, i.e., the trans-lunar transfer opportunities for $LL_2$ point can be produced by the procedure developed for $LL_1$ point. The initial conditions are: the initial lunar phasic angle is $\theta_{s0}$=0$^0$, and the solar phasic angle $\beta$ ranges from 0$^0$ to 360$^0$, and the initial values to integrate forwards and backwards Eq.(4) are equally [$x_{LL_2}$, 0, 0, 0, 0, 0]$^T$. The trans-lunar transfer opportunities is illustrated in Fig.15.

Different from the only type of cislunar trajectories transiting $LL_1$ point, all the transfer trajectories transiting $LL_2$ point are classified as the inner cislunar trajectories and the outer WSB trans-lunar ones. The former is essentially the cislunar transfer trajectories passing through $LL_1$ point, and costs more fuels than the cislunar trajectories transiting $LL_1$ point. While the latter has the same geometrical shape in the inertial frame as Belbruno's theory and is named after outer trans-lunar WSB trajectories in this paper (**Belbruno and Miller 1993**; **Belbruno 2004**), which can be considered as the patched connection between the invariant manifolds near $EL_1$ (or $EL_2$) point and unstable manifolds near $LL_2$ point (**Koon et al. 2001**). Therefore, the former is not a fuel-efficient Earth-to-Moon transfer, and only the trans-lunar WSB trajectories are employed in this paper to transit $LL_2$ point. Due to the harsh conditions of two patched manifolds, only a few of intervals can be used to construct the whole WSB transfer trajectories from the Earth to $LL_1$ point and then to the Moon, which are bounded by the vertical dashed lines in Fig.15, i.e., $\beta \in [21.8^0, 23.3^0] \cup [201.5^0, 203^0]$. Compared to the cislunar transfer opportunities listed in Fig.11, the WSB transfers have fewer opportunities to transit $LL_2$ point.

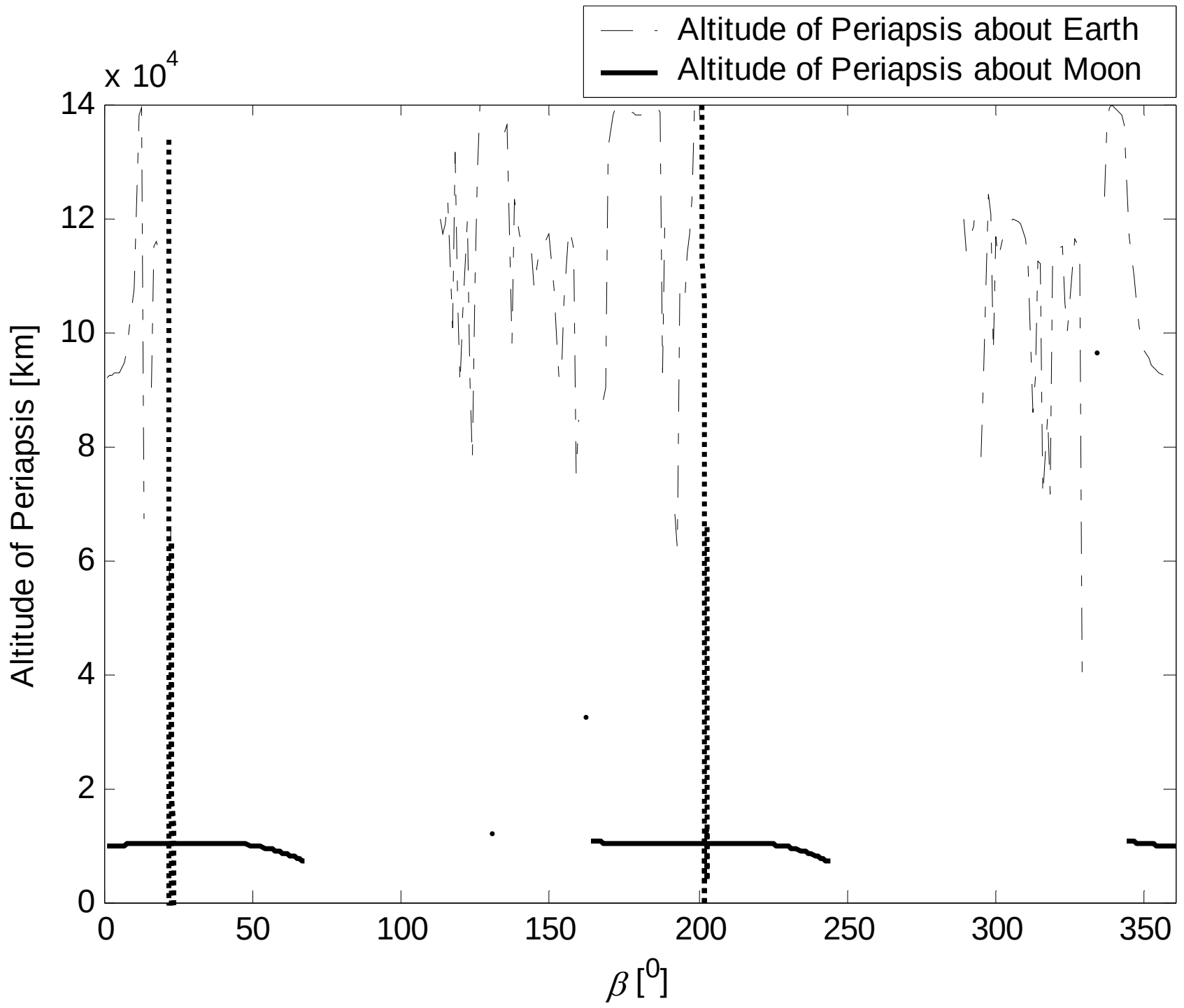

**Fig.15 WSB Transfer opportunities by transiting $LL_2$ point measured by the solar phasic angle $\beta$**: the solid lines illustrate the available altitudes of periapsis about the Earth, and the dash-dotted lines illustrate the available altitudes of periapsis about the Moon; the intersection of the intervals transiting from the Earth to $LL_1$ point and from $LL_1$ point to the Moon is bounded by the vertical dashed lines, i.e., $[21.8^0, 23.3^0] \cup [201.5^0, 203^0]$, which drives the trajectories to orbit successively the Earth and Moon; the gaps between the altitudes of periapsis are caused by some $\beta$s intervals failing the invariant manifolds in transiting $LL_2$ point.

The procedures to produce transfer trajectories transiting $LL_1$ point in the rotating $S_{E\text{-}M}$ frame and the inertial $\mathbf{I}_{\mathbf{E\text{-}M}}$ frame can be employed to produce the two types of inner trans-lunar and outer trans-lunar transfer trajectories transiting $LL_2$ point, as shown in Figs.16, 17, 19 and 20. For the four figures, the initial lunar phasic angle at the epoch time ($t$=0) is $\theta_{s0}=0^0$, and the integral initial value is $[x_{LL_2}, 0, 0, 0, 0, 0]^T$; for Figs.16 and 17, the initial solar phasic angle $\beta$ at the epoch time ($t$=0) is $193^0$; and for Figs.19 and 20, the initial solar phasic angle $\beta$ at the epoch time ($t$=0) is $202^0$.

Furthermore, the procedures to create the osculating semi-major axis and eccentricity for a cislunar transfer trajectory in Section 3.1 can be used to deal with the $LL_2$ point case. The time history of the orbital elements is presented in Figs.18 and 21 respectively for typical inner trans-lunar and outer WSB trans-lunar transfer trajectories transiting $LL_2$ point. The initial lunar phasic angle and the integral initial value are $\theta_{s0}=0^0$ and $[x_{LL_2}, 0, 0, 0, 0, 0]^T$ for the two figures, and the initial solar phasic angle $\beta$ of Fig.18 is $193^0$, and the initial solar phasic angle $\beta$ of Fig.21 is $202^0$.

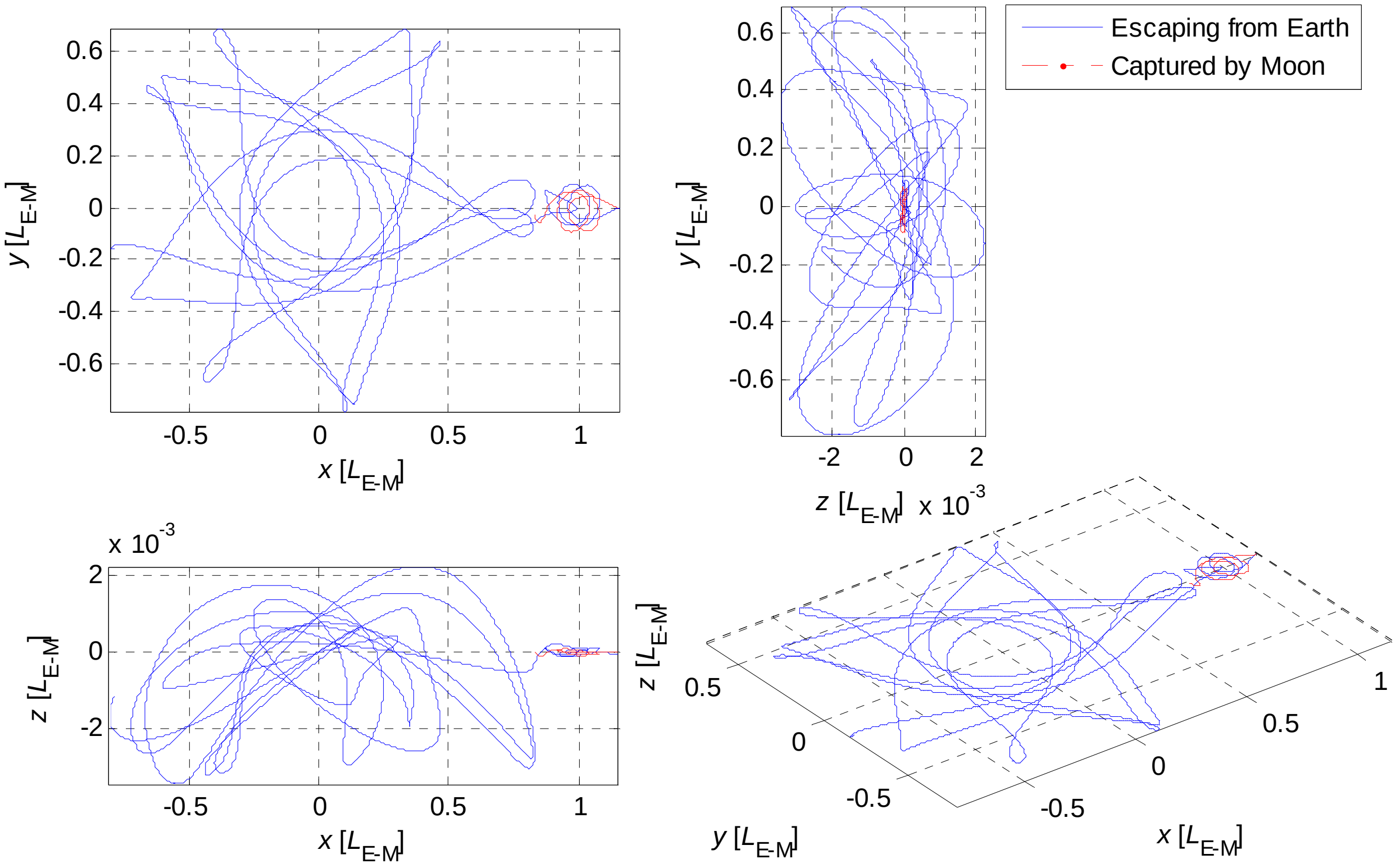

**Fig.16 Typical inner trans-lunar transfer trajectory transiting $LL_2$ point in the rotating $S_{E\text{-}M}$ frame for $\beta$=193$^0$**: the blue solid lines illustrate the segment escaping from the Earth, and the red dash-dotted lines illustrate the segment captured by the Moon; the $z$ component ranges between $-3\times10^{-3}$~$+3\times10^{-3}$ ($L_{E\text{-}M}$), while the $x$ and $y$ components range between -0.9~+1.2 and -0.8~+0.8 ($L_{E\text{-}M}$), which indicates the spatial perturbation has few effects on the low-energy transfer.

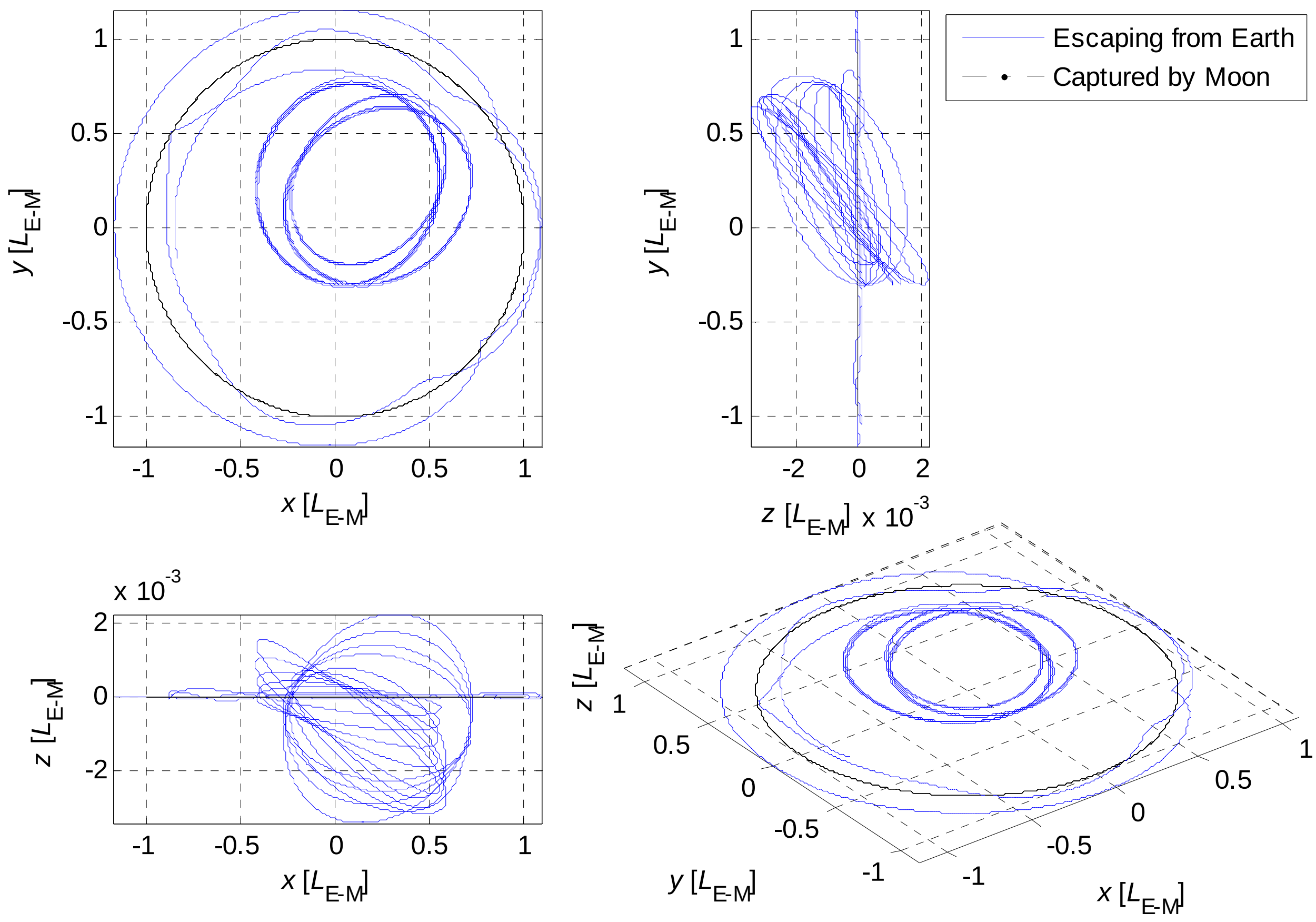

**Fig.17 Typical inner trans-lunar transfer trajectory transiting $LL_2$ point in the inertial $\mathrm{I}_{\text{E-M}}$ frame for $\beta$=193$^0$**: the blue solid lines illustrate the segment escaping from the Earth, and the red dash-dotted lines illustrate the segment captured by the Moon; the $z$ component ranges between $-3\times10^{-3}$~$+3\times10^{-3}$ ($L_{\text{E-M}}$), while the $x$ and $y$ components range between -0.9~+1.2 and -0.8~+0.8 ($L_{\text{E-M}}$), which indicates the spatial perturbation has fewer effects on the low-energy transfer.

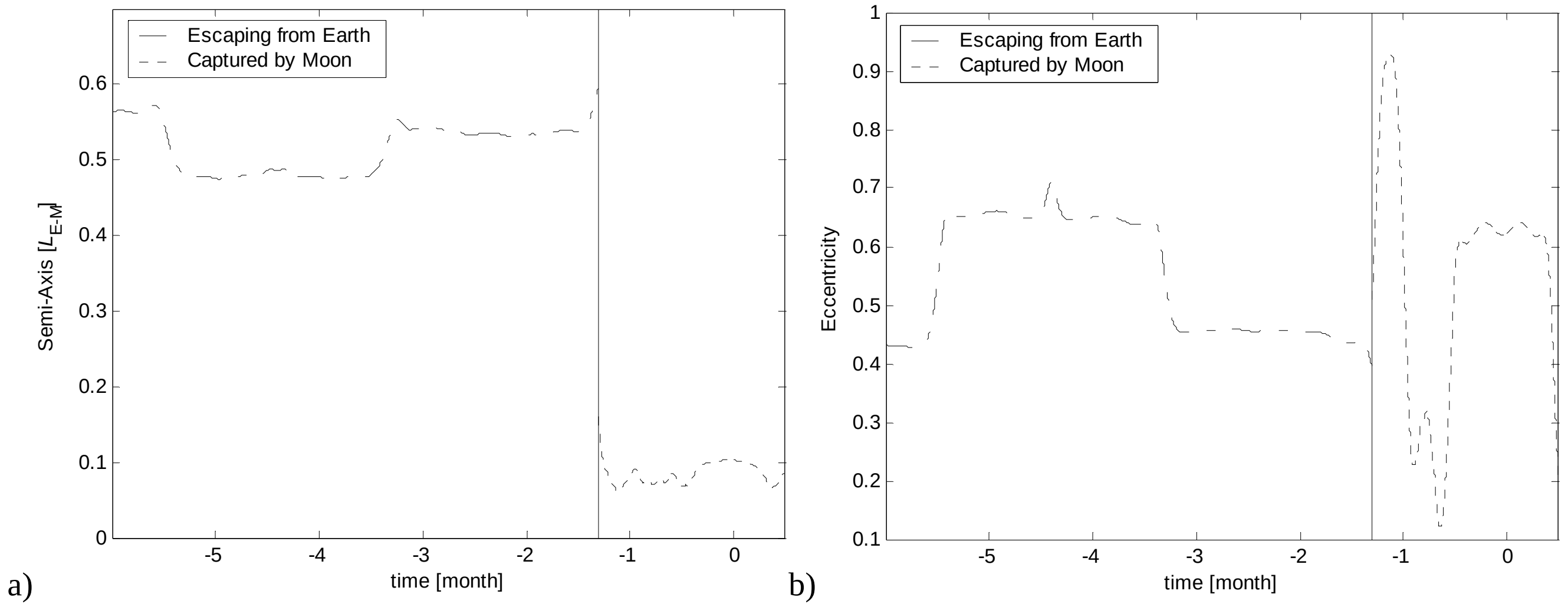

**Fig.18 Osculating semi-major axis and eccentricity of a cislunar transfer trajectory for $\beta$=193$^0$:** ***a***) the history of the osculating semi-major axis; ***b***) the history of the osculating eccentricity. The dash-dotted lines illustrate the elements during escaping from the Earth, and the dashed lines illustrate the elements during captured by the Moon, and the vertical solid lines illustrate the time epoch ($t$=0); the osculating semi-major axis and eccentricity before the epoch ($t$<0) are conversed from the position and velocity in the Earth-center inertial frame, and the osculating elements after the epoch ($t$<0) are conversed in the Moon-center inertial frame; compared with the Moon before the epoch, the Earth has more perturbation on the osculating orbital elements conversed in the Moon-center inertial frame; this cislunar transiting trajectory is classified as low-energy transfer because the eccentricities before and after the epoch are less than 1, compared to the hyperbolical velocity captured by the Moon in classical Hohmann transfer (like Apollo (NASA) and Chang'E (China) missions).

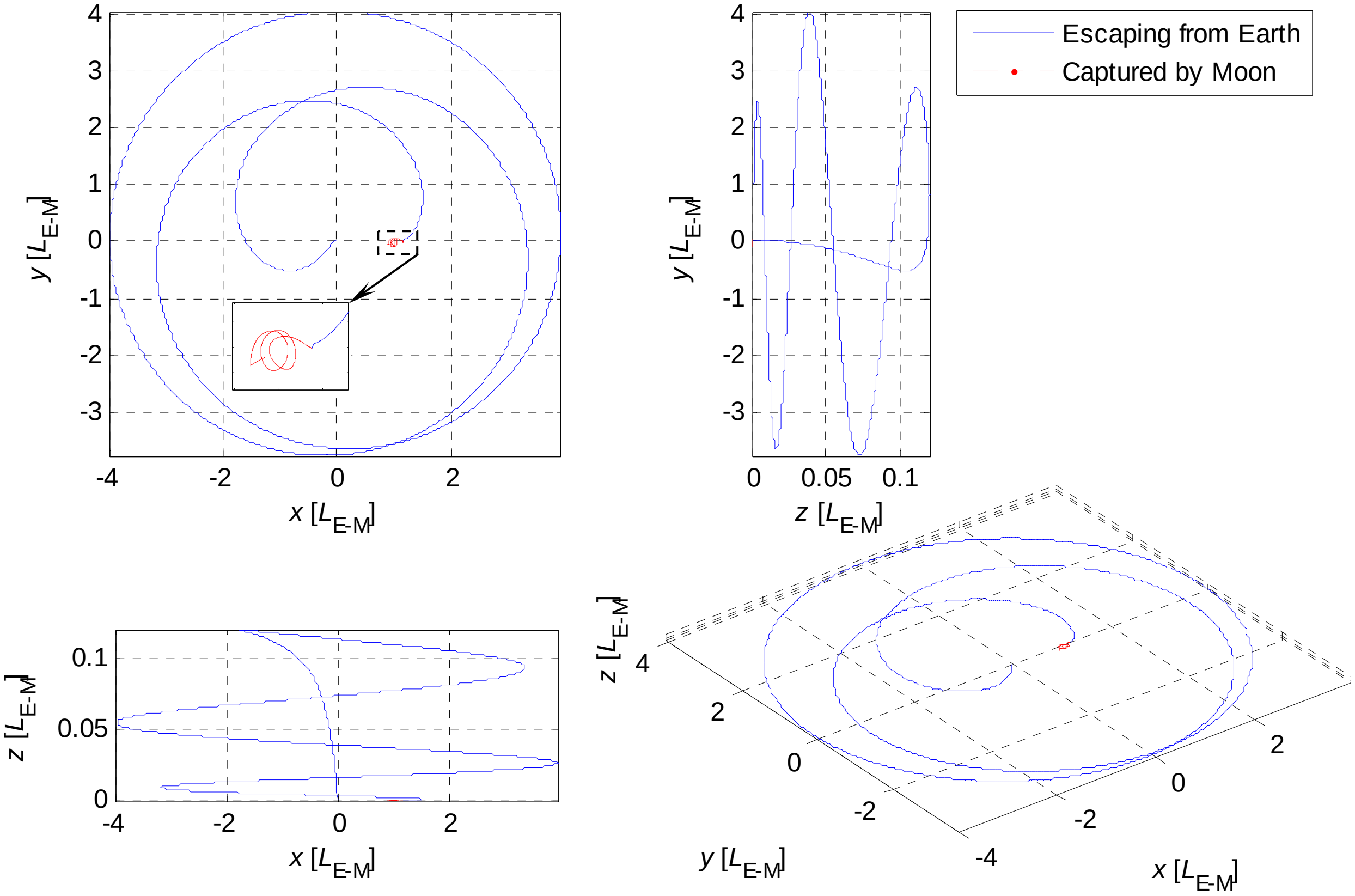

**Fig.19 Typical outer trans-lunar transfer trajectory transiting $LL_2$ point in the rotating $S_{E-M}$ frame for $\beta$=202$^0$:** the blue solid lines illustrate the segment escaping from the Earth, and the red dash-dotted lines illustrate the segment captured by the Moon; the $z$ component ranges between 0~+0.12 ($L_{E-M}$), while the $x$ and $y$ components range between -4~+4 and -4~+4 ($L_{E-M}$), which indicates the spatial perturbation has few effects on the low-energy transfer.

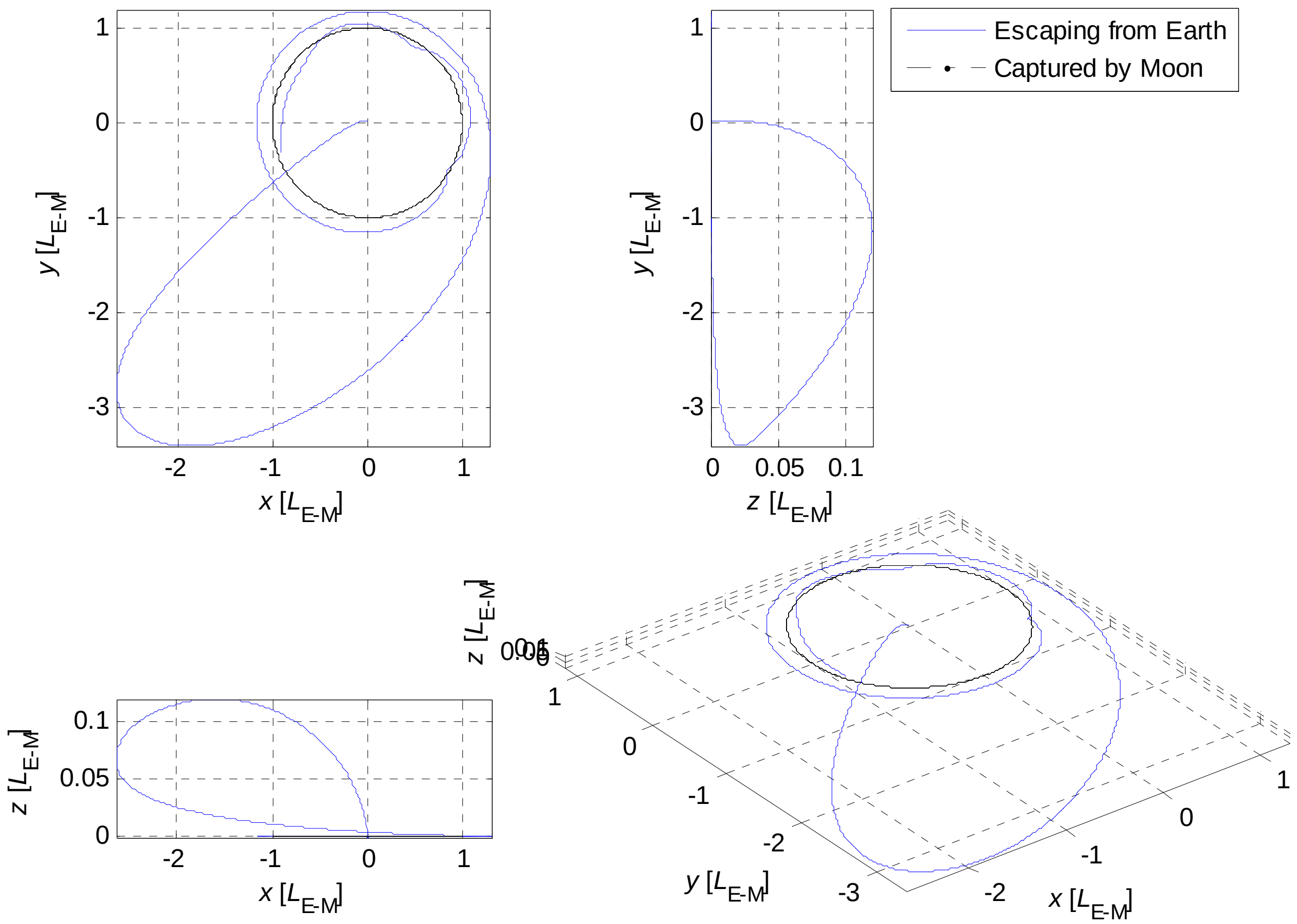

**Fig.20 Typical outer trans-lunar transfer trajectory transiting $LL_2$ point in the inertial $I_{E-M}$ frame for $\beta$=202$^0$**: the blue solid lines illustrate the segment escaping from the Earth, and the red dash-dotted lines illustrate the segment captured by the Moon; the $z$ component ranges between 0~+0.12 ($L_{E-M}$), while the $x$ and $y$ components range between -2.7~+1.3 and -3.4~+1.2 ($L_{E-M}$), which indicates the spatial perturbation has few effects on low-energy transfer; this type of transfer trajectories in the inertial frame has the same geometrical shape as Belbruno's WSB theory (**Belbruno and Miller 1993**; **Belbruno 2004**), which is renamed as outer WSB trajectories as well.

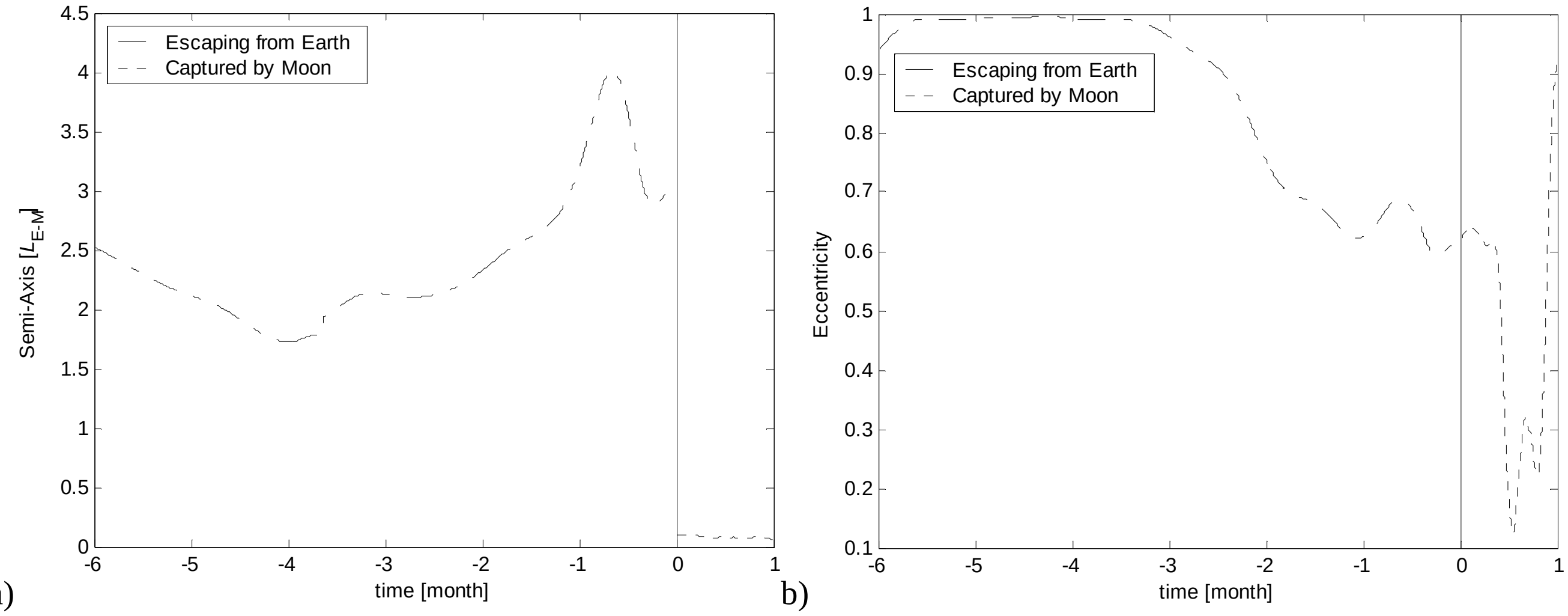

a) b)

**Fig.21 Osculating semi-major axis and eccentricity of a cislunar transfer trajectory for *β*=202$^0$: *a*)** the history of the osculating semi-major axis; ***b***) the history of the osculating eccentricity. The dash-dotted lines illustrate the elements during escaping from the Earth, and the dashed lines illustrate the elements during captured by the Moon, and the vertical solid lines illustrate the time epoch ($t$=0); the osculating semi-major axis and eccentricity before the epoch ($t$<0) are conversed from the position and velocity in the Earth-center inertial frame, and the osculating elements after the epoch ($t$<0) are conversed in the Moon-center inertial frame; compared with the lunar perturbation before the epoch, the Earth has more perturbation on the osculating orbital elements conversed in the Moon-center inertial frame; this cislunar transiting trajectory is classified as low-energy transfer because the eccentricities before and after the epoch are less than 1, compared with the hyperbolical velocity captured by the Moon in classical Hohmann transfer (like Apollo (NASA) and Chang'E (China) missions).

The cislunar transfer trajectories transiting $LL_1$ point have a total opportunities measured by $\Delta\beta$=89$^0$, while the outer WSB trans-lunar trajectories have much fewer opportunities of $\Delta\beta$=3$^0$. Thus, an effective way to increase the transfer opportunities for the WSB trajectories is to transit a halo orbit near $LL_2$ point instead of itself.

## 4. Low-energy Transfers by Transiting Halo Orbits

Compared with the only variable (i.e., $\beta$) to design a transfer trajectory transiting the libration point, the halo orbit is employed to increase the transfer opportunities by introducing another variable (i.e., serial points of halo orbit). Subsequently, a global investigation on the Earth-escaping and the Moon-captured opportunities is implemented respectively for transiting $LL_1$ and $LL_2$ points in this section.

A halo orbit is a periodic three-dimensional orbit near $LL_1$ and $LL_2$ points, and is symmetrical about the $x$-$z$ plane in the rotating $\mathbf{S}_{\text{E-M}}$ frame (**Xu and Xu 2012**), shown in Fig.22. A halo orbit in the Earth-Moon system can be characterized by the maximum of its $y$ component or its orbital period $T_H$. Thus, all points on a specified halo orbit can be marked by the phase of halo orbit $\tau = i/N$, where $i$ is the serial number of this point measured clockwise from the starting point which is located closest to the Earth on the $x$ axis, and $N$=360 is the total number of evenly spaced points in time selected in this paper. Even through there is no halo orbit under the solar perturbation in SBCM model, the periodic orbit is still acting as a powerful tool to investigate the transfer trajectories in this paper, because both the cislunar and the trans-lunar trajectories are transiting it rather than staying on it (**Koon et al. 2001**; **Koon et al. 2007**). The algorithm to produce halo orbit is beyond the scope of this paper, which can be found in the references (**Richardson 1980**; **Xu et al. 2013**). The maximal values of the $y$ components of halo orbits in Fig.22 are respectively ±40142.16km near $LL_1$ point and ±33818.07km near $LL_2$ point.

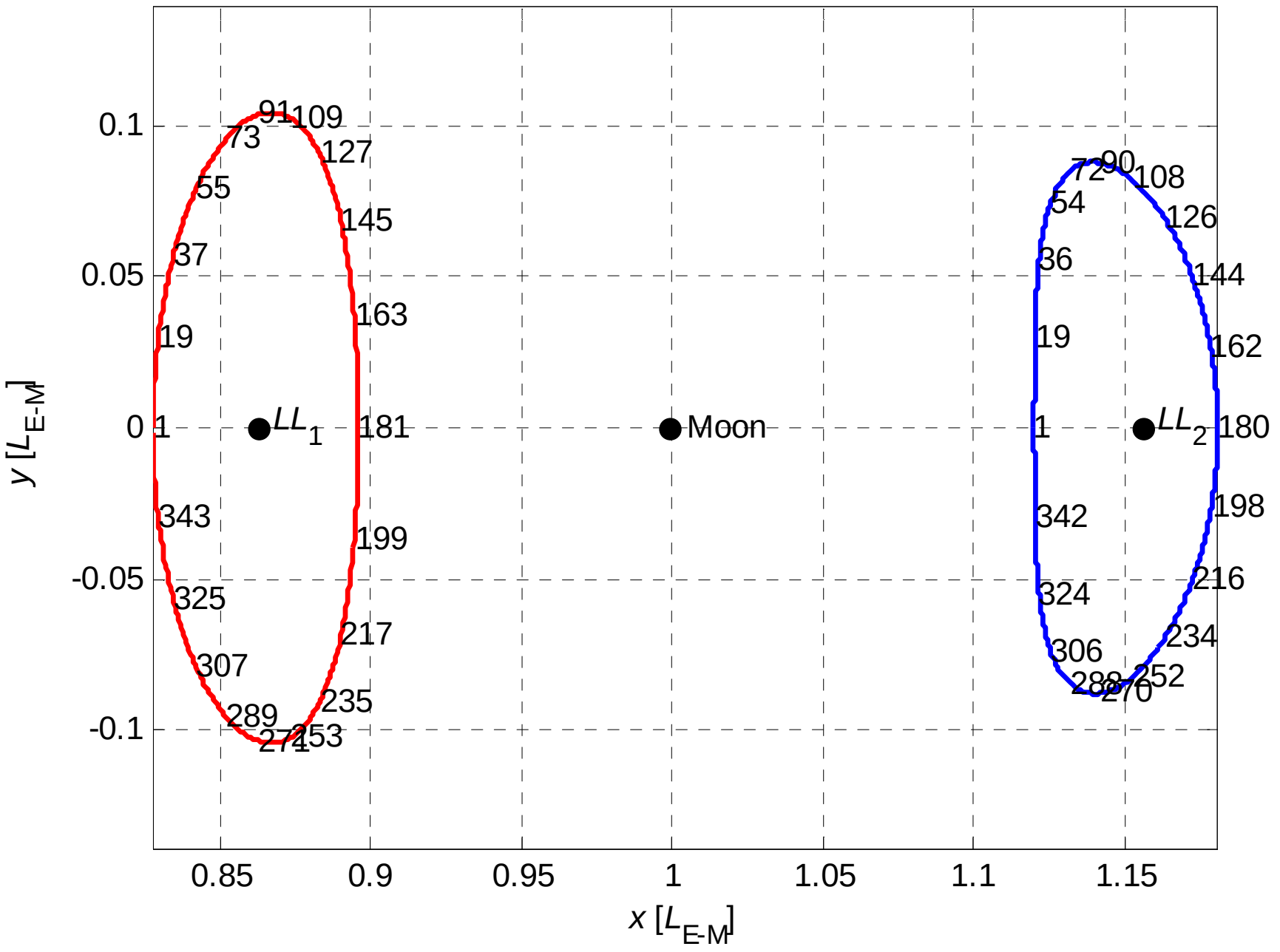

**Fig.22 Halo orbits near $LL_1$ and $LL_2$ points and serial points on the orbits**: the maximal values of the $y$ components of halo orbits are respectively ±40142.16km near $LL_1$ point and ±33818.07km near $LL_2$ point; the serial evenly spaced points in time are selected clockwise from the starting point; the starting point locates closest to the Earth on the $x$ axis.

## 4.1 Low-energy transfers by transiting halo obits near $LL_1$ point

In CR3BP model, the invariant manifolds of a halo orbit near $LL_1$ point can be classified into four branches as $W_E^S$, $W_E^u$, $W_M^S$ and $W_M^u$, where the subscript $E$ and $M$ indicate this branch leaves from/to the Earth and Moon respectively, and the superscript $s$ and $u$ indicate the branch approaches the halo orbit forwards and backwards respectively. Thus, the branches $W_M^S$ and $W_E^u$ construct a whole Moon-to-Earth transfer trajectory labeled by light-colored lines in the right subgraph of Fig.23, while $W_E^S$ and $W_M^u$ construct a whole Earth-to-Moon transfer trajectory labeled by dark-colored lines in the right subgraph of Fig.23; however, both of the two trajectories are not practical due to the infinite durations. The algorithm to produce invariant manifolds of halo orbit is beyond the scope of this paper, which can be found in the references (**Howell et al. 1997**; **Howell et al. 2006**). The maximal values of the $y$ components of halo orbits in this figure are respectively ±40142.16km near $LL_1$ point and ±33818.07km near $LL_2$ point.

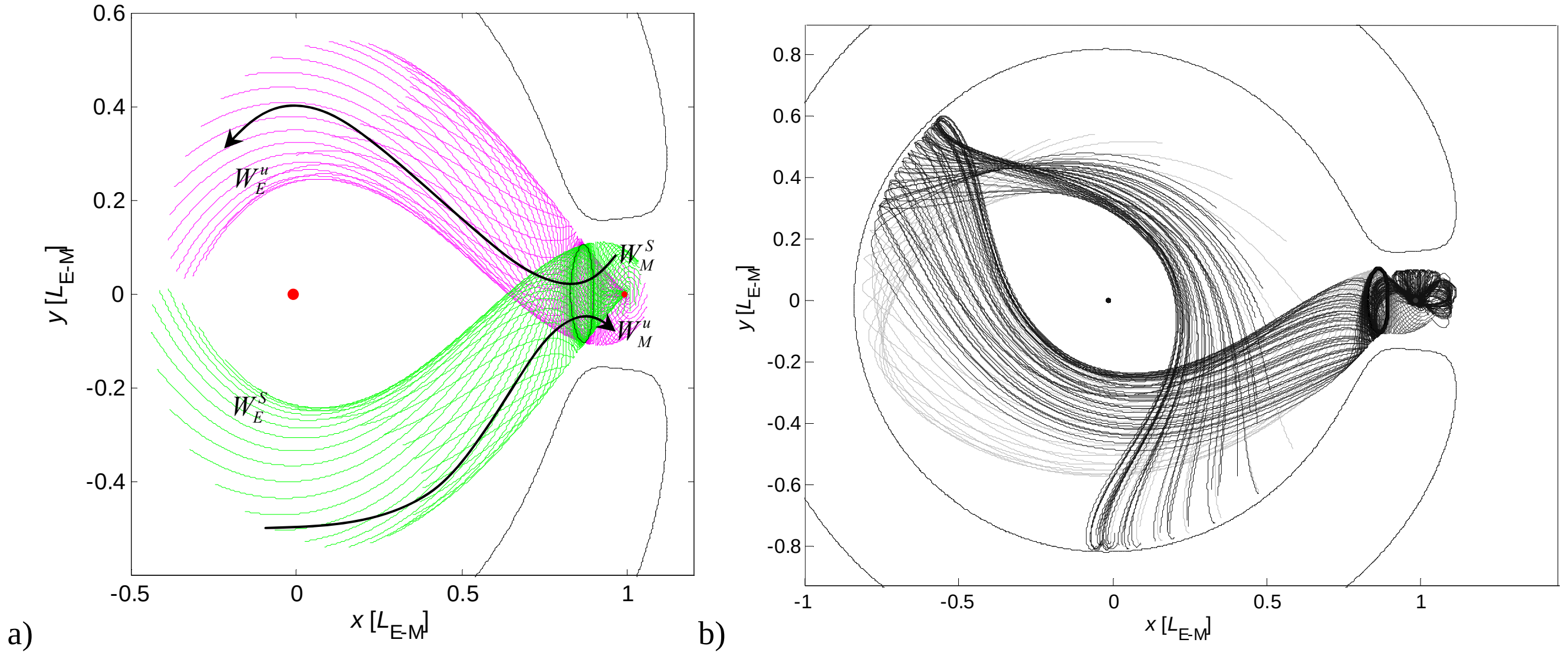

**Fig.23 Invariant manifolds of a halo orbit and cislunar transfer trajectories constructed from these manifolds**: a) $W_E^S$ illustrates the stable branch leaving from the Earth to the halo orbit, and $W_E^u$ illustrates the unstable branch leaving from the halo orbit to the Earth, and $W_M^S$ illustrates the stable branch leaving from the Moon to the halo orbit, and $W_M^u$ illustrates the unstable branch leaving from the halo orbit to the Moon; b) the branches $W_M^S$ and $W_E^u$ construct a whole Moon-to-Earth transfer trajectory labeled by light-colored lines, while $W_E^S$ and $W_M^u$ construct a whole Earth-to-Moon transfer trajectory labeled by dark-colored lines.

Fortunately, the perturbation of the solar gravity employed by the SBCM model may cut down the infinite durations of some branches to finite ones, which is quite practical for Earth-to-Moon transfers. For the pairs $(\beta, \tau)\in[0, 2\pi]\times[0, 1]$ not transiting halo orbit near $LL_1$ point, the perturbed manifolds will lose the phase of orbiting the Earth or the Moon, i.e., there is no periapsis about the Earth or the Moon in this case. Only the intersecting pairs of transiting from the Earth to halo orbit and another intervals transiting from halo orbit to the Moon, i.e., $([95^0, 150^0]U[262^0, 345^0])\times[0, 1]$ and $([100^0, 200^0]U[270^0, 30^0])\times([0.16, 0.26]U[0.47, 0.58])$, can drive the trajectories to orbit successively the Earth and Moon, which is considered as the cislunar transfer opportunities shown in Fig.24.

The procedure to produce the transfer opportunities is presented as following: (*i*) vary $\beta$ in the interval of $[0^0, 360^0]$ and the phase of serial points on halo orbit $\tau\in[0, 360]/360$ to integrate the SBCM dynamics formulized by the differential Eq. (4) backwards to yield the Earth-escaping segment, and integrate forwards to yield the Moon-captured segment; (*ii*) collect the altitudes of periasis when the Earth-escaping or the Moon-captured segments reach their first periasis, and then draw them by the contour-map of $r_p$, which are considered as cislunar or trans-lunar transfer opportunities; (*iii*) the two integrations (forwards and backwards) have the same initial condition of $\boldsymbol{X}= [\boldsymbol{r}, \ \dot{\boldsymbol{r}}]|_\tau$ of a serial point on the halo orbit. Only some subintervals of $\beta$ and $\tau$ can make the integrated trajectories closer to the Earth or Moon. The initial conditions are: the initial lunar phasic angle is $\theta_{s0}=0^0$, and the solar phasic angle $\beta$ ranges from $0^0$ to $360^0$, and the phase of serial points on halo orbit $\tau$ ranges from 0 to 1, and the maximal $y$ component of halo orbit to integrate forwards and backwards Eq.(4) is $\pm$40142.16km.

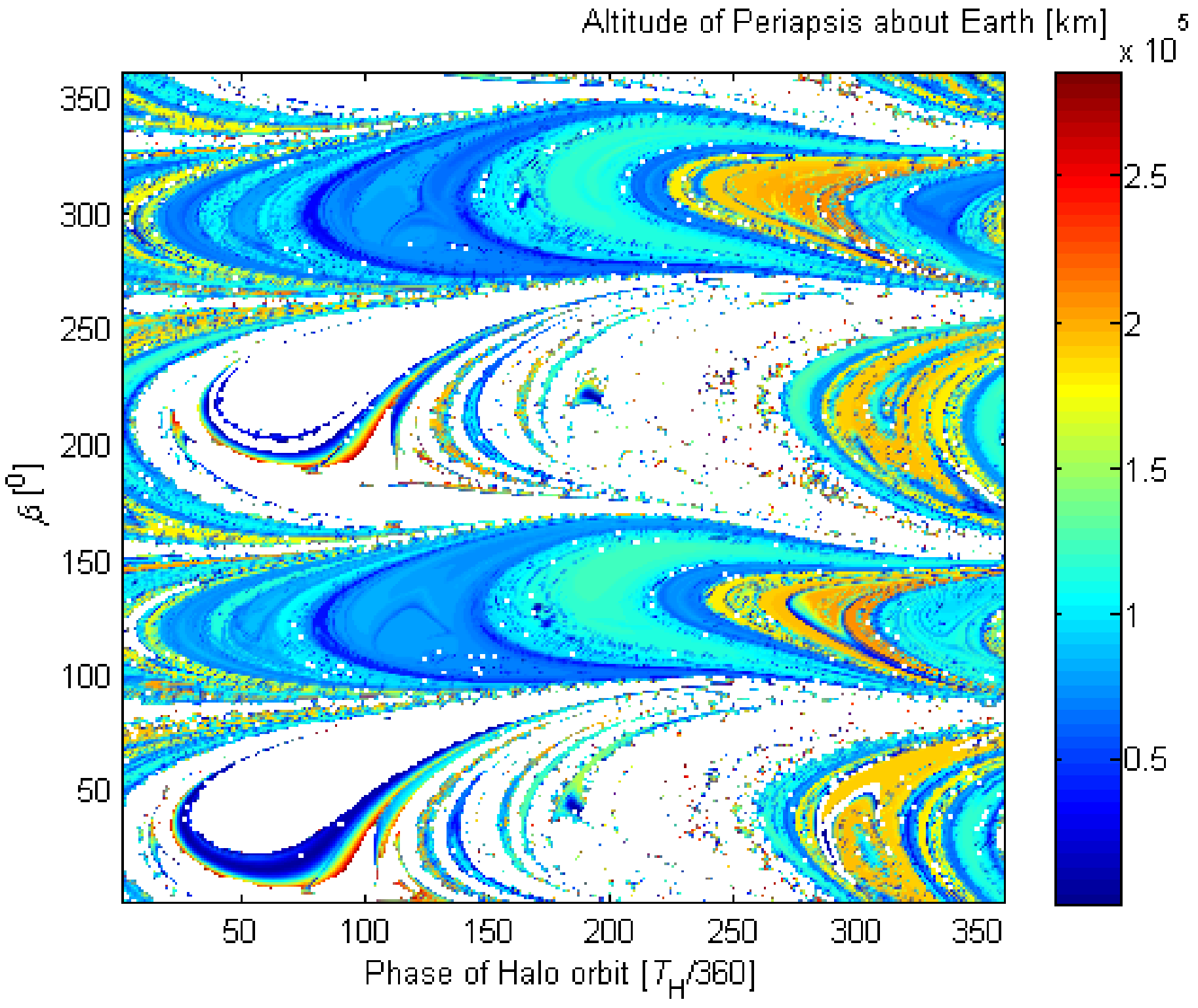

a)

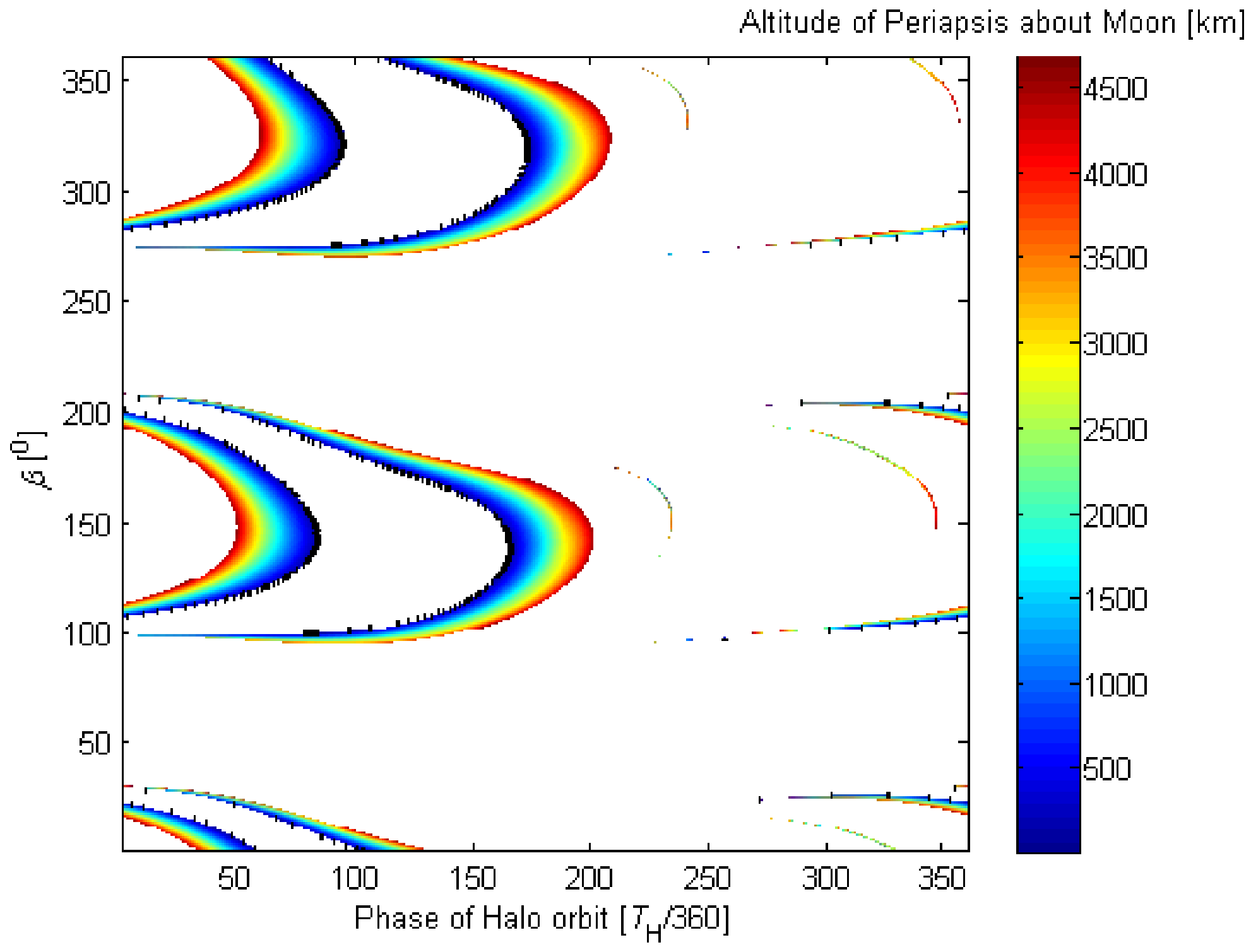

b)

**Fig.24 Contour-map of transfer opportunities for trans-lunar WSB trajectories**: a) transfer opportunities for the Earth-escaping segments; b) transfer opportunities for the Moon-captured segments; the solar phasic angle $\beta$ has more effects on the existence of the trajectories than the phase of halo orbit, and most of them

are located within the ($\beta$, $\tau$) pairs of ([$95^0$, $150^0$]U[$262^0$, $345^0$])×[0, 1] (for the Earth-escaping segment) and ([$100^0$, $200^0$]U[$270^0$, $30^0$])×([0.16, 0.26]U[0.47, 0.58]) (for the Moon-captured segment); all the cislunar trajectories mapped from the contour-map have the similar geometrical shape with the typical trajectories shown in Fig.25.

The procedures to produce transfer trajectories transiting $LL_1$ point in the rotating $S_{E\text{-}M}$ frame can be employed to produce the cislunar transfer trajectories transiting halo orbit in Fig.25: for some specified pair ($\beta$, $\tau$), integrate Eq.(4) backwards from a halo orbit to obtain the Earth-escaping segment and forwards to achieve the Moon-captured segment in the rotating $\mathbf{S}_{E\text{-}M}$ frame. The initial conditions are listed as following: the initial lunar and solar phasic angle at the epoch time ($t$=0) is $\theta_{s0}=0^0$ and $\beta=150^0$ respectively, and the two integrations (forwards and backwards) have the same initial condition $\boldsymbol{X}=[\boldsymbol{r},\ \dot{\boldsymbol{r}}]|_{\tau=82/360}$ of a serial point on the halo orbit with its maximal $y$ component equal to ±40142.16km.

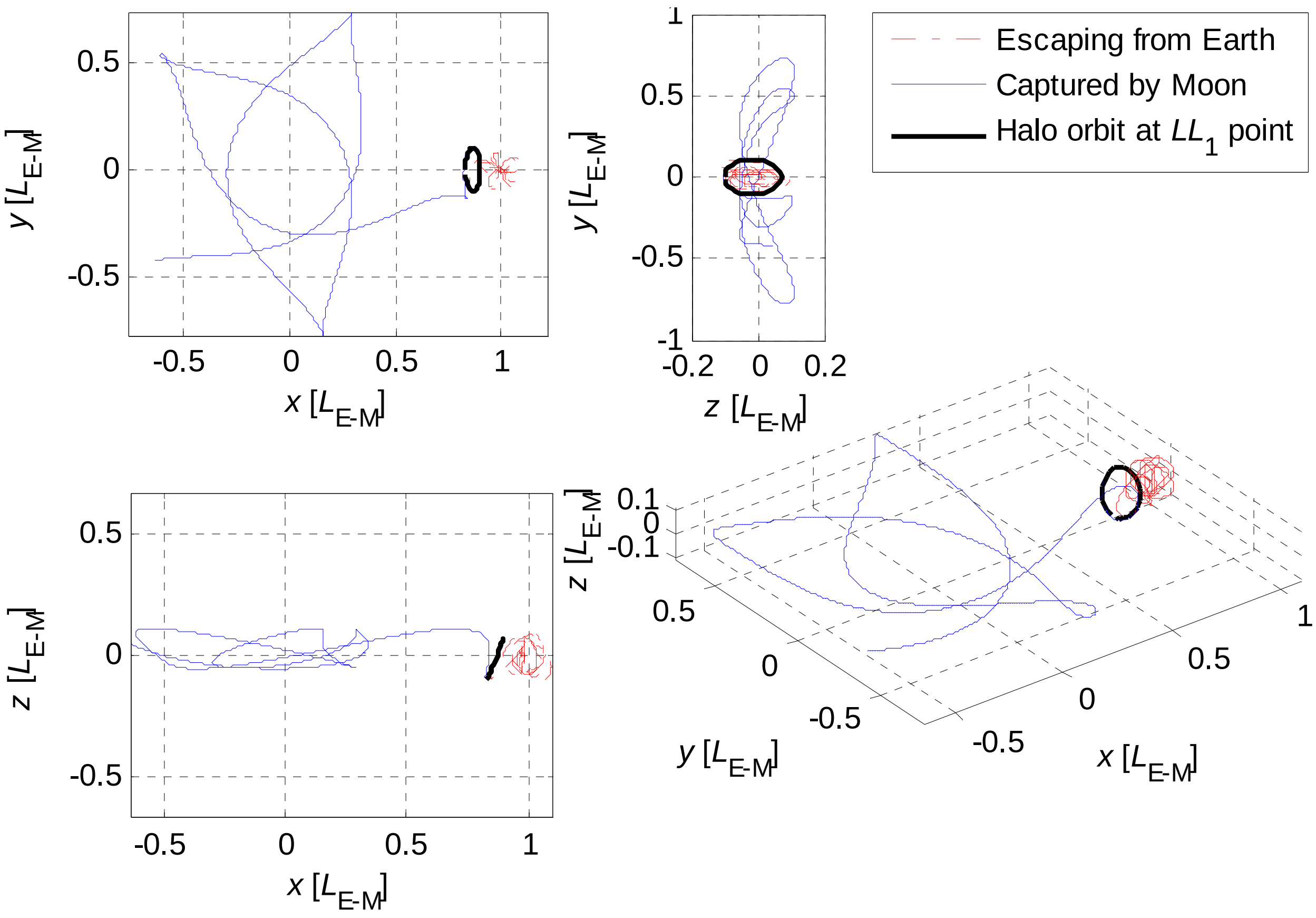

**Fig.25 Typical cislunar transfer trajectory transiting a halo obit near $LL_1$ point in the rotating $S_{E\text{-}M}$ frame for $\beta=150^0$ and $\tau$=82/360**: the blue solid lines illustrate the segment escaping from the Earth, and the red dash-dotted lines illustrate the segment captured by the Moon, and the black thick lines illustrate the halo orbit; the $x$ and $y$ components range between -0.9~+1.2 and -0.8~+0.8 ($L_{E\text{-}M}$), while the $z$ component

ranges between -1.1~+1.1 ($L_{E\text{-}M}$), which is larger than the $z$ component of a trajectory transiting $LL_1$ point; the maximal $y$ component of the halo obit is ±40142.16km.

It is worth mentioning that the numerical simulations indicate all the cislunar trajectories mapped from the contour-map have the similar geometrical shape with the typical trajectories shown in Fig.25.

## 4.2 Low-energy transfers by transiting halo obits near $LL_2$ point

Similar to transiting $LL_2$ point, the inner transfer trajectories transiting halo orbit near $LL_2$ point is essentially the cislunar transfer trajectories passing through $LL_1$ point, and costs more fuels than the cislunar trajectories achieved in Section 4.1. Hence, only the trans-lunar WSB trajectories are employed in this paper so as to construct some practical transfer trajectories. According to the work of **Koon et al.** (**2001**), there are smooth-patched manifolds on a Poincaré section to drive the spacecraft flying from the Earth to another halo orbit near $EL_1$ (or $EL_2$) point and then to the targeting halo orbit near $LL_2$ point. Thus, the following investigation will verify that both the invariant manifolds of halo orbits near $EL_1$ and $EL_2$ points can be used to construct the whole trans-lunar WSB trajectories.

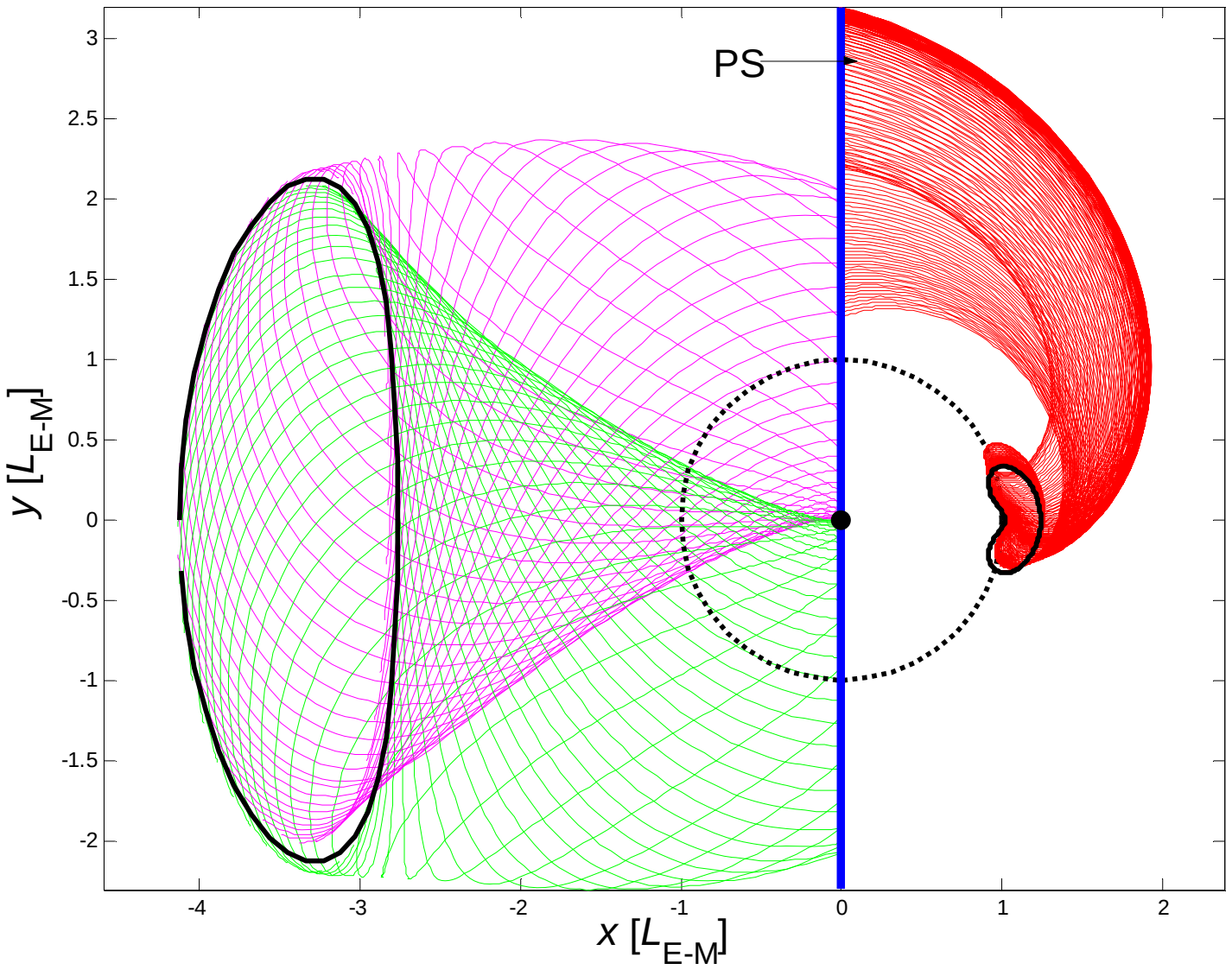

**Fig.26 The conceptual geometry of two smooth-patched manifolds on a Poincaré section**: both the invariant manifolds of halo orbits near $EL_1$ and $EL_2$ points can be used to construct the whole trans-lunar WSB trajectories.

The procedure developed for the transfer opportunities in the above section can also be used to create the colorful Porkchop-like contour-maps of transfer opportunities for the Earth-escaping and the Moon-captured segments transiting halo orbit near $LL_1$ point. The initial conditions are: the initial lunar phasic angle is $\theta_{s0}=0^0$, and the solar phasic angle $\beta$ ranges from $0^0$ to $360^0$, and the phase of serial points on halo orbit $\tau$ ranges from 0 to 1. The maximal $y$ component of halo orbit to integrate forwards and backwards Eq.(4) is ±33818.07km.

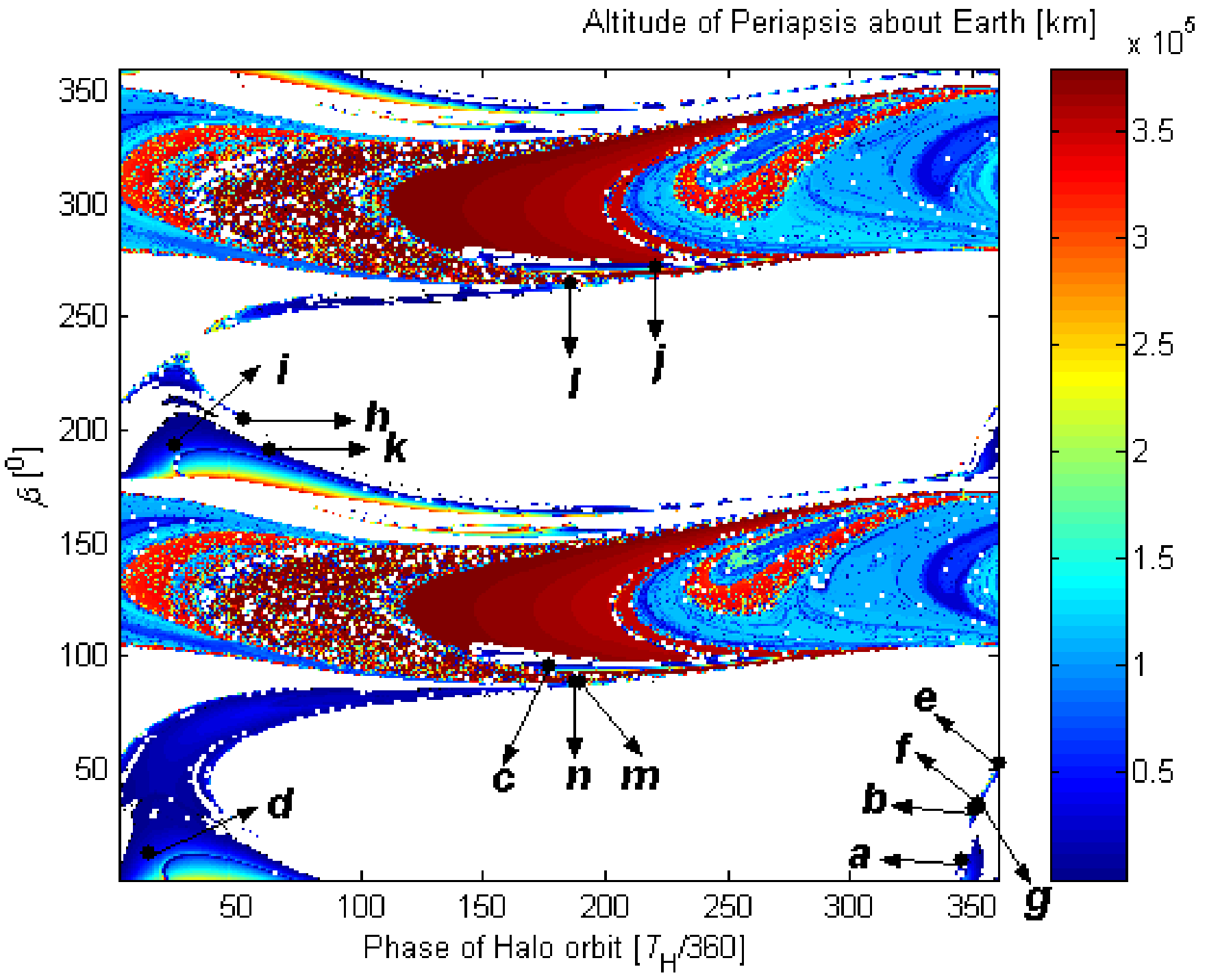

a)

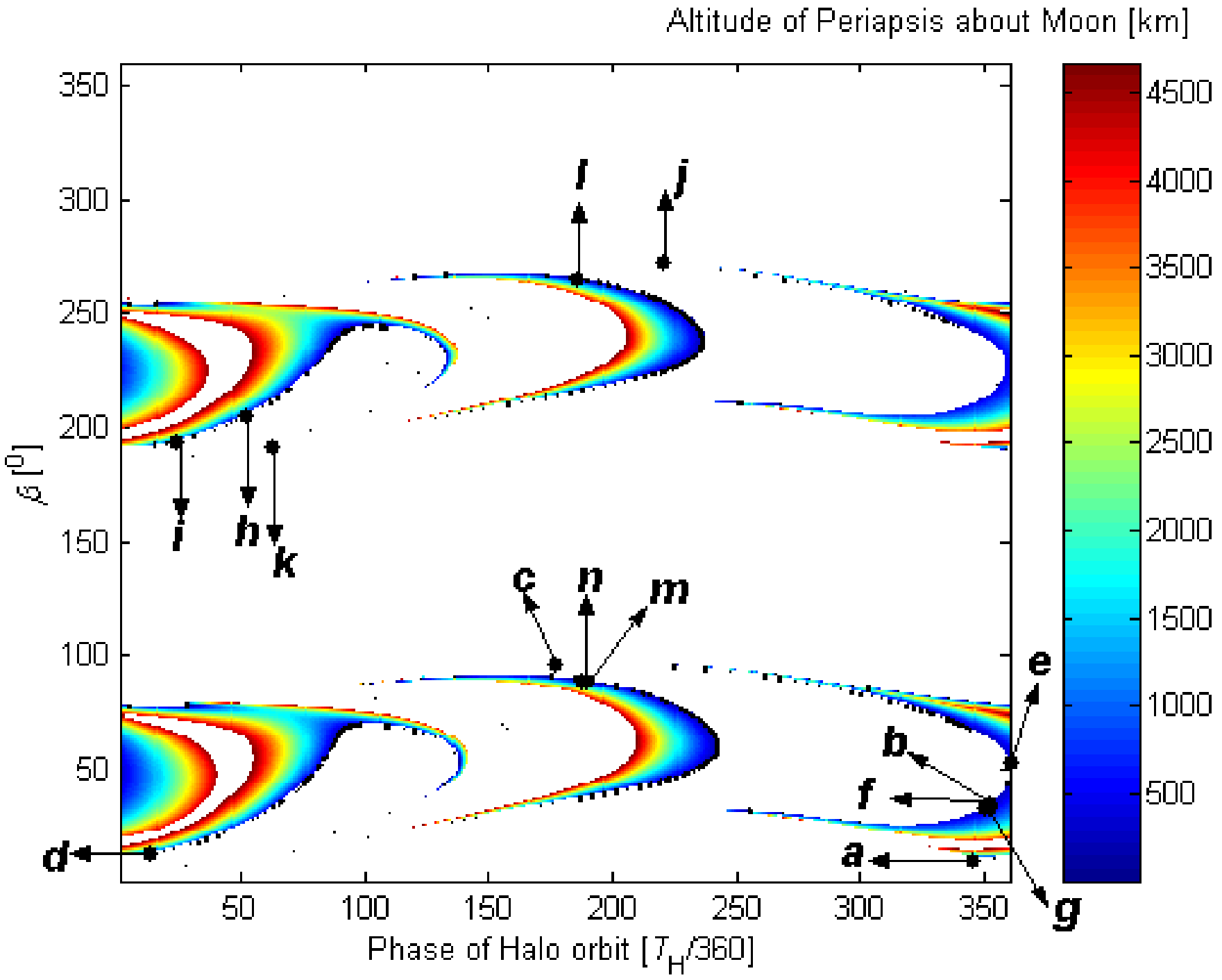

b)

**Fig.27 Contour-map of transfer opportunities for trans-lunar WSB trajectories**: a) transfer opportunities for the Earth-escaping segments; b) transfer opportunities for the Moon-captured segments; the solar phasic angle $\beta$ has more effects on the existence of the trajectories than the phase of halo orbit, and most of them

are located in the $\beta$s intervals of $[85^0, 165^0]\text{U}[262^0, 330^0])\times[0, 1]$ (for the Earth-escaping segment) and $([10^0, 92^0]\text{U}[192^0, 268^0])\times([0, 0.26]\text{U}[0.57, 0.66])$ (for the Moon-captured segment); the phase of halo orbit has more effects on the altitudes of periapsis about the Earth and Moon; the points labeled as ***a***, ***b***, ..., ***n***, are mapped into 14 typical trans-lunar WSB trajectories, as shown in the rotating $\mathbf{S}_{\text{S-E/M}}$ frame in Fig. 28; the maximal $y$ component of the halo obit is ±33818.07km.

Only the intersecting pairs of transiting from the Earth to halo orbit and another intervals transiting from halo orbit to the Moon, i.e., $([85^0, 165^0]\text{U}[262^0, 330^0])\times[0, 1]$ and $([10^0, 92^0]\text{U}[192^0, 268^0])\times([0, 0.26]\text{U}[0.57, 0.66])$, can drive the trajectories to orbit successively the Earth and Moon, and is considered as the cislunar transfer opportunities shown in Fig.27. The points labeled as ***a***, ***b***, ..., ***n***, are mapped into 14 typical trans-lunar WSB trajectories in the rotating $\mathbf{S}_{\text{S-E/M}}$ frame, which can be produced by the following procedure: (*i*) integrate Eq.(4) backwards to obtain the Earth-escaping segment and forwards to achieve the Moon-captured segment in the rotating $\mathbf{S}_{\text{E-M}}$ frame both from the same initial condition of a serial point on the halo orbit; (*ii*) the transfer trajectories $\boldsymbol{R}_I$ in the $\mathbf{S}_{\text{S-E/M}}$ frame are converted from the integrated trajectories $\boldsymbol{r}$ in the rotating $\mathbf{S}_{\text{E-M}}$ frame, based on the following transition matrix of $\boldsymbol{R}_{\text{I}}=\boldsymbol{R}_x(-i)\boldsymbol{R}_x(-\beta)\boldsymbol{r}+\boldsymbol{A}_S$.

The initial conditions are listed as follows. The initial lunar phasic angle at the epoch time ($t$=0) is $\theta_{s0}=0^0$, and the initial condition $\boldsymbol{X}=[\boldsymbol{r},\dot{\boldsymbol{r}}]|_\tau$ is selected from a serial point with its phase $\tau$ on the halo orbit. All the subgraphs are produced by the halo orbit with its maximal $y$ component equal to ±33818.07km. The initial phase of serial points $\tau$ and the solar phasic angle $\beta$ at the epoch time ($t$=0) are: ***a***) $\tau$=345/360 and $\beta=10^0$; ***b***) $\tau$=350/360 and $\beta=33^0$; ***c***) $\tau$=177/360 and $\beta=96^0$; ***d***) $\tau$=13/360 and $\beta=13^0$; ***e***) $\tau$=360/360 and $\beta=53^0$; ***f***) $\tau$=352/360 and $\beta=35^0$; ***g***) $\tau$=352/360 and $\beta=33^0$; ***h***) $\tau$=52/360 and $\beta=205^0$; ***i***) $\tau$=24/360 and $\beta=194^0$; ***j***) $\tau$=220/360 and $\beta=13^0$; ***k***) $\tau$=62/360 and $\beta=191^0$; ***l***) $\tau$=51/360 and $\beta=205^0$; ***m***) $\tau$=187/360 and $\beta=89^0$; ***n***) $\tau$=189/360 and $\beta=89^0$.

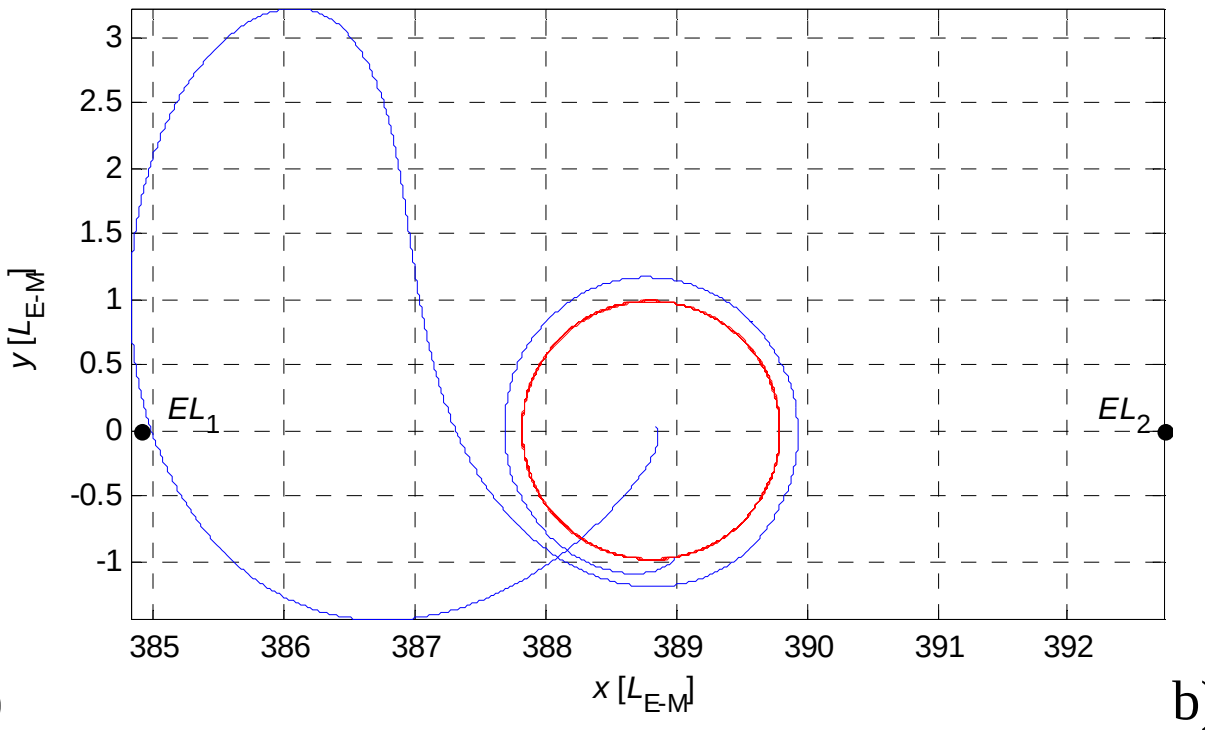

a)

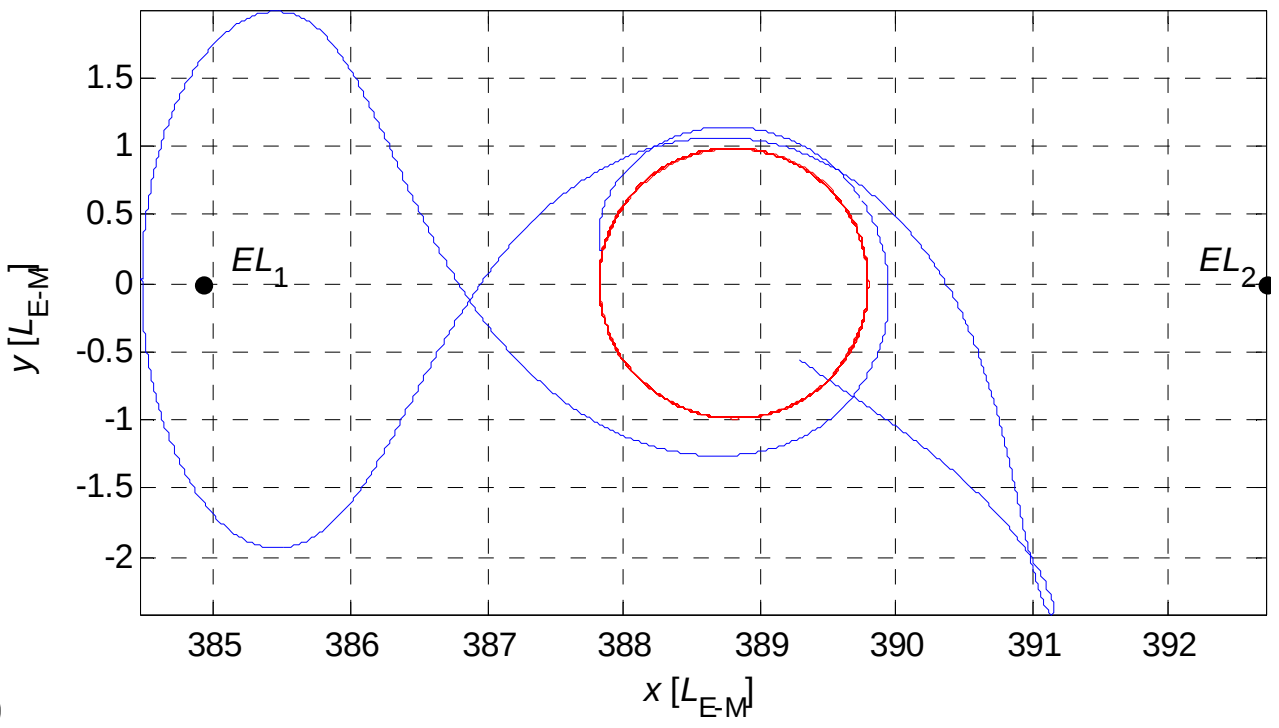

b)

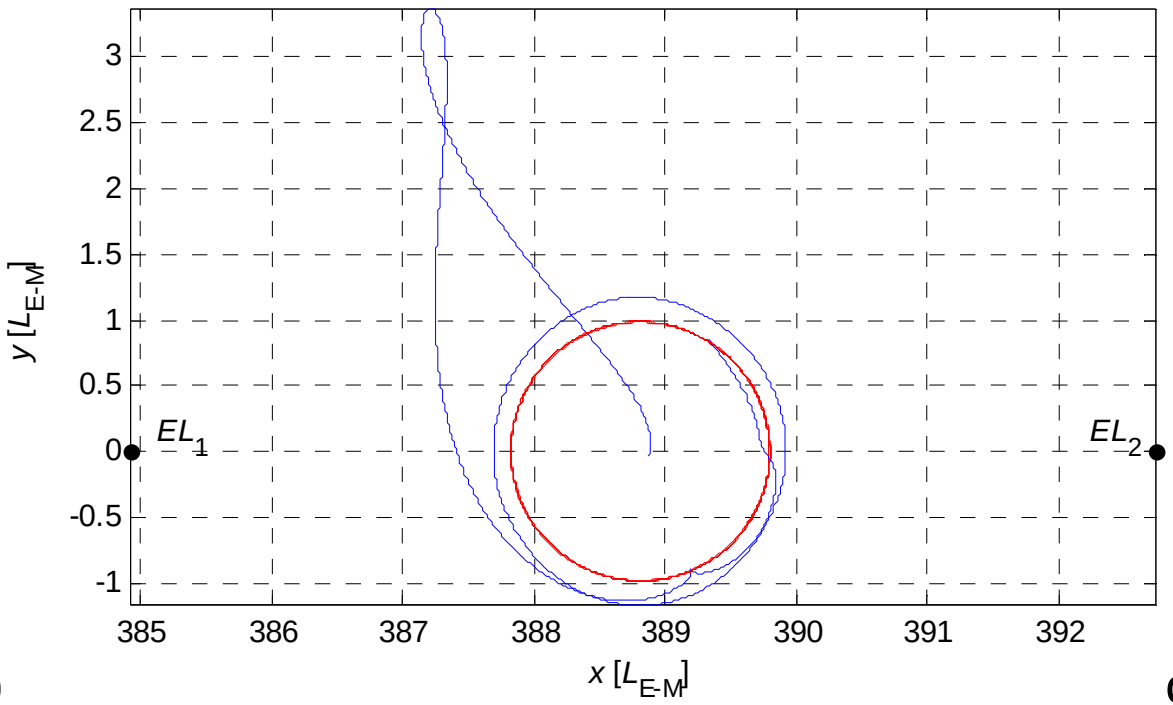

3
2.5
2
1.5
1
0.5
0
-0.5
-1
y [L_E-M]
EL1
EL2
385 386 387 388 389 390 391 392
x [L_E-M]

c)

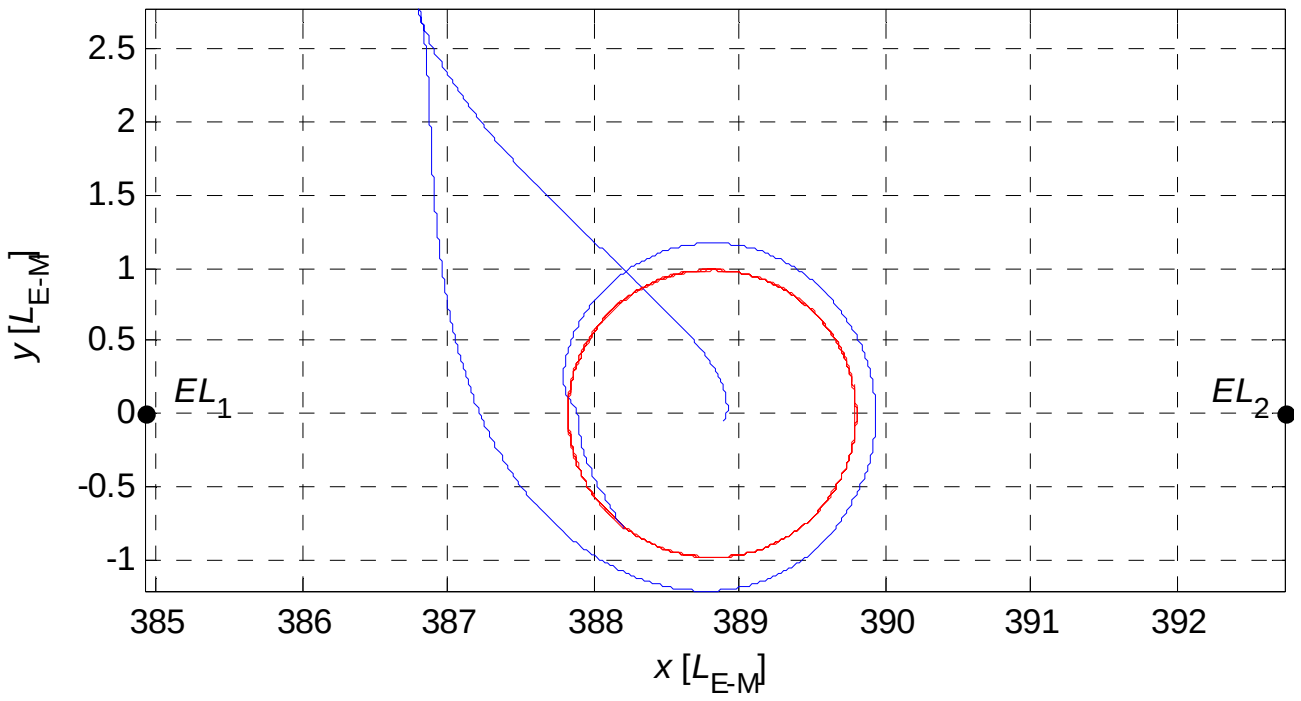

2.5
2
1.5
1
0.5
0
-0.5
-1
y [L_E-M]
EL1
EL2
385 386 387 388 389 390 391 392
x [L_E-M]

d)

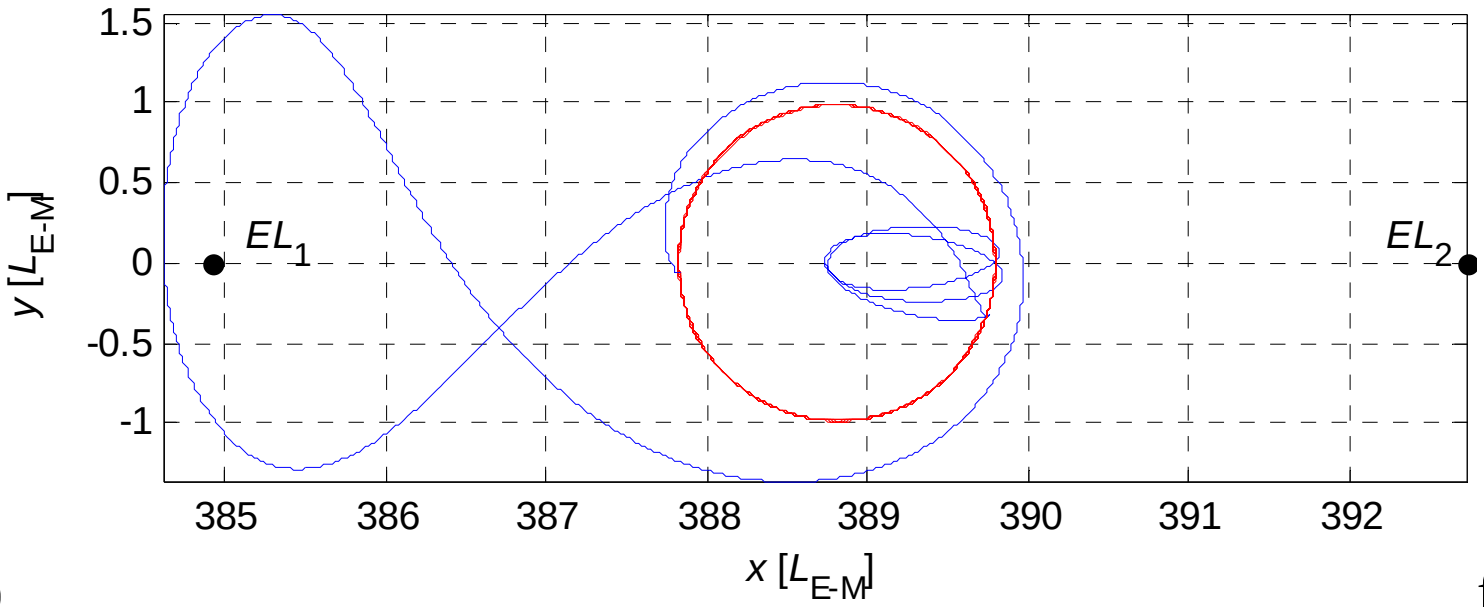

1.5
1
0.5
0
-0.5
-1
y [L_E-M]
EL1
EL2
385 386 387 388 389 390 391 392
x [L_E-M]

e)

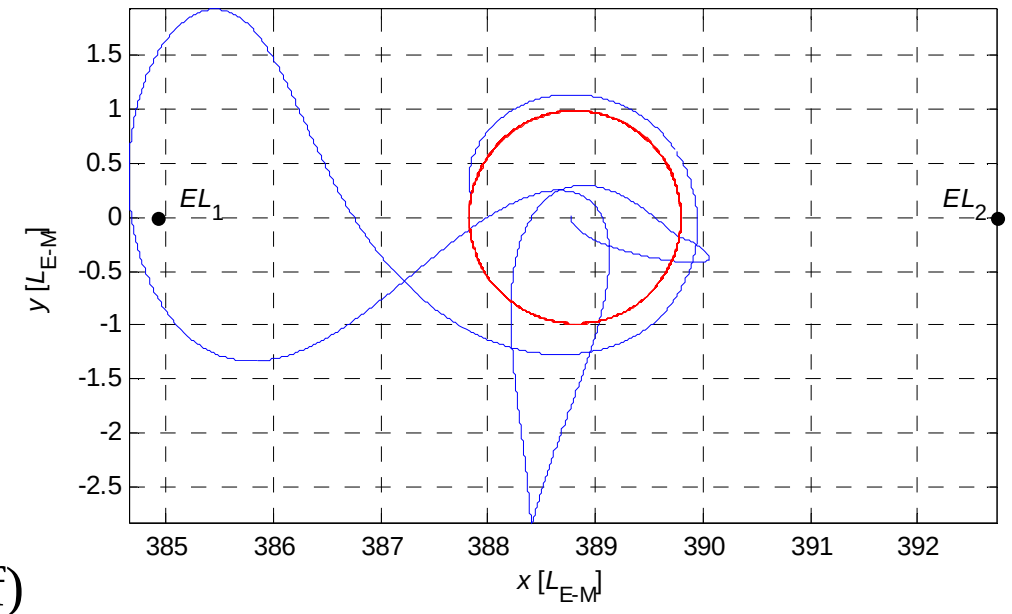

1.5
1
0.5
0
-0.5
-1
-1.5
-2
-2.5
y [L_E-M]
EL1
EL2
385 386 387 388 389 390 391 392
x [L_E-M]

f)

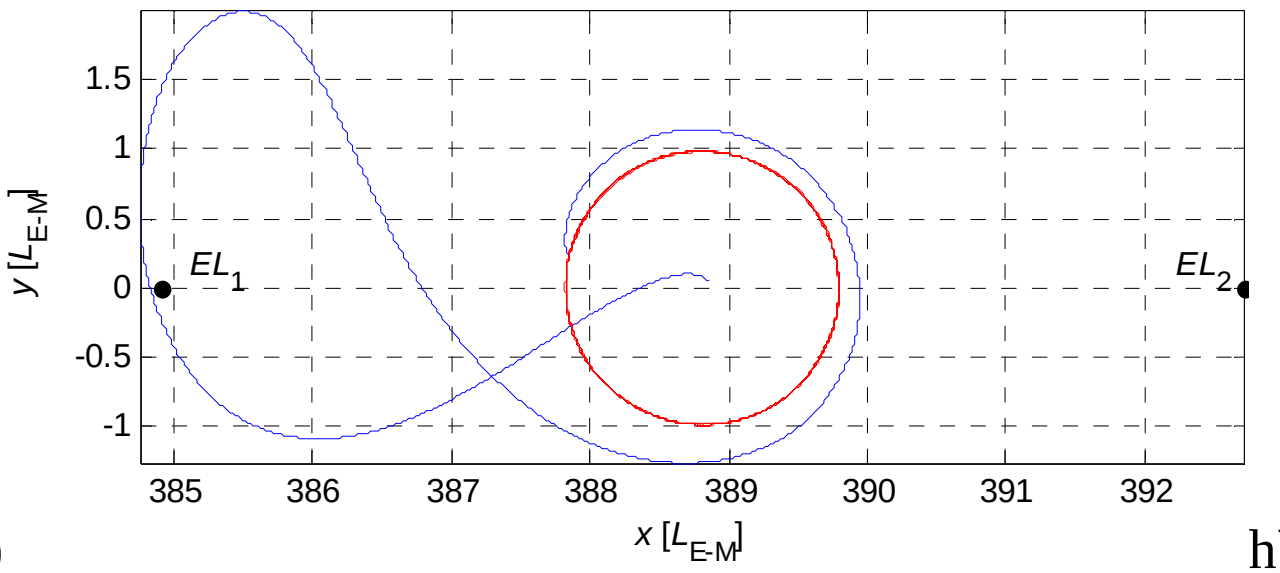

1.5
1
0.5
0
-0.5
-1
y [L_E-M]
EL1
EL2
385 386 387 388 389 390 391 392
x [L_E-M]

g)

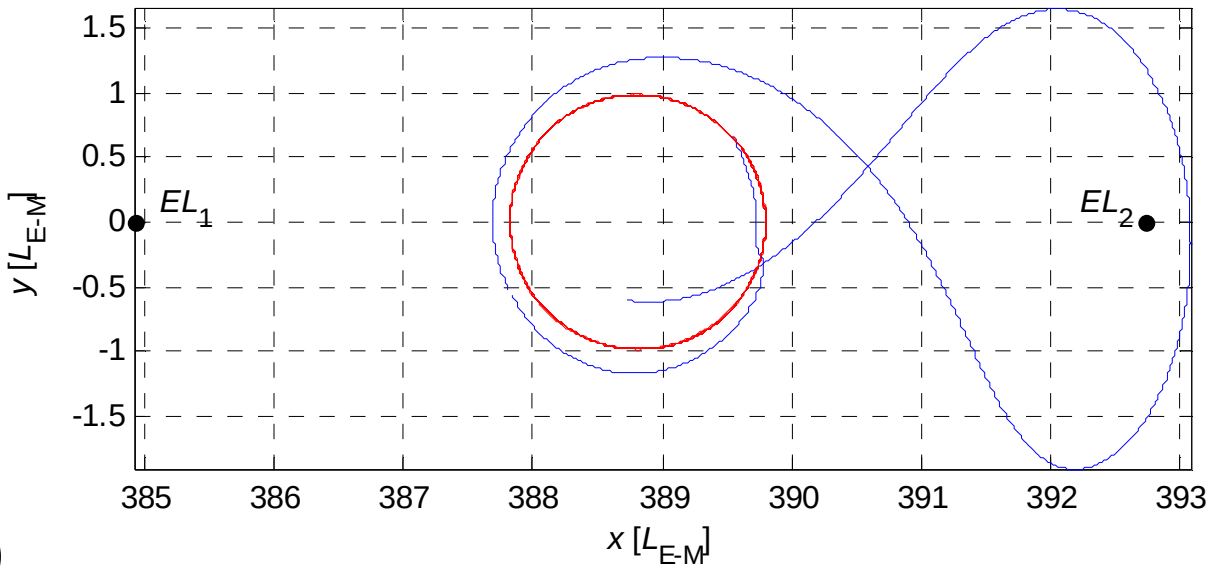

1.5
1
0.5
0
-0.5
-1
-1.5
y [L_E-M]
EL1
EL2
385 386 387 388 389 390 391 392 393
x [L_E-M]

h)

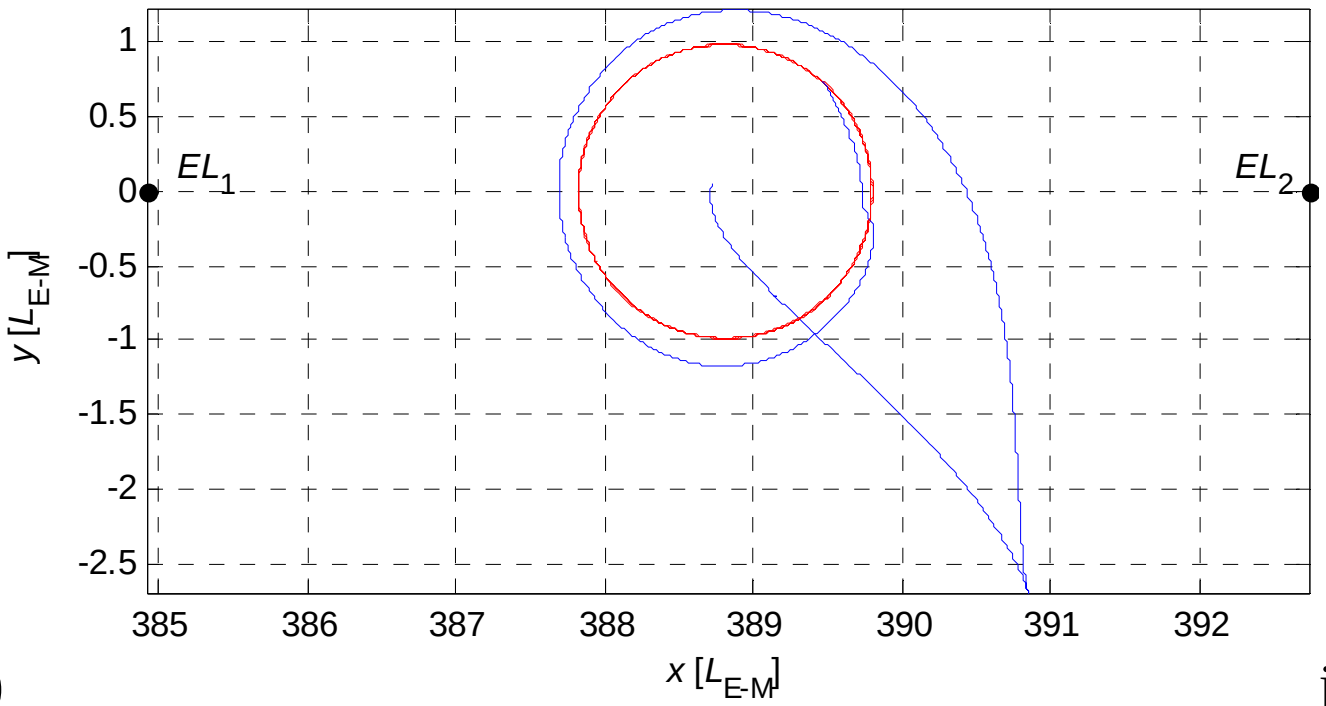

1
0.5
0
-0.5
-1
-1.5
-2
-2.5
y [L_E-M]
EL1
EL2
385 386 387 388 389 390 391 392
x [L_E-M]

i)

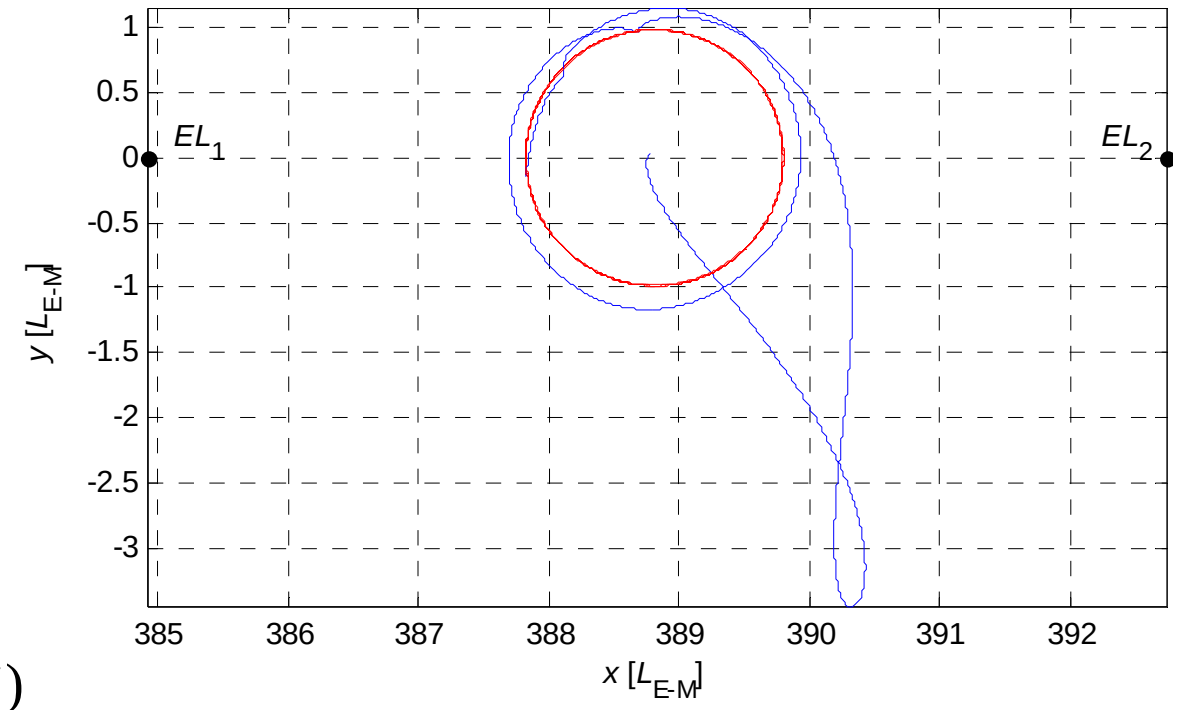

1
0.5
0
-0.5
-1
-1.5
-2
-2.5
-3
y [L_E-M]
EL1
EL2
385 386 387 388 389 390 391 392
x [L_E-M]

j)

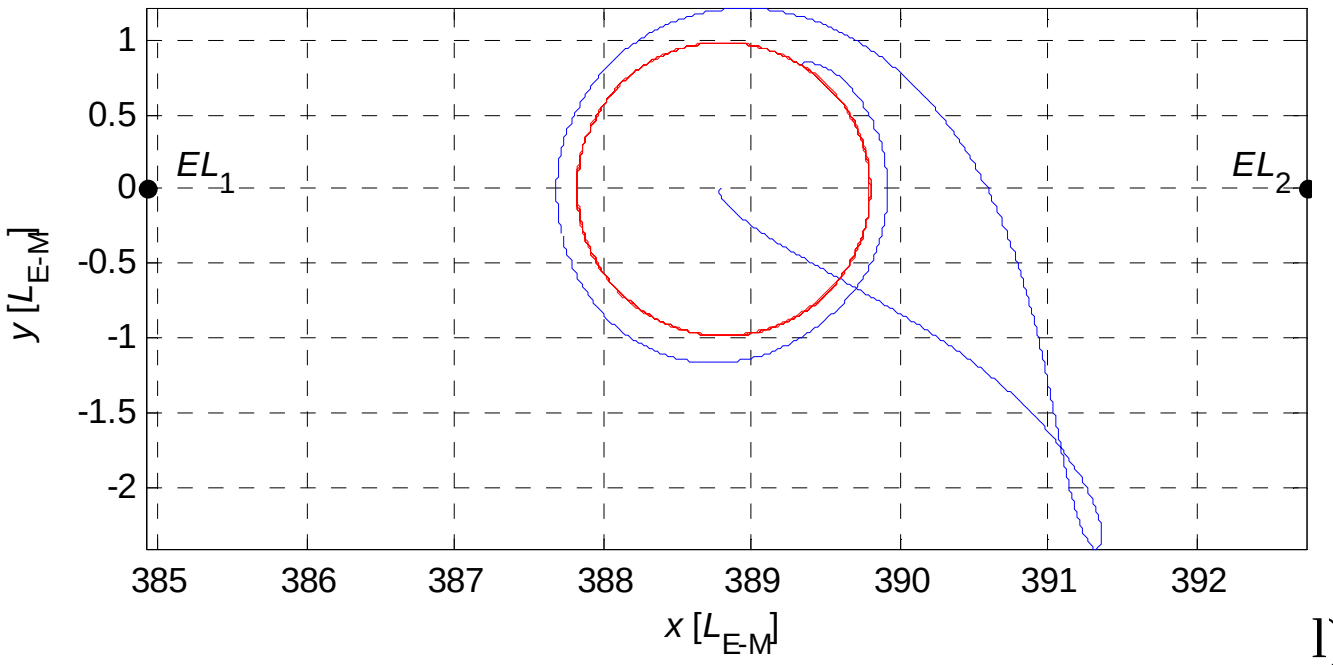

1
0.5
0
-0.5
-1
-1.5
-2
y [L_E-M]
EL1
EL2
385 386 387 388 389 390 391 392
x [L_E-M]

k)

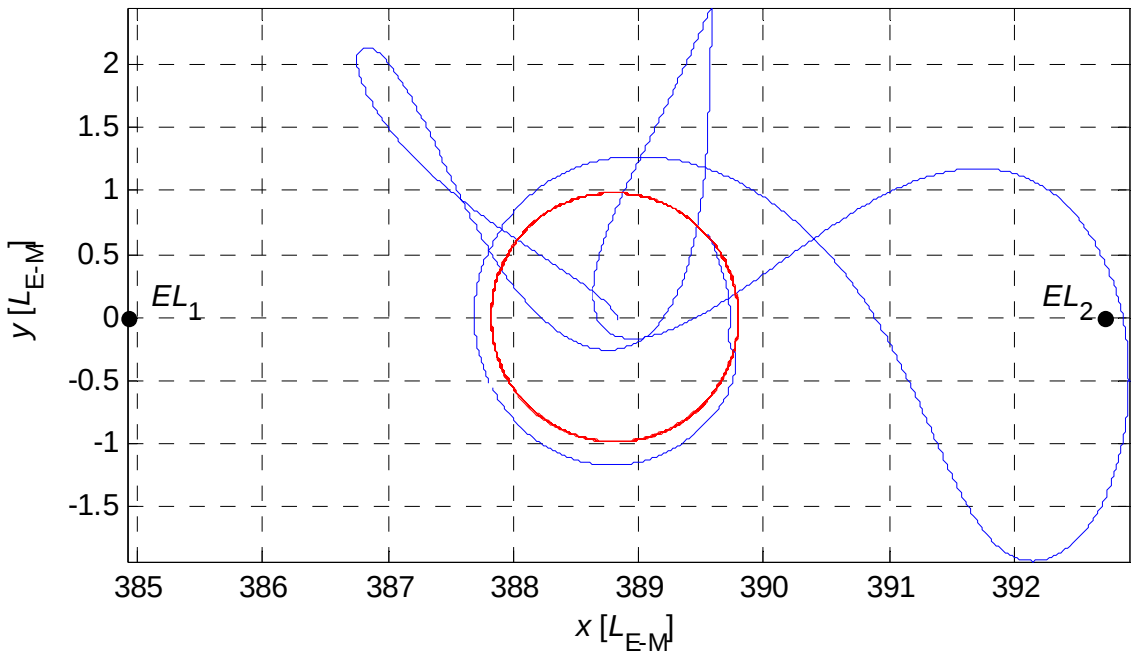

2
1.5
1
0.5
0
-0.5
-1
-1.5
y [L_E-M]
EL1
EL2
385 386 387 388 389 390 391 392
x [L_E-M]

l)

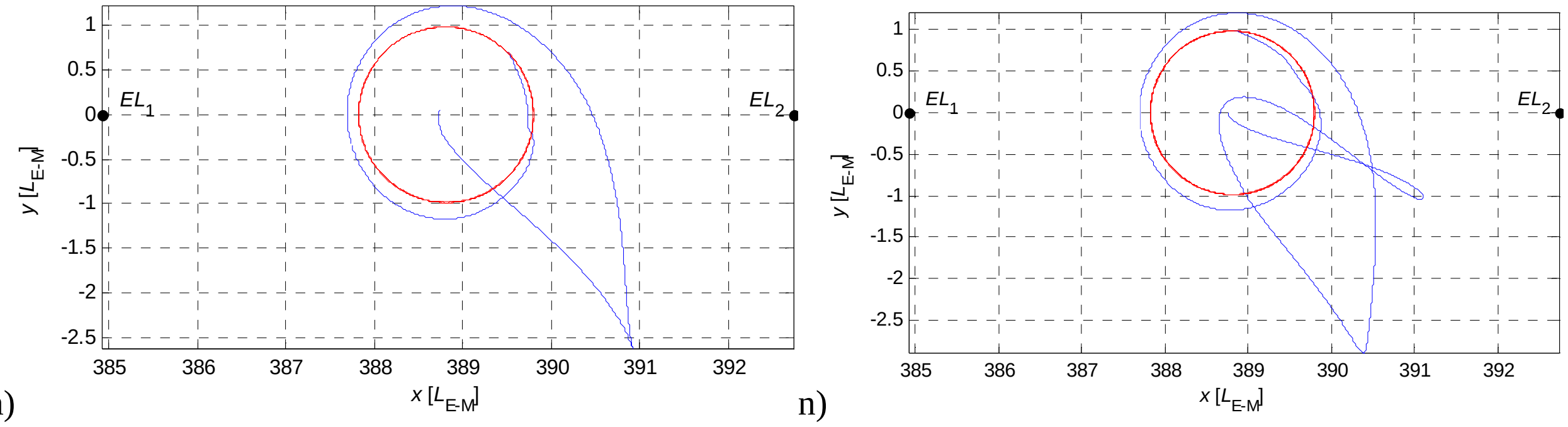

**Fig.28 Typical trans-lunar WSB trajectories in the $\mathbf{S_{S\text{-}E/M}}$ frame**: all the subgraphes correspond to the points labeled in Figs. 20 and 21; the phase of halo orbit and the solar phasic angle are selected as: ***a***) $\tau$=345/360 and $\beta$=10$^0$; ***b***) $\tau$=350/360 and $\beta$=33$^0$; ***c***) $\tau$=177/360 and $\beta$=96$^0$; ***d***) $\tau$=13/360 and $\beta$=13$^0$; ***e***) $\tau$=360/360 and $\beta$=53$^0$; ***f***) $\tau$=352/360 and $\beta$=35$^0$; ***g***) $\tau$=352/360 and $\beta$=33$^0$; ***h***) $\tau$=52/360 and $\beta$=205$^0$; ***i***) $\tau$=24/360 and $\beta$=194$^0$; ***j***) $\tau$=220/360 and $\beta$=13$^0$; ***k***) $\tau$=62/360 and $\beta$=191$^0$; ***l***) $\tau$=51/360 and $\beta$=205$^0$; ***m***) $\tau$=187/360 and $\beta$=89$^0$; ***n***) $\tau$=189/360 and $\beta$=89$^0$. ***a***, ***b***, ..., ***g*** are classified as the trans-lunar WSB trajectories passing through $EL_1$ point, and ***h***, ***i***, ..., ***n*** are classified as the ones passing through $EL_2$ point; the circles indicate the lunar surface; the maximal $y$ component of the halo obit is ±33818.07km.

Because of the conclusion in Section 2.2 that all the cislunar and trans-lunar trajectories have the $z$ component much smaller than the other components, only the $x$-$y$ view is presented for the labeled points, ***a***, ***b***, ..., ***n***. The 14 typical trajectories are classified as the ones passing through $EL_1$ point (i.e., ***a***, ***b***, ..., ***g***) and the others passing through $EL_2$ point (i.e., ***h***, ***i***, ..., ***n***). Thus, both $EL_1$ and $EL_2$ points can be employed to join with the unstable manifolds of a halo or a Lyapunov orbit near $LL_2$ point in driving the spacecraft from the Earth to the Moon, which is according to the Koon et al.'s conclusion (**Koon et al. 2001**). In the $\mathbf{S_{S\text{-}E/M}}$ frame, the Earth, $EL_1$ and $EL_2$ points are located respectively on the $\xi$ axis at 388.810, 384.918 and 392.728 based on the length unit normalization $L_{\text{E-M}}$ mentioned in Section 2.1. The temporarily captured segment of the trans-lunar trajectory has fewer loops orbiting the Moon but requires more energy than the cislunar one. Even so, the deceleration from the temporary capture to the permanent capture is small than Hohmann transfer.

## 5. Conclusion

The low-energy cislunar and WSB trajectories are investigated in this paper from the viewpoint of the cislunar libration point ($LL_1$) and trans-lunar libration point ($LL_2$), respectively. According to the geometry of instantaneous Hill's boundary, the equivalent $LL_1$ point is defined as the critical point connecting the two gravitational fields around the Earth and Moon, while the equivalent $LL_2$ point is defined as the critical point connecting interior and forbidden regions. The locations of the equivalent equilibria and their Hamiltonian values are solved from the partial derivative of the potential function with respect to the $x$ component.

The systematical discussion on the Moon-captured energy in the frame of a spatial analytical four-body model (i.e., SBCM) is implemented by the numerical Poincaré mapping, which is only focusing on the statistical features of the fuel cost and captured elements (like altitude of periapsis and eccentricity) rather than a specified Moon-captured segment.

The minimum-energy cislunar and trans-lunar trajectories are yielded by transiting $LL_1$ and $LL_2$ points respectively in Chapter 3. The trajectories transiting $LL_2$ point are classified into the inner cislunar type essentially passing through $LL_1$ point, and the outer WSB type connecting the invariant manifolds near $EL_1$ (or $EL_2$) point and unstable manifolds near $LL_2$ point. Moreover, it is demonstrated that the solar phasic angle $\beta$ has positive affects on the transfer opportunities: for the cislunar case transiting $LL_1$ point, a whole Earth-to-Moon transfer trajectory can be achieved only within the $\beta$s interval $[77^0, 109^0]U[285^0, 342^0]$; for the outer WSB case transiting $LL_2$ point, a whole Earth-to-Moon transfer trajectory can be achieved only within the $\beta$s interval $[21.8^0, 23.3^0]U[201.5^0, 203^0]$.

Compared with the only variable (i.e., $\beta$) to construct a transfer trajectory transiting the libration point, the halo orbit is employed to increase the transfer opportunities by introducing another variable (i.e., serial points of halo orbit). Subsequently, a global investigation on the Earth-escaping and Moon-captured opportunities is implemented for practical transfer trajectories transiting halo orbits near $LL_1$ and $LL_2$ points respectively. For the cislunar case transiting halo orbit near $LL_1$ point, a whole Earth-to-Moon transfer trajectory can be achieved only within the pairs $(\beta, \tau)$ of $([95^0, 150^0]U[262^0, 345^0])\times[0, 1]$ and $([100^0, 200^0]U[270^0, 30^0])\times([0.16, 0.26]U[0.47, 0.58])$; for the outer WSB case transiting halo orbit near $LL_2$ point, a whole Earth-to-Moon transfer trajectory can be achieved only within the pairs $(\beta, \tau)$ of $([85^0, 165^0]U[262^0, 330^0])\times[0, 1]$ and $([10^0, 92^0]U[192^0, 268^0])\times([0, 0.26]U[0.57, 0.66])$.

## 6 Acknowledgment

The authors are very grateful to the anonymous reviewer for helpful comments and suggestions on revising the manuscript. The research is supported by the National Natural Science Foundation of China (11172020), the National High Technology Research and Development Program of China (863

Program: 2012AA120601), Talent Foundation supported by the Fundamental Research Funds for the Central Universities, Aerospace Science and Technology Innovation Foundation of China Aerospace Science Corporation, and Innovation Fund of China Academy of Space Technology.